\documentclass[aps,floats,twocolumn,prb,showpacs,10pt,floatfix]{revtex4-1}
\usepackage{graphicx}
\usepackage{epstopdf}
\usepackage{hyperref}
\usepackage{amsmath}
\usepackage{amssymb}
\usepackage{grffile}
\usepackage{bm}
\usepackage{color}
\usepackage{enumitem}
\usepackage{todonotes}
\usepackage{dsfont}

\newcommand{\nop}[1]{}
\def\pM{\ensuremath{\genfrac{}{}{0pt}{1}{+}{\scriptstyle(\kern-1pt-\kern-1pt)}}}

\begin{document}

\title{\texorpdfstring{Non-perturbative landscape of the Mott-Hubbard transition:\\Multiple divergence lines around the critical endpoint}{Non-perturbative landscape of the Mott-Hubbard transition}}

\author{T. Sch\"afer$^a$, S. Ciuchi$^b$, M. Wallerberger$^a$, P. Thunstr\"om$^a$,\\O. Gunnarsson$^c$, G. Sangiovanni$^d$, G. Rohringer$^{a,e}$, and A. Toschi$^a$}

\affiliation{$^a$Institute of Solid State Physics, TU Wien, 1040 Vienna, Austria\\
             $^b$Dipartimento di Scienze Fisiche e Chimiche, Universit\`a dell'Aquila, and Istituto dei Sistemi Complessi, CNR, Via Vetoio I-67010 Coppito-L'Aquila, Italy\\
             $^c$Max Planck Institute for Solid State Research, D-70569 Stuttgart, Germany\\
             $^d$Institut f\"ur Theoretische Physik und Astrophysik, Universit\"at W\"urzburg, Am Hubland, D-97074 W\"urzburg, Germany\\
						$^e$Russian Quantum Center, Novaya street, 100, Skolkovo, Moscow region 143025, Russia}

\date{ \today }

\begin{abstract}
We analyze the highly non-perturbative regime surrounding the
Mott-Hubbard metal-to-insulator transition (MIT) by means of dynamical
mean field theory (DMFT) calculations at the two-particle level. By extending
the results of Sch\"afer, {\sl et al.} [Phys. Rev. Lett. {\bf 110}, 246405 (2013)] we show the existence of
{\sl infinitely many} lines in the phase diagram of the Hubbard model
where the local Bethe-Salpeter equations, and the related irreducible vertex functions, become singular in the charge as well
as the particle-particle channel. By comparing our numerical data for the Hubbard model 
with analytical calculations for exactly solvable systems of increasing complexity [disordered binary mixture (BM), Falicov-Kimball (FK) and atomic limit (AL)], 
we have (i) identified two different kinds of divergence lines; (ii) classified them in terms of the frequency-structure of the associated singular 
eigenvectors; (iii) investigated their relation to the emergence of multiple branches in the Luttinger-Ward functional.  
In this way, we could distinguish the situations where the multiple divergences simply reflect the emergence 
of an underlying, {\sl single} energy scale $\nu^*$ below which perturbation theory is no longer applicable,
 from those where the {\sl breakdown} of perturbation theory affects, not trivially, different energy regimes.
Finally, we discuss the
implications of our results on the theoretical understanding of the
non-perturbative physics around the MIT and for future developments of
many-body algorithms applicable in this regime.
\end{abstract}

\maketitle

\let\n=\nu \let\o =\omega \let\s=\sigma


\section{Introduction}

Many-particle perturbation theory and its graphical representation through the Feynman diagrammatics 
is particularly suited for the description of problems in quantum electrodynamics (QED)\cite{QED}, due to the natural identification of a small parameter 
for the perturbative expansion, the fine structure constant $\alpha=\frac{1}{137}$.
Despite significant formal similarities, the situation is conceptually different for the Feynman diagrammatics of
many electron systems in condensed matter. Here the bare Coulomb interaction between electrons and/or between electrons and ions does not define {\sl per se}  a small scale with respect to the electronic bandwidth. Thus, the applicability of perturbation expansions  for several classes of materials is  only justified
because of the efficiency of electronic density fluctuations in screening the bare Coulomb interaction: This is, ultimately, the microscopic reason behind the success of 
the Landau Fermi liquid (FL)\cite{LandauFermi} description of metals and of the $GW$\cite{GW} approach for semiconductors.  For the same reason, however, 
in all cases where electronic screening works poorly, 
such as, e.g.,  for the electrons of $3d$ and $4f$ transition metal oxides, a {\sl breakdown} of the perturbation expansion 
must be expected.

In spite of the importance of the phenomena occurring in these systems, ranging from Mott-Hubbard metal-insulator transitions (MITs)\cite{MH,ImadaREV}, 
to unconventional superconductivity\cite{Damascelli2003,Lee2006,Scalapino2012}, quantum criticality\cite{QCP}, emergence of kinks in spectral functions\cite{Byczuk2007,Raas2009} and specific heat\cite{Toschi2009,Held2013}, etc.,
only little is known about {\sl how} the breakdown of the  many-electron perturbation expansion actually manifests itself in the corresponding Feynman diagrammatics and  {\sl how} it is reflected in physical observables. 
This is far from being a merely academic issue,  because many\cite{DGA,DF,Kusunose,Multiscale,DB,1PI,DMF2RG,TRILEX,FLEXDMFT,QUADRILEX} of the cutting edge approaches recently developed to treat strongly correlated electrons in finite (three or two dimensional) systems largely exploit the Feynman diagrammatics and/or the Luttinger-Ward functional formalism.
First specific reports about unexpected theoretical consequences induced by the breakdown of the many-body 
perturbation expansion have been recently presented by several groups. Such effects range from the occurrence\cite{divergence,YangarXiv,Janis2014,Ribic2016} of low-frequency divergences of
 two-particle irreducible vertex functions in the Hubbard\cite{Hubbard} and Falicov-Kimball\cite{FKoriginal} models to the multivaluedness\cite{Kozik2015,EderComment,Stan2015,Berger2014,Lani2012,Rossi2015,Rossi2016} of the electronic self-energy
 expressed as a functional of the (interacting) Green's function or, equivalently, of the Luttinger-Ward functional. 
 All these manifestations of non-perturbativeness are interconnected and represent different aspects of the same problem, as suggested in some of the abovementioned works. 

In this paper, we aim at making progress towards a full understanding of such perturbation-theory breakdowns, by analyzing their several manifestations, mutual 
interplays, and physical meanings. For this goal, we will extend the pioneering work of Refs.~[\onlinecite{divergence,YangarXiv,Janis2014}], where, by applying the dynamical mean field theory (DMFT)\cite{DMFTREV}, singularities in the Bethe-Salpeter equations (and, thus, divergencies in the corresponding irreducible vertex functions)  were found in the case of  the single-band Hubbard and  Falicov-Kimball models.
More precisely, it was demonstrated that the local generalized susceptibility $\chi$, associated with the auxiliary Anderson impurity model (AIM) of the DMFT self-consistent solution, becomes a singular (i.e., not invertible) matrix in the (fermionic) Matsubara frequency space. This induces a divergence in the irreducible vertex $\Gamma$, defined by the inversion of the corresponding local Bethe-Salpeter equation. Such singularities were found in the charge (as well as particle-particle) scattering channel(s) of the FK (Hubbard) model, defining one (two) lines in their phase-diagrams.  Remarkably, for both models, the minimal values of the bare local Coulomb interaction $U$ where the singularities appear are, at low temperatures, significantly {\sl lower} than those for which the MIT is found in DMFT, i.e., they occur well inside the metallic phase of the models. It is worth noticing that the occurrence of analogous divergences has been recently reported\cite{Gunnarsson2016} also for the case of dynamical cluster approximation calculations (DCA)\cite{DCAREV} for the two-dimensional Hubbard model, demonstrating that their presence cannot be ascribed to artifacts of the purely local physics described by DMFT.

 Because of the intrinsic complexity of the subject, in this work we adopt the following strategy: We will analyze systematically, by means of DMFT, the occurrence of singularities in the generalized susceptibilities of different  model systems, where the correlated physics is introduced progressively, until, eventually, the full complexity of the Hubbard Hamiltonian is recovered. In particular, we will consider the cases of (i) disordered binary mixture (BM) and Falicov-Kimball (FK) models, where correlation effects are effectively induced by averaging over the impurity configurations, (ii) the atomic limit (AL) of the Hubbard model and (iii) the Hubbard model itself. This way, we exploit the opportunity of treating analytically several aspects of the divergences occurring in the simplified situations to gain a general understanding of their properties and mutual relations, such as their connection to the recently found multivaluedness of the Luttinger-Ward functional. In particular, we will demonstrate that frequency-localized divergences are associated with an underlying, unique, energy scale $\nu^{*}$, while this is not the case for the non-localized singularities.
\\\\
The paper is organized as follows. In Sec.~\ref{sec:model} all models
studied in this work are introduced in a general framework. Thereafter, we will recall the quantum many-body formalism at the
two-particle level, focusing on the specific equations needed for our analysis. In particular, at the end of this section, we 
will discuss the formal relation between irreducible vertex divergences and singularities of the generalized susceptibility matrix, as well as the
 properties of the corresponding singular eigenvectors. Moreover, we establish a formal connection between these singularities and a multivaluedness of the self-energy as a functional of the single-particle Green's functions. 
The analytical DMFT solution of  disordered models (i.e., binary mixture, Falicov-Kimball and distributed disorder) are discussed in Sec.~III, by investigating and classifying the corresponding divergences and their relations with the multivaluedness of the Luttinger-Ward formalism.
The analytically solvable atomic limit of the Hubbard model and its corresponding vertex divergences are analyzed in Sec.~\ref{sec:atomic}.
In Sec.~\ref{sec:hubbard} our DMFT numerical data for the whole phase diagram of the Hubbard
model are presented, extending the analysis of  Ref.~\onlinecite{divergence}.  Then, on the basis of the comparison with the results of the previous sections,  possible physical and algorithmic implications of the irreducible vertex divergences in the Hubbard model are discussed, before we, in Sec.~\ref{sec:conclu} draw our conclusions. Technical details about the
numerical calculations as well as analytical derivations are reported in several Appendices.

\section{Models and formalism}
\label{sec:model}
The starting point for our studies is the following lattice Hamiltonian, which encodes all models analyzed in the paper: 
\begin{equation}
 H=\!-\!\sum\limits_{<ij>,\sigma}t_{\sigma}{c^{\dagger}_{i\sigma}c_{j\sigma}} + \sum_{i,\sigma} \epsilon_i {c^{\dagger}_{i\sigma}c_{i\sigma}}+U\sum\limits_{i}{n_{i\uparrow}n_{i\downarrow}},
 \label{eqn:FKhubb}
\end{equation}
where $t_\sigma$ is the (spin-dependent) hopping amplitude for an electron with spin $\sigma$ between nearest neighbor lattice sites $i$ and $j$, $\epsilon_i$ is a random potential at site $i$, $U$ is the local Coulomb interaction, and $c^{\dagger}_{i\sigma}$ ($c_{i\sigma}$) creates (annihilates) an electron with spin $\sigma$ on site $i$. $n_{i\sigma}=c^{\dagger}_{i\sigma}c_{i\sigma}$ and $\beta=1/T$ denotes the inverse temperature. All calculations throughout this paper have been performed for the paramagnetic phase with $n=1$ electrons per site. Note that this filling corresponds to the most strongly correlated situation for the Hamiltonian in Eq.~(\ref{eqn:FKhubb}).

In the following $i\nu$ and $i\nu'$  will indicate fermionic and $i\Omega$ bosonic Matsubara frequencies, respectively. In particular, for a better readability, the imaginary unit $i$ will be explicitly written only in the argument of one-particle quantities, while it will be omitted in two-particle objects. Real frequencies of analytically-continued quantities will be denoted with $\omega$.  Finally, if correlation functions in a general situation are considered, we will indicate their frequency argument with the complex variable $z$, which can be either $i\nu$ or $\omega$.

\subsection{Choice of parameters}
\label{sec:2a}

The models of interest for this work are defined by  considering four distinct choices of the parameters $t_{\sigma}$, $\epsilon_i$ and $U$ in the Hamiltonian in Eq.~(\ref{eqn:FKhubb}): 

\noindent(i) the binary mixture disorder problem (BM) where $t_{\uparrow}=t_{\downarrow}=t$, $\epsilon_i=\pm W/2$ randomly distributed with equal probability (see Sec. \ref{sec:fk} and Appendix \ref{app:Anderson}) and $U=0$;

\noindent(ii) the (repulsive) Falicov-Kimball (FK) model\cite{FKoriginal} $t_\uparrow=t$,$t_\downarrow=0$, $\epsilon_i=0$ and $U>0$;

\noindent(iii) the atomic limit (AL) of the Hubbard model in which both $t_\uparrow=t_\downarrow=0$, $\epsilon_i=0$ and $U>0$;

\noindent(iv) the standard Hubbard model $t_\uparrow=t_\downarrow=t$, $\epsilon_i=0$ and $U>0$.

For (i) and (ii) (BM and FK) we derive mostly general expressions valid for an arbitrary non-interacting density of states (DOS, i.e., for any type of underlying lattice) of unitary half-bandwidth. In the specific situations, where explicit expression are needed, we have assumed the semi-elliptic DOS of a (infinite dimensional) Bethe-lattice where the unitary half-bandwidth corresponds to $2t\!=\!1$. Our DMFT calculations for the Hubbard model (iv) have been performed, consistent with previous studies\cite{divergence}, for a square lattice of unitary half-bandwidth ($4t=1$). This choice ensures that the non-interacting  DOS in $2d$ has the same standard deviation of our Bethe-lattice. For (i), (ii) and (iv) all energy scales will be given in units of the half-bandwidth, i.e., twice the standard deviation of the non-interacting DOS. 

We will study the Hamiltonian in Eq.~(\ref{eqn:FKhubb}) by means of DMFT for all four choices of the parameters (i)-(iv) [which for (ii), the AL, corresponds to the exact solution]. This way all purely local (temporal) fluctuations are included while non-local spatial correlations are treated on a mean-field level only. Nevertheless, this is sufficient to describe, non-perturbatively, the Mott-Hubbard transition \cite{DMFTREV,DMFTMIT,DMFTPT}. In standard DMFT applications typically one-particle (1P) quantities such as the self-energy and the spectral function allow for a clear-cut identification of the boundaries of the first-order Mott-Hubbard transition \cite{Bluemer_PhD, Zitzler, Bulla}. Here, however, we will focus mainly on the DMFT analysis of local {\sl two-particle} (2P) correlation functions, whose behavior is also strongly affected by the Mott transition\cite{RVT,georgthesis,Wentzel2016}. Hence, in the following, we will recapitulate  in Sec.~\ref{sub:2b} the basic definitions of all (local) two-particle quantities of interest in the framework of DMFT, adopting the same notation introduced in Ref.~\onlinecite{RVT} and in the supplementary material of Ref~\onlinecite{Gunnarsson2015}. 

As for the concrete evaluation of the local one- and two-particle correlation functions in the four different cases considered the calculations have been done analytically in the cases (i), (ii) and (iii) while for (iv) a Hirsch-Fye quantum Monte Carlo (HF-QMC)\cite{HirschFye} impurity solver has been adopted. Moreover, for obtaining data at extremely low temperatures, a continuous time quantum Monte-Carlo (CT-QMC) algorithm in hybridization expansion has been used\cite{Gull}. The accuracy of both codes has been also tested in selected cases
by means of a comparison with exact-diagonalization (ED) calculations.

\subsection{\texorpdfstring{The irreducible vertex $\Gamma_r$}{The irreducible vertex}}
\label{sub:2b}
The vertex function $\Gamma_r$, which is two-particle irreducible in a given scattering channel $r$, represents the central object of interest for this paper. It can be derived in two different ways: (i) One can calculate it from the two-particle Green's function (or, more precisely, from the generalized susceptibility) of the system by an inversion of the corresponding Bethe-Salpeter equation. (ii) Alternatively, it can be also obtained from purely one-particle quantities within the framework of the Luttinger-Ward functional\cite{LuttingerWard} formalism as the {\sl functional derivative} of the self-energy (functional) with respect to the Green's function of the system. Both methods (and their equivalence) have been extensively discussed in the literature\cite{Tremblay,Bickersbook}. However, as these two approaches provide different mathematical and physical insights into the mechanisms responsible for the divergence of the irreducible vertex $\Gamma_r$, we will briefly recapitulate them in the following two sections (for more details see Ref. \onlinecite{RVT}). 

\subsubsection{Generalized susceptibilities and Bethe-Salpeter equations}
\label{sec:2b1}
Let us start by defining the (purely local) generalized susceptibility in DMFT as \\
\begin{figure*}[t!]
	\centering
		\includegraphics[width=1.00\textwidth]{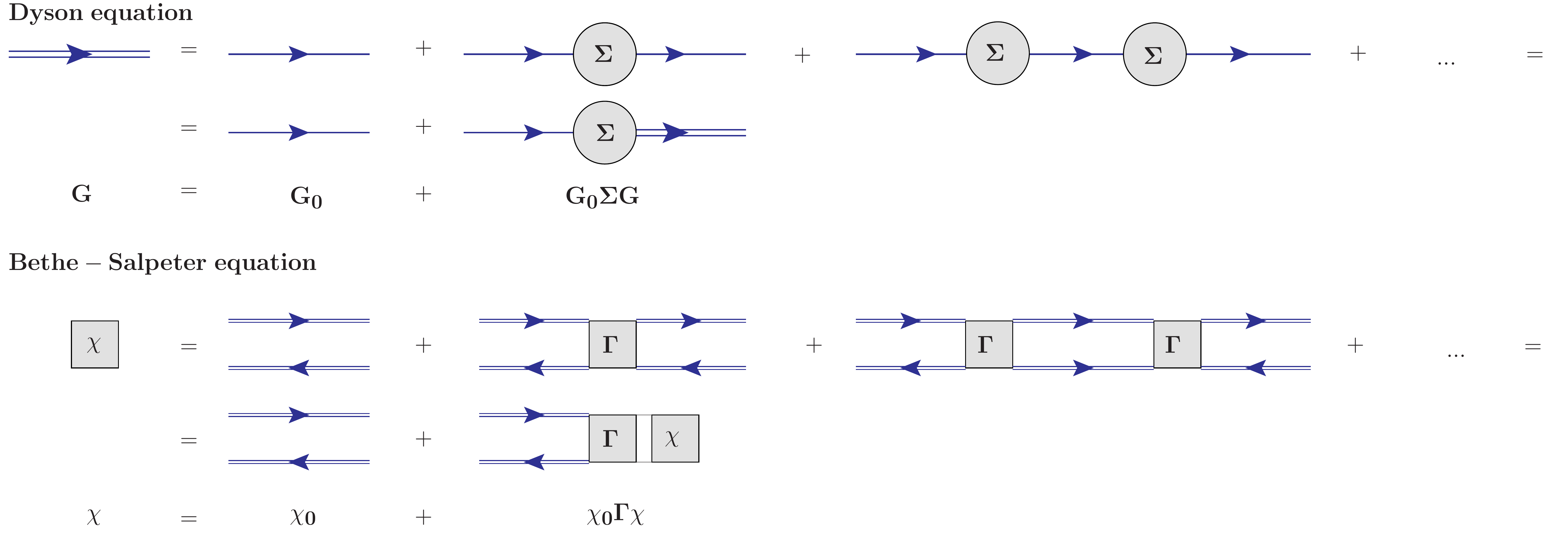}
	\caption{Dyson equation (upper rows) and Bethe-Salpeter equation (lower rows) represented in terms of Feynman diagrams and simplified, symbolic notation. The blue double lines with arrows represent interacting Green's functions ($G$), whereas single blue lines with arrows are non-interacting Green's functions ($G_{0}$). The self-energy is denoted by $\Sigma$. At the two-particle level, $\chi$ denotes the interacting generalized susceptibility and $\Gamma$ is the irreducible vertex in the selected channel of the Bethe-Salpeter equation.}
	\label{fig:dyson_bse}
\end{figure*}
\begin{equation}
 \begin{aligned}
\chi^{\nu\nu'\Omega}_{\text{ph},\sigma\sigma'}&=\int\limits_{0}^{\beta}{d\tau_{1}d\tau_{2}d\tau_{3} \, e^{-i\nu\tau_{1}}e^{i(\nu+\Omega)\tau_{2}}e^{-i(\nu'+\Omega)\tau_{3}}} \\
 &\times \big[\big<T_{\tau}c_{\sigma}^{\dagger}(\tau_{1})c_{\sigma}(\tau_{2})c_{\sigma'}^{\dagger}(\tau_{3})c_{\sigma'}(0)\big> \\
 &- \big<T_{\tau}c_{\sigma}^{\dagger}(\tau_{1})c_{\sigma}(\tau_{2})\big>\big<T_{\tau}c_{\sigma'}^{\dagger}(\tau_{3})c_{\sigma'}(0)\big>\big].
  \end{aligned}
 \label{eqn:ph}
\end{equation}
It can be directly computed through the impurity solver of choice. Here $T_{\tau}$ denotes the time-ordering operator and $\langle\ldots\rangle=\frac{1}{Z}\mbox{Tr}(e^{-\beta H}\ldots)$, where $Z=\mbox{Tr}(e^{-\beta H})$ is the thermal expectation value and $\beta=1/T$ is the inverse temperature. Note that the frequencies in Eq.~(\ref{eqn:ph}) are chosen according to the so-called particle-hole ($ph$) frequency-convention, which refers to the description of a scattering event between a particle (with frequency $\nu$) and a hole (with frequency $\nu'$) with a  bosonic transfer frequency $\Omega$ (see Ref.~\onlinecite{RVT}). Analogously, one can express the generalized susceptibility $\chi$ in its particle-particle representation which describes the scattering between two particles with a transferred frequency $\Omega$. The corresponding explicit functional form can be obtained from the $ph$-one in Eq.~(\ref{eqn:ph}) by a mere frequency shift, i.e., $\Omega\rightarrow\Omega-\nu-\nu'$ and, hence, $\chi_{pp,\sigma\sigma'}^{\nu\nu'\Omega}=\chi_{ph,\sigma\sigma'}^{\nu\nu'(\Omega-\nu-\nu')}$.

Summing the generalized susceptibilities over their fermionic Matsubara frequencies $\nu$ and $\nu'$ yields the physical response functions. Since such a response is defined w.r.t.~a physical perturbation, i.e., a chemical potential, a (staggered) magnetic or a pairing field, we consider in the following the generalized susceptibilities in the charge, spin and particle-particle-singlet scattering channels (or, more precisely, spin combinations)\cite{ppsusc}
\begin{subequations}
\label{subequ:channeldef}
\begin{align}
 \label{equ:channeldefph}
 &\chi^{\nu\nu'\Omega}_{\text{c(s)}}=\chi^{\nu\nu'\Omega}_{\text{ph},\uparrow\uparrow}\pM\chi^{\nu\nu'\Omega}_{\text{ph},\uparrow\downarrow}, \\
 \label{equ:channeldefpp}
 &\chi_{{\text{pair}}}^{\nu\nu'\Omega}=\chi_{pp,\uparrow\downarrow}^{\nu(\Omega-\nu')\Omega}-\chi_{pp,0}^{\nu\nu'\Omega},
\end{align}
\end{subequations}

where the bare susceptibility (``bubble'' term) is defined as
\begin{subequations}
\label{subequ:defchi0}
\begin{align}
 &\chi_{ph,0}^{\nu\nu'\Omega}=-\beta{G(\nu)G(\nu+\Omega)}\delta_{\nu\nu'},
 \label{eqn:chi0}\\
&\chi_{pp,0}^{\nu\nu'\Omega}=\chi_{ph,0}^{\nu\nu'(\Omega-\nu-\nu')}=-\beta G(\nu)G(\Omega-\nu)\delta_{\nu\nu'}\label{eqn:chi0_pp}.
\end{align}
\end{subequations}

The above defined generalized susceptibilities fulfill the so-called Bethe-Salpeter (BS) equations which read
\begin{equation}
 \pm\chi_{r}^{\nu\nu'\Omega}=\chi_{r,0}^{\nu\nu'\Omega}-\frac{1}{\beta^{2}}\sum\limits_{\nu_{1}\nu_{2}}{\chi_{r,0}^{\nu\nu_{1}\Omega}\Gamma_{r}^{\nu_{1}\nu_{2}\Omega}\chi_{r}^{\nu_{2}\nu'\Omega}},
 \label{eqn:bse_dm}
\end{equation}
where on the l.h.s. of this equation the $+$ sign accounts for the particle-hole channels ($r=c,s$) while the $-$ sign has to be considered for the particle-particle-singlet scattering channel ($r=\text{pair}$). 
The irreducible vertex $\Gamma_{r}^{\nu\nu'\Omega}$ represents the effective interaction between the electrons in the given scattering channel. 
To some extent, the BS equation can be regarded as a {\sl two-particle} level analogon of the Dyson equation 
\begin{equation}
 G(\nu) = G_{0}(\nu) + G_{0}(\nu)\Sigma(\nu)G(\nu),
 \label{eqn:dyson}
\end{equation}
with $G_{0}$ being the non-interacting Green's function and $\Sigma$ is the self-energy (irreducible {\sl one-particle} ``vertex''). Figure \ref{fig:dyson_bse} represents 
both equations in terms of Feynman diagrams and symbolic notations without indices and sums.

Considering $\chi_{r}$, $\chi_{r,0}$ and $\Gamma_{r}$ as matrices w.r.t. the (discrete) fermionic Matsubara frequencies $\nu$ and $\nu'$, e.g., $\boldsymbol{\chi_r}^{\Omega}\equiv\chi_r^{\nu\nu'\Omega}$, Eq.~(\ref{eqn:bse_dm}) can be represented as
\begin{equation}
 \label{equ:bse_mm}
 \pm\boldsymbol{\chi_r}^{\Omega}=\boldsymbol{\chi_{r,0}}^{\Omega}-\frac{1}{\beta^2}\boldsymbol{\chi_{r,0}}^{\Omega}\cdot\boldsymbol{\Gamma_r}^{\Omega}\cdot\boldsymbol{\chi_r}^{\Omega},
\end{equation}
where $\cdot$ denotes the matrix multiplication w.r.t. the fermionic frequencies. Obviously, $\boldsymbol{\Gamma_r}^{\Omega}$ can be obtained from $\boldsymbol{\chi_r}^{\Omega}$ by multiplying Eq.~(\ref{equ:bse_mm}) from the left with the inverse of $\boldsymbol{\chi_{r,0}}^{\Omega}$ and from the right with the inverse of $\boldsymbol{\chi_r}^{\Omega}$. This yields
\begin{equation}
 \label{equ:gammainv}
 \boldsymbol{\Gamma_r}^{\Omega}=\beta^2\left(\left[\boldsymbol{\chi_r}^{\Omega}\right]^{-1}\mp\left[\boldsymbol{\chi_{r,0}}^{\Omega}\right]^{-1}\right),
\end{equation}
where the $-$ sign has to be taken for $r=c,s$ while the $+$ sign corresponds to $r=\text{pair}$. $[\ldots]^{-1}$ denotes a matrix inversion w.r.t. $\nu$ and $\nu'$.

Eq.~(\ref{equ:gammainv}) already allows for some general statements about the occurrence of divergences in the irreducible vertex $\Gamma_r^{\nu\nu'\Omega}$: (i) Let us first note that $\boldsymbol{\chi_{r,0}}^{\Omega}$ is a diagonal matrix in $\nu,\nu'$ whose (diagonal) elements are given by the product of two Green's functions [see Eqs. (\ref{eqn:chi0},\ref{eqn:chi0_pp})] and, hence, are always $\ne 0$ in the metallic phase at finite $T$ (except for the trivial limit $\nu,\nu'\rightarrow \infty$). Consequently, a divergence of $\Gamma_r^{\nu\nu'\Omega}$ corresponds to a singular behavior of the matrix $\boldsymbol{\chi_r}^{\Omega}$, i.e., to a vanishing  eigenvalue of this matrix. (ii) The eigenvector $V_{0,\nu}^{\Omega}$ which corresponds to such a vanishing eigenvalue $\lambda_0^{\Omega}=0$ provides some general information about {\sl how} $\Gamma_r^{\nu\nu'\Omega}$ diverges. To this end we consider the spectral representation\cite{noteMarkus}  of the inverse of $\boldsymbol{\chi_r}^{\Omega}$
\begin{equation}
 \label{equ:chiinvspec}
 [\boldsymbol{\chi_r^{\Omega}}]^{-1}_{\nu\nu'}=\sum_{i} (V_{i,\nu'}^{\Omega})^*(\lambda_i^{\Omega})^{-1} V_{i,\nu}^{\Omega},
\end{equation} 
where the sum over $i$ runs over all eigenvectors/eigenvalues of $\boldsymbol{\chi_r}^{\Omega}$. Clearly, for $\lambda_0^{\Omega}\!=\!0$ the inverse of $\boldsymbol{\chi_r}^{\Omega}$ and, hence, $\Gamma_r^{\nu\nu'\Omega}$, diverges. The fermionic frequencies, however, for which such a singularity can be observed in the irreducible vertex is determined by the structure of the eigenvectors. Specifically, according to Eq.~(\ref{equ:chiinvspec}) the divergence of $(\lambda_0^{\Omega})^{-1}$ will lead to a corresponding divergence in $\Gamma_r^{\nu\nu'\Omega}$ only at frequencies $\nu, \nu'$ for which the corresponding eigenvector has {\sl finite} weight, i.e., where $V_{0,\nu}^{\Omega}\ne 0$ and $V_{0,\nu'}^{\Omega}\ne 0$. Consequently, if the singular eigenvector is localized in frequency space, that is $V_{0,\nu}^{\Omega}\ne 0$ only for a finite set $\mathds{D}=\{\nu|V_{0,\nu}^{\Omega}\!\ne\!0\}$ of Matsubara frequencies, $\Gamma_r^{\nu\nu'\Omega}$ diverges only at those frequencies: We thus observe a  frequency {\sl localized} divergence, occurring only for $\nu,\nu'\!\in\!\mathds{D}$. In the opposite case, i.e., for a ``full-entry'' eigenvector with non-zero components at {\sl all} Matsubara frequencies, a {\sl global} divergence will affect, simultaneously, the {\sl entire} $\Gamma_r^{\nu\nu'\Omega}$ (at the given value of $\Omega$). \\

\subsubsection{\texorpdfstring{$\Gamma_r$ from a functional derivative}{The irreducible vertex from a functional derivative}}
\label{sec:2b2}

Complementary to the derivation in the previous section, the irreducible vertex $\Gamma_r$ can be also obtained as the functional derivative of the self-energy w.r.t.~the one-particle Green's function. To this end, the single-particle Green's function $G$ {\sl must} be calculated, in principle, in the most general situation by including symmetry-breaking external fields, which violate, e.g., time translational invariance and SU(2) symmetry and, hence, give rise to self-energies and Green's functions depending on {\sl two} (rather than one) fermionic Matsubara frequencies, {\sl two} spin indices, etc. After the functional derivative is performed, the resulting irreducible vertex function will be evaluated at zero external field which restores all conservation laws of the problem. Hence, we can write in a (partly) symbolic notation\cite{FreericksRMP}
\begin{equation}
 \label{equ:funcder}
 \Gamma_{\sigma\sigma'}^{\nu\nu'\Omega}=\beta\frac{\delta\Sigma_{\sigma}(\nu,\nu+\Omega)}{\delta G_{\sigma'}(\nu',\nu'+\Omega)}.
\end{equation}
This way one obtains the irreducible vertices $\Gamma_{\uparrow\uparrow}^{\nu\nu'\Omega}$ and $\Gamma_{\uparrow\downarrow}^{\nu\nu'\Omega}$ which can be combined to the corresponding irreducible vertices in the spin and charge channel, respectively, according to Eq.~(\ref{equ:channeldefph}). 

Similarly as for the particle-hole channels, i.e., charge and spin, one can calculate the irreducible particle-particle vertex $\Gamma_{\text{pair}}$ by means of a functional derivative. To this end one has to introduce an external pairing field (which breaks gauge symmetry and, hence, particle conservation) and perform the derivative of the anomalous self-energy w.r.t. the anomalous Green's function evaluated at zero field\cite{Bickersbook}.

Let us note that for the very simple disorder systems discussed in Sec. \ref{sec:bmfk} (BM and FK models), we can consider a simplified version of Eq.~(\ref{equ:funcder}) which avoids the cumbersome introduction of symmetry breaking fields and, hence, non-diagonal self-energies and Green's functions: $\Sigma(\nu,\nu+\Omega)\rightarrow \Sigma(\nu,\nu)\equiv \Sigma(\nu)$ and the same for $G$. Evidently, with this restriction, we can only compute the irreducible vertex for $\Omega=0$. However, in this work we are mainly interested to study the divergences of $\Gamma_r^{\nu\nu'\Omega}$ at $\Omega=0$, and, hence, we  can indeed exploit the simpler relation
\begin{equation}
 \label{equ:funcdersimple}
 \Gamma_{\sigma\sigma'}^{\nu\nu'(\Omega=0)}=\beta\frac{\delta\Sigma_{\sigma}(\nu,\nu)}{\delta G_{\sigma'}(\nu',\nu')}=\beta\frac{\delta\Sigma_{\sigma}(\nu)}{\delta G_{\sigma'}(\nu')},
\end{equation}
for the determination of the irreducible vertex for the disorder models in Sec. \ref{sec:bmfk}. Let us, however, stress that for more complex, fully interacting systems as the AL (Sec. \ref{sec:atomic}) and the Hubbard model (Sec. \ref{sec:hubbard}), Such a simplification is {\sl not} applicable, which prevents, there, an easy calculation of $\Gamma_{\text{r}}$ as a functional derivative.

Finally, we present here an intuitive general argument how the vertex divergences can be directly related to a {\sl multivaluedness} of the functional $\Sigma[G]$ (and, hence, of the Luttinger Ward functional). Adopting a mere symbolic notation (neglecting also the factor $\beta$), we realize that $\Gamma\!=\!\delta\Sigma/\delta G\!=\!\infty$ implies that the inverse functional $G[\Sigma]$ has a zero derivative at this point, i.e., $\delta G/\delta\Sigma\!=\!\Gamma^{-1}\!=\!0$, where the functionals are to be evaluated at their respective physical arguments after taking the derivatives. Therefore, if $\Gamma$ diverges, $G[\Sigma]$ is stationary at the physical $\Sigma$. Excluding the possibility of a saddle point, $G[\Sigma]$ will, hence, not be injective in the vicinity of the physical self-energy. As a consequence $\Sigma[G]$ cannot be single-valued. These considerations establish a generic connection between divergences of the irreducible vertices and a multivaluedness of the functional $\Sigma[G]$. 

\section{binary mixture and Falicov-Kimball model}
\label{sec:bmfk}

We start our analysis of the occurrence, the general properties and the possible implications of the irreducible vertex divergences in DMFT by considering their simplest realization in disordered models. In fact, while these describe non-interacting electrons in the presence of impurity scattering centers, the action of averaging over different (random or ensemble) distributions of the impurities, renders the electrons subjected to scattering events, which can be interpreted to some extent in terms of a simplified interacting problem.

In particular, in this section, we will consider the infinite dimensional\cite{Dinfty} (DMFT) solution for the  binary mixture (BM) and the Falicov-Kimball (FK) model, see Eq.~(\ref{eqn:FKhubb}) in Sec.~II and the related discussion.

Let us, at this point, briefly elaborate on the similarities and differences between these two models. In fact, this issue has sometimes caused confusion in the literature, but it is of significant importance to our study. In particular, the immobile $\downarrow$-electrons in the FK model can be considered just as scattering potentials for the mobile $\uparrow$ electrons. The former, hence, act in a similar way as the random scattering potential $\epsilon_i$ in Eq.~(\ref{eqn:FKhubb}). In fact, when identifying $\epsilon_i=\pm U/2$ (i.e., $U=W$) one obtains the same {\sl one-particle} Green's function for both models in the infinite dimensional limit which corresponds to the coherent potential approximation (CPA)\cite{CPA} for the BM and DMFT for the FK model. 
However, let us point out that, in the BM case, the scattering potential is {\sl purely external} and, hence, independent of the system itself. On the contrary, in the FK model, the scattering potential is generated by the immobile $\downarrow$-electrons which are in thermal equilibrium with the mobile $\uparrow$-electrons and, hence, not independent of the latter.

As a consequence, at the {\sl two-particle} level specific differences emerge between the two models due to the intrinsic differences in the nature of the disorder: Specifically, the isothermal response of the systems, and hence its static susceptibility ($\Omega=0$), differs\cite{Wilcox}  due the absence/presence of a thermal-ensemble averaging of the scattering potential in the BM/FK (quenched/annealed disorder), respectively. In spite of this difference,  we will demonstrate that the specific vertex anomalies found in the BM at $\Omega=0$ can be also observed in the FK model. The latter, however, allows in addition for the realization of a {\sl second} class of divergences not present in the BM.  

In the rest of the section, we will prove: 

\noindent(i) that the DMFT solution of the $T-W(U)$-phase-diagrams, of both BM and FK models, displays infinitely many lines along which $\Gamma_c^{\nu\nu'\Omega}$ diverges for $\Omega=0$ (i.e., where the matrix $\boldsymbol{\chi_c}^{\Omega=0}$ is singular, see Sec. \ref{sec:2b1}); 

\noindent(ii) that at $T=0$ all divergence lines accumulate at a unique interaction value ($W=U=\frac{1}{\sqrt{2}}$ for the Bethe lattice\cite{divergence,Janis2014}), definitely {\sl lower than the one of the corresponding MIT ($W=U=1$)};

\noindent(iii) that, for all divergence lines of BM (and of the first kind in FK), these singularities appear in $\Gamma_c^{\nu\nu'(\Omega=0)}$ only at one given Matsubara frequency (and its negative value) $\pm\bar{\nu}$ and are, according to the discussion in Sec. \ref{sec:2b1}, associated with an eigenvector of the generalized susceptibility with weight only at exactly these values ($\pm\bar{\nu}$), which allows for a classification of the singularities in terms of this frequency;

\noindent(iv) that all divergence lines of the BM (and of first kind in FK) model do collapse onto a {\sl unique} line by rescaling them with the Matsubara-factor $2n-1$, defining a {\sl unique energy scale} $\nu^*(W(U))$ associated with all divergences;

\noindent(v) that this energy scale coincides with the one at which the physical branch of the (multivalued) self-energy functional $\Sigma[G]$ no longer coincides with the perturbative one and where, hence, the physical branch of the (multivalued) Luttinger-Ward functional is no longer approximated by a perturbation expansion in $U$\cite{Kozik2015};

\noindent(vi) that $\nu^{*}$ exactly coincides with the frequency for which Im~$G(i\nu)$ is stationary (i.e., $\frac{d}{d\nu}$Im~$G(i\nu)=0$), suggesting a direct relation between the divergences of $\Gamma_{\text{c}}$ and the tendency towards a spectral gap formation.

\subsection{DMFT divergences in the binary mixture model}
\label{sec:BMdiv}

\subsubsection{\texorpdfstring{The irreducible vertex $\Gamma_c$ for the BM}{The irreducible vertex for the BM}}
\label{sec:irrvertBM}

In order to derive analytical expressions for the irreducible vertex divergences of the BM model, we will exploit the definition of $\Gamma_c$ as a functional derivative of the self-energy $\Sigma$  w.r.t.~$G$ (see Sec.~\ref{sec:2b2}). This is possible here because of the simple CPA expression for the DMFT Green's function of the BM model, which, at half-filling, is related to the local $G_0(z)$ via
\begin{equation}
G(z)=\frac{1}{2}\left ( \frac{1}{G^{-1}_0(z)+\frac{W}{2}}+\frac{1}{G^{-1}_0(z)-\frac{W}{2}}\right ).
\label{eq:FKGG0}
\end{equation}
where $G_0(z)=[z-\Delta(z)]^{-1}$, with $\Delta(z)$ being the (self-consistently determined) DMFT hybridization function. As mentioned in Sec.~\ref{sec:model}, $z$ corresponds either to an imaginary (Matsubara) or a real frequency, i.e., $z=i\nu$ or $\omega$. The chemical potential has been set to $\mu=0$ corresponding to the half-filled solution. We also note that all the equations derived in this section, unless explicitly stated, are generally valid for an arbitrary non-interacting DOS.

The corresponding (local) self-energy $\Sigma(z)$ is related to the local Green's function by the Dyson equation
\begin{equation}
G=\frac{1}{G^{-1}_0-\Sigma}.
\label{eq:FKGSigma}
\end{equation}
where we have omitted the frequency argument $z$ in all quantities.

Combining Eqs. (\ref{eq:FKGG0}) and (\ref{eq:FKGSigma}) we get
\begin{equation}
\Sigma[G_0]=\frac{W^2}{4} G_0
\label{eq:FKSigmaG0}
\end{equation}
showing that the self-energy is a single-valued functional of $G_0$.

On the other hand, eliminating $G_0$ from Eqs. (\ref{eq:FKGG0}) and (\ref{eq:FKGSigma}) leads to a second order equation for $\Sigma$ which reads
\begin{equation}
 \label{equ:sig2ndord}
 \Sigma^2+G^{-1}\Sigma-\frac{W^2}{4}=0.
\end{equation}
This equation has {\sl two} independent solutions\cite{Stan2015}:
\begin{equation}
\Sigma^{\pm}[G]= \frac{\pm\sqrt{1+W^2G^2}-1}{2G}.
\label{eq:FKSigmaG}
\end{equation}
This shows that -in general- {\sl two} different self-energies $\Sigma$ correspond to a single given $G$, i.e., $\Sigma[G]$ is not single-valued. We will discuss this issue in detail in Sec.~\ref{sec:mult} showing its intimate relation to divergences of the irreducible charge vertex $\Gamma_c$. For the moment we will ignore this non-single-valuedness of $\Sigma[G]$ and proceed with the calculation of $\Gamma_c$ considering both solutions for $\Sigma[G]$. 

For the BM, $\Gamma_c$ can be easily obtained by exploiting the Luttinger-Ward formalism [see Eqs.~(\ref{equ:funcder}) and (\ref{equ:funcdersimple})]. In fact, Eq.~(\ref{eq:FKSigmaG}) represents the exact functional $\Sigma[G]$ for the CPA solution of the BM model. In the specific situation at hand, the functional $\Sigma[G]$ is local in both, the frequency (either on the imaginary or the real axis) and the spin domains, i.e., reduces to just a mere {\sl function} of $G$. Hence, the resulting vertex will be proportional to $\delta_{\nu\nu'}\delta_{\sigma\sigma'}$. This means that the $\uparrow\downarrow$ vertex function vanishes and, according to Eq.~(\ref{equ:channeldefph}), $\Gamma_c=\Gamma_{\uparrow\uparrow}$. Performing now the derivative of $\Sigma^{\pm}[G]$ in Eq.~(\ref{eq:FKSigmaG}) w.r.t. to $G$ explicitly, we obtain:
\begin{equation}
 \label{eq:FKGamma}
 \Gamma_{c,\pm}^{\nu\nu'(\Omega=0)}=\beta\delta_{\nu\nu'}\frac{\sqrt{1+W^2G^2}\mp 1}{2G^2\sqrt{1+W^2G^2}},
\end{equation}
which has been written here for imaginary Matsubara frequencies, i.e., $G=G(i\nu)$. The corresponding result for real frequencies can be easily obtained by replacing $i\nu\rightarrow \omega+i\delta$ and $\beta\delta_{\nu\nu'}\rightarrow 2\pi\delta(\omega-\omega')$. Similarly as for the self-energy, we obtain two solutions $\Gamma_{c,+}$ and $\Gamma_{c,-}$ whose properties will be discussed later in Sec. \ref{sec:mult}.

\begin{figure}[t!]
	\centering
        \includegraphics[width=0.50\textwidth]{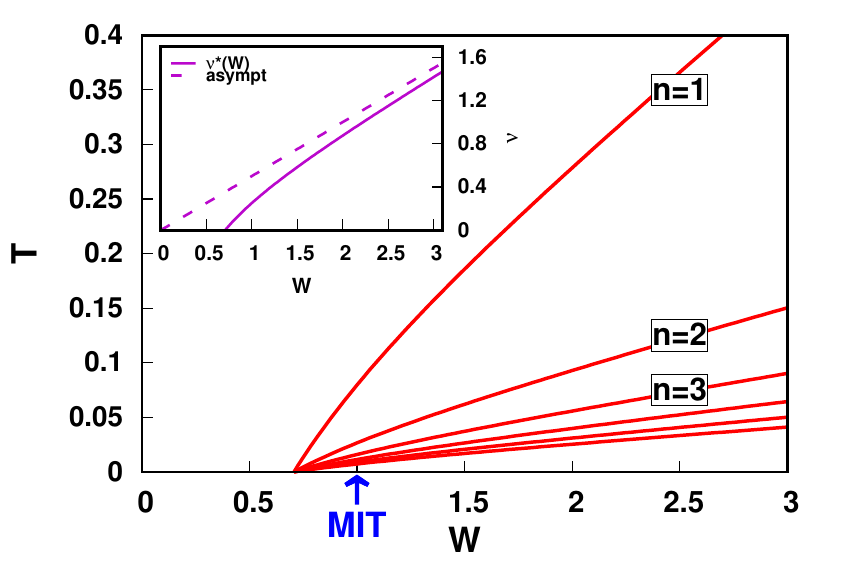}
	\caption{Phase diagram of the BM model (with a Bethe-lattice DOS), showing the divergence lines for $\Gamma_{c,\pm}$. The index $n$ corresponds to the index of the Matsubara frequency $\nu=\pi T(2n+1)$ for which the divergence appears at a given temperature $T$ and disorder strength $W$. Inset: The energy scale $\nu^*(W)$ at which the divergence appears for a given $W$. Note that the same phase-diagram is also applicable to the frequency-localized divergences of the FK model.}
	\label{fig:phase_diagram_FK}
\end{figure}

From Eq.~(\ref{eq:FKGamma}) it is clear that if
\begin{equation}
1+W^2G^2=0,
\label{eq:FKdivergence}
\end{equation}
the vertex diverges. Note that this happens for both $\Gamma_{c,+}$ and $\Gamma_{c,-}$. It should be also stressed that the divergence of $\Gamma_{c,\pm}$ in Eq.~(\ref{eq:FKdivergence}) is local in the frequency domain, which is a consequence of the local dependence of this irreducible vertex on $G$. Hence, the divergence of the vertex will occur at a {\sl single} (positive and the corresponding negative) Matsubara frequency for which condition (\ref{eq:FKdivergence}) is fulfilled\cite{note1}. Moreover, let us recall that the single-particle Green's function for the BM (and the FK model at half-filling) within CPA (DMFT) does {\sl not} depend on the temperature (apart from the explicit T-dependence of the Matsubara frequencies): This means that for a given value of $W$ the CPA (DMFT) Green's function of the system is a {\sl universal} function $G(z)$ which, for a given temperature, must be evaluated at the Matsubara frequencies $z=i\nu=i\pi T(2n-1)$. Therefore, for each $W$, the condition (\ref{eq:FKdivergence}) is fulfilled at a single energy $z=i\nu^*$ which, thus, defines an energy scale $\nu^*(W)$ associated with the divergence of $\Gamma_{c,\pm}$. Exploiting the DMFT result for the BM on a Bethe lattice this energy scale, i.e., the function $\nu^*(W)$, can be determined analytically (for details see Appendix \ref{sec:scaleanalytic}), yielding 
\begin{equation}
 \label{equ:scale}
 \nu^*(W)=\frac{2W^2-1}{4W},
\end{equation}
in energy units of the half-bandwidth. Let us stress that Eq.~(\ref{equ:scale}) is valid for positive values of $\nu^*$ only (see Appendix \ref{sec:scaleimag}): If the r.h.s. of this relation becomes smaller than $0$ the energy scale -and with it the divergence of $\Gamma_{c,\pm}$- disappears as it can be seen in the inset of Fig.~\ref{fig:phase_diagram_FK}.

Consequently, a ``critical'' $W$-value ($\widetilde{W}$) is found for the existence of a vertex divergence and a corresponding scale $\nu^*$. The value of $\widetilde{W}$ is obtained by setting $\nu^*=0$ in Eq.~(\ref{equ:scale}), which yields a critical disorder strength of $\widetilde{W}=1/\sqrt{2}$ for the Bethe-lattice, consistent with previous studies\cite{divergence, Janis2014, Stan2015} at $T=0$. Please note that $\widetilde{W}<W_{\text{MIT}}=1$, which supports the physical interpretation\cite{divergence,Janis2014} of the vertex divergences as a precursor of the MIT. 

The above discussion makes clear that, at finite temperatures, $\Gamma_{c,\pm}$ diverges whenever a Matsubara frequency $\nu$ becomes equivalent to $\nu^*$ and the divergence occurs exactly at this frequency. Indeed, this fully explains the structure of the divergence lines observed in the BM model, depicted in the main panel of Fig.~\ref{fig:phase_diagram_FK}. Let us now, for a fixed value of $W>\widetilde{W}$, consider a rather high temperature where all Matsubara frequencies are larger than the energy scale $\nu^*(W)$. Upon decreasing the temperature the corresponding values of the single Matsubara frequencies decrease and hence, the first, second, third, $\ldots$ Matsubara frequency will successively match the energy scale $\nu^*(W)$. This proves the existence of an infinite number of divergences for each $W>\widetilde{W}$ (accompanied by a divergence of $\Gamma_{\text{c},\pm}$ at this frequency). Considering the evolution of this infinite number of divergences with $W$, one obtains an infinite number of divergence {\sl lines}, defined by the condition $\nu=\pi T(2n-1)=\nu^*$, yielding in the case of a Bethe-lattice DOS
\begin{equation}
\label{equ:divlines}
 T_n(W)=\frac{1}{2n-1}\frac{2W^2-1}{4\pi W}.
\end{equation}
The corresponding lines $T_n(W)$ are shown in the phase-diagram in Fig.~\ref{fig:phase_diagram_FK}. At the critical value $W=\widetilde{W}=1/\sqrt{2}$ they collapse to the same point at $T=0$. Moreover, as they reflect the existence of a single energy scale, they are {\sl exactly rescalable} to a single line by multiplying the r.h.s. of Eq.~(\ref{equ:divlines}) by the Matsubara factor $2n-1$.

As discussed in Sec.~(\ref{sec:2b1}), a divergence of $\Gamma_c$ is intrinsically related to a vanishing eigenvalue of the generalized susceptibility $\chi_c^{\nu\nu'\Omega}$, from which the irreducible vertex can be calculated via Eq.~(\ref{equ:gammainv}). For the BM, similarly as $\Gamma_{c,\pm}$, also the corresponding susceptibility is diagonal in the fermionic frequencies $\nu$ and $\nu'$. A straightforward evaluation of Eq.~(\ref{eqn:ph}) for the BM in CPA yields [together with Eq.~(\ref{eq:FKSigmaG})], for $\chi_{c}=\chi_{\uparrow\uparrow}$:
\begin{equation}
\chi_{c,\pm}^{\nu\nu'(\Omega=0)}=\underset{E_{\text{BM}}(\nu)}{\underbrace{-\frac{2\beta}{W^2}\sqrt{1+W^2G^2}\left(\sqrt{1+W^2G^2}\mp 1\right)}}\delta_{\nu\nu'}.
\label{eq:BMchi}
\end{equation}  
Let us mention that, similar as the self-energy $\Sigma^{\pm}$ and the irreducible vertex $\Gamma_{c,\pm}$, $\chi_{c,\pm}$ in Eq.~(\ref{eq:BMchi}) is also multivalued, since we have expressed it as a function of $G$ (rather than $\Sigma$ or $G_0$), see Sec. \ref{sec:mult}.

Because of the diagonal form of $\chi_{c,\pm}^{\nu\nu'(\Omega=0)}$, the prefactor of $\delta_{\nu\nu'}$ on the r.h.s. of Eq.~(\ref{eq:BMchi}) represents exactly the eigenvalues $E_{BM}(\nu)$ of this generalized susceptibility. If the $n$-th Matsubara frequency $\nu$ matches the energy scale $\nu^*(W)$ the related eigenvalue $E_{\text{BM}}(\nu=\nu^*)$ vanishes as $1+W^2G^2(\nu^*)=0$ [see Eq.~(\ref{eq:FKdivergence})]. This corresponds exactly to the condition for the divergence of $\Gamma_c$ in Eq.~(\ref{eq:FKdivergence}) illustrating the simultaneous realization of both phenomena (i.e., the vanishing of an eigenvalue of $\chi_{c,\pm}$ and divergence of $\Gamma_{c,\pm}$ discussed in Sec. \ref{sec:2b1}). The corresponding eigenvectors are -due to the diagonal form of $\chi_{c,\pm}$- completely localized in frequency space. According to the discussion in Sec. \ref{sec:2b1}, this is indeed consistent with the fact that $\Gamma_{c,\pm}$ diverges at a {\sl single} (positive and the corresponding negative) Matsubara frequency. Evidently, because of the even symmetry of $E_{\text{BM}}(\nu)= E_{\text{BM}}(-\nu)$ for the half-filled system, the eigenvalues are always twofold degenerate. Specifically, the eigensubspace spanned by the eigenvectors corresponding to a specific eigenvalue $E_{\text{BM}}(\bar{\nu})$ [for a fixed $\bar{\nu}=(2\bar{n} -1)\pi T$] can be defined by the linear combination $A \delta_{\nu\bar{\nu}} + B\delta_{-\nu\bar{\nu}}$ (with $A^2+B^2=1$).

\subsubsection{Multivaluedness of correlation functions on the Matsubara axis}
\label{sec:mult}

\begin{figure*}[t!]
    \includegraphics[width=0.45\textwidth]{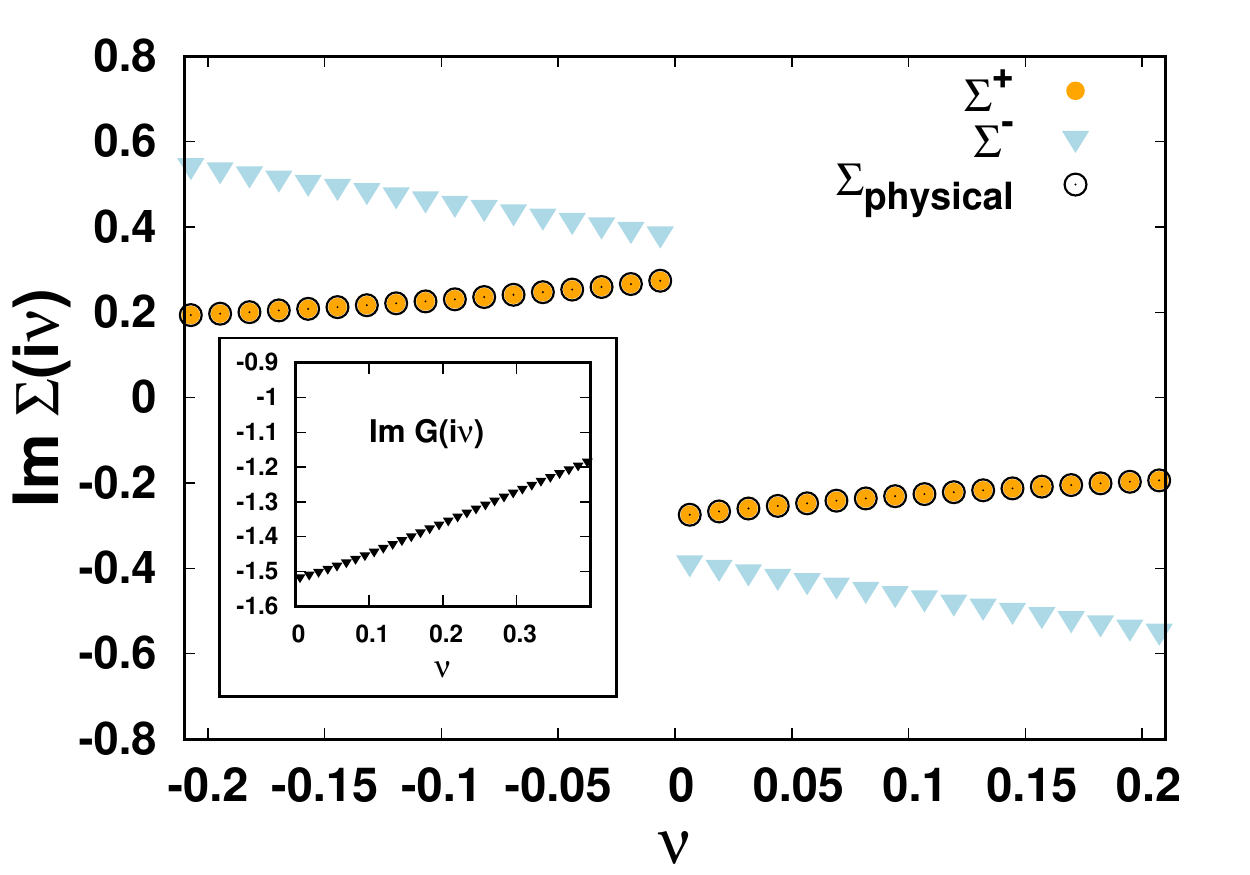} \hspace{2mm} \includegraphics[width=0.45\textwidth]{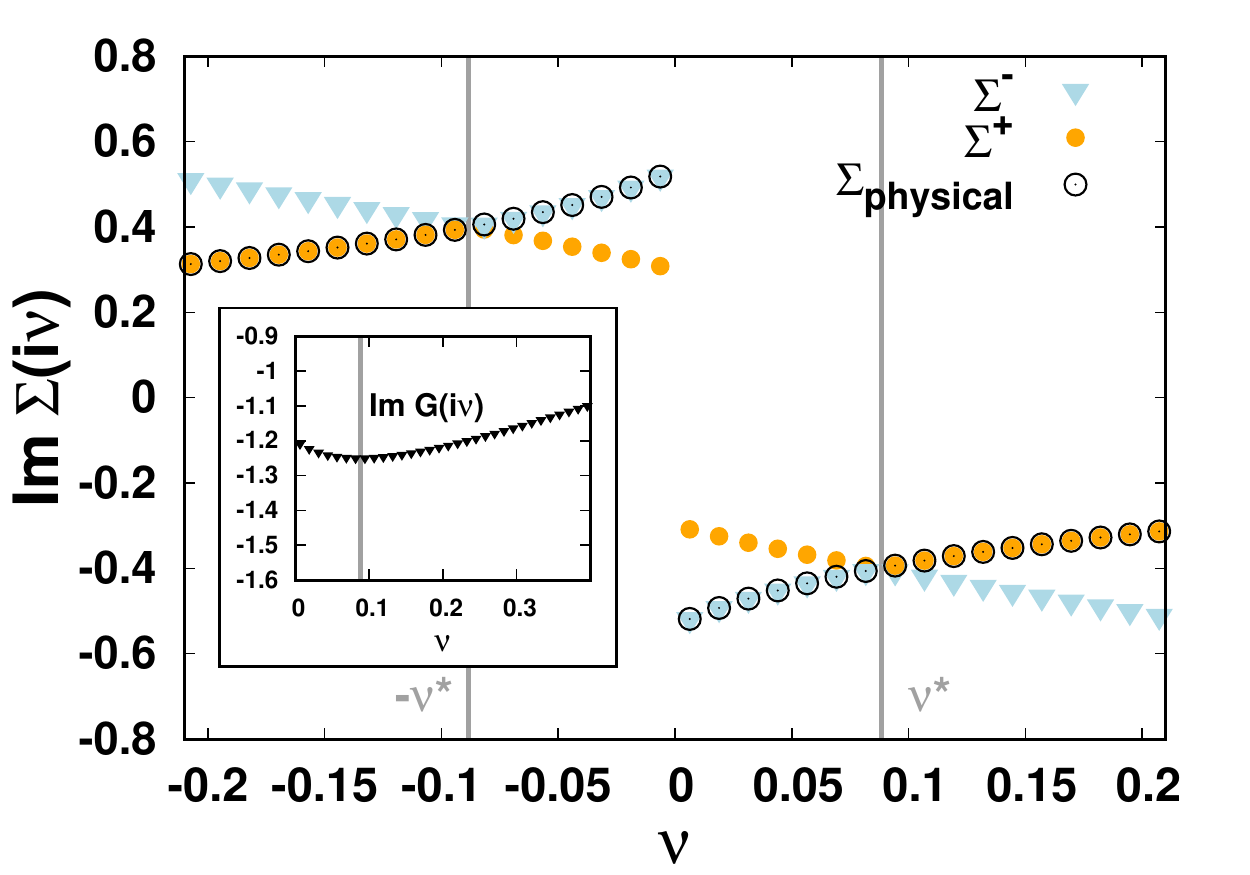}
    \caption{Imaginary parts of $\Sigma^\pm(i\nu)$ and $\Sigma(i\nu)=\Sigma_{\text{phys}}(i\nu)$ for the half-filled BM (and FK) model on the Bethe lattice at $T=0.002$. Left panel: $W=0.65<\widetilde{W}$; Right panel: $W=0.8>\widetilde{W}$. For $W>\widetilde{W}$, at the frequency $\nu^*(W)$, the physical self-energy changes from $\Sigma^-$ to $\Sigma^+$. Insets: Imaginary parts of the corresponding (physical) Green's functions $G(i\nu)$. For $W<\widetilde{W}$, $G(i\nu)$ is monotonic (for $\nu>0$) while for $W>\widetilde{W}$ it exhibits a well-defined minimum (or maximum in absolute values $\lvert G(i\nu)\rvert$) at $\nu=\nu^*$.}
    \label{fig:FKsigma}
\end{figure*}

Let us now discuss the multivaluedness of $\Sigma$, $\Gamma_c$ and $\chi_c$ found in the previous section as well as their relation to the vertex divergences. In particular, we aim at identifying the physical branch of the two solutions for these correlation functions.

 To achieve this  goal,  we compute the two possible determinations of the self-energy, i.e., $\Sigma^+$ and $\Sigma^-$, provided by  Eq.~(\ref{eq:FKSigmaG}) and compare them to the physical one, which can be readily extracted from Eqs.~(\ref{eq:FKGSigma}) and (\ref{eq:FKSigmaG0}). In the left panel of Fig.~\ref{fig:FKsigma}  the three corresponding quantities, i.e., $\Sigma^+$, $\Sigma^-$ and $\Sigma=\Sigma_{\text{phys}}$, are shown as a function of the Matsubara frequencies for $W=0.65<\widetilde{W}$. In this case the physical self-energy coincides (at all frequencies) with $\Sigma^+$. This is indeed consistent with the expectations from perturbation theory at this small value $W$: According to Eq.~(\ref{eq:FKSigmaG}) $\Sigma^+$ vanishes for $W\rightarrow 0$ while $\Sigma^-$ approaches the {\sl finite} value $-1/(2G)$ in this limit.  

However, the plausible expression for $\Sigma^+$ does not always
correspond to the physical self-energy of the BM. In fact, for $W>\widetilde{W}$
it starts to be gradually replaced as physical solution by the non-perturbative
$\Sigma^-$. As one can see (right panel of Fig. \ref{fig:FKsigma}) the
latter becomes indeed the physical self-energy for small Matsubara frequencies.
For a given $W>\widetilde{W}$ the frequency at which
$\Sigma^+$ and $\Sigma^-$ exchange their roles as physical self-energy
corresponds exactly to the energy scale $\nu^*(W)$ at which the irreducible
vertex $\Gamma_c$ diverges. This constitutes a definite proof that the vertex
divergences and the multivaluedness of the self-energy (and, hence, also of the
Luttinger-Ward) functional, recently found in Refs.~\onlinecite{Kozik2015,Stan2015}, are intimately related as already anticipated in Sec. \ref{sec:2b2}. Let us stress that for
frequencies $\nu>\nu^*$, the physical solution is still given by the
perturbative branch, i.e., by $\Sigma^+$. This is consistent with the fact that
the asymptotic behavior of $\Sigma(i\nu)$ is always determined by perturbation
theory \cite{Kunes2011,Gunnarsson2016,georgthesis,SchaeferPhD}. 

While the detailed analytic derivation and a deeper investigation of the relation between the vertex divergences and the multivaluedness of $\Sigma[G]$ is presented in Appendix \ref{sec:scaleanalytic}, we discuss here another important feature which characterizes the singularities in $\Gamma_c$ and the change between $\Sigma^+$ and $\Sigma^-$ as physical self-energy. In the insets of Fig. \ref{fig:FKsigma} the imaginary part of the local (physical) Matsubara Green's function of the BM is shown for $\nu \geq 0$. For $W=0.65<\widetilde{W}$ (left panel), where no divergences are observed, $G(i\nu)$ is a monotonous function reaching its minimum at the lowest Matsubara frequency. Instead, for $W=0.8>\widetilde{W}$ (right panel), a {\sl minimum} is observed in $G(i\nu)$ {\sl exactly} at $\nu=\nu^*(W)$ [for an analytic proof see Appendix \ref{sec:scaleimag}] rendering $G(i\nu)$ {\sl non-monotonous} for $\nu \geq 0$. In this respect, let us recall that in the Mott phase of the BM (i.e., for $W>W_{\text{MIT}}=1$) $G(i\nu)$ exhibits qualitatively the same (non-monotonous) behavior reflecting the complete opening of the Mott spectral gap ($G(0)=0$ at $T=0$). Hence, the up-bending of the Green's function for $\nu\rightarrow 0$ signals the process of the formation of the Mott gap. Remarkably, this upturn starts at {\sl exactly} the same energy scale $\nu^*(W)$ where also the vertex divergence and the interchange between the perturbative $\Sigma^+$ and the non-perturbative $\Sigma^-$ as physical self-energy occur.
This indicates that all these (three) phenomena are intimately related, manifesting the breakdown of perturbation theory just from different perspectives.

\begin{figure}[t!]
    \centering
    \includegraphics[width=0.5\textwidth]{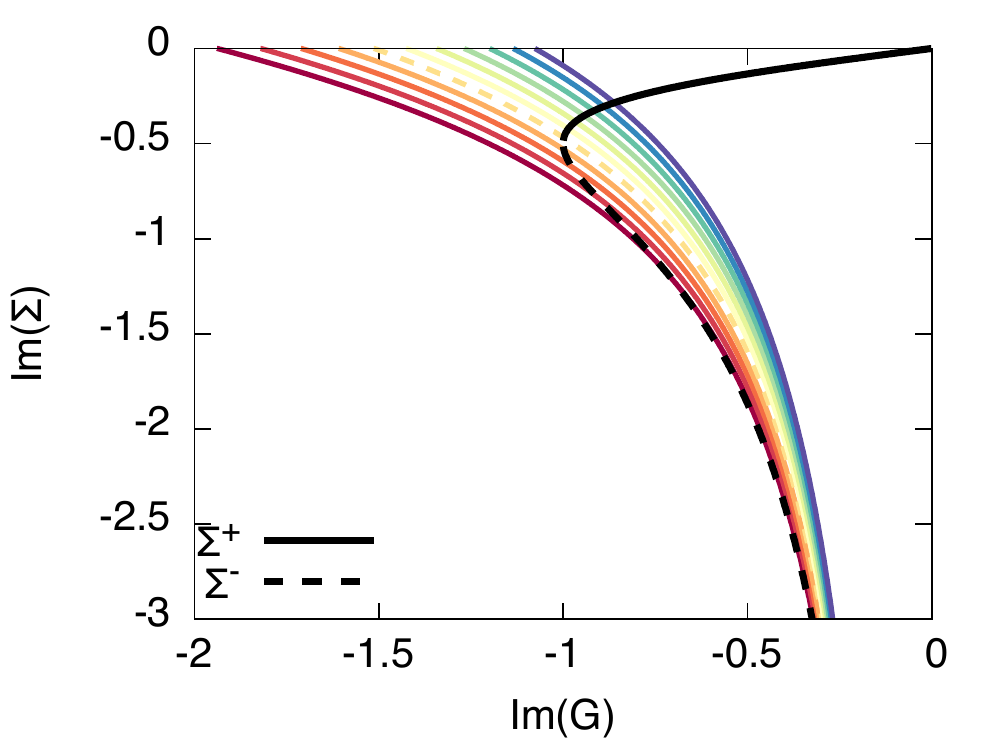}
    \caption{Colored lines: $\Sigma[G]$ as given in Eq.~(\ref{eq:FKselfcons2}) for $T=0.01$ and $W=1$. Different colors correspond to different Matsubara frequencies $\nu=\pi T(2n+1)$, from $n=1$ (brick red) to $n=11$ (dark blue). Black (full and dashed) line: $\Sigma^{\pm}[G]$ as given in Eq.~(\ref{eq:FKSigmaG}). The intersection between the colored and the black lines determines the physical self-energy and Green's function at the given frequency $\nu$. The dashed yellow curve is that at $\nu=\nu^*$.}
    \label{fig:FKRainbow}
\end{figure}

In the following, we want to provide the reader a more pictorial understanding of the relation between the vertex divergences and the multivaluedness of $\Sigma[G]$. To this end we consider explicitly the DMFT self-consistency condition
\begin{equation}
\label{eq:FKselfcons}
G(i\nu)=\int d\epsilon\text{ } {\cal N}(\epsilon) \frac{1}{i\nu+\mu-\epsilon-\Sigma(i\nu)}.
\end{equation}
where ${\cal N}(\epsilon)$ is the lattice-dependent non-interacting DOS.
In the Bethe lattice case ${\cal N}(\epsilon) \propto \sqrt{1-\epsilon^{2}}$, comparing Eq.~(\ref{eq:FKselfcons}) and Eq.~(\ref{eq:FKGSigma}),
we obtain the well-known result that $G^{-1}_0(i\nu)=i\nu-\frac{1}{4}G(i\nu)$ (for the case of half-filling, where $\mu\!=\!0$).
Therefore, to determine the physical self-energy, we have to fulfill both, Eq.~(\ref{eq:FKSigmaG}) and
\begin{equation}
\Sigma(i\nu)=i\nu-\frac{1}{4}G(i\nu)-G^{-1}(i\nu)
\label{eq:FKselfcons2}
\end{equation}
for each frequency $\nu$. The graphical solution of this set of equations is shown in Fig. \ref{fig:FKRainbow}. The colored lines represent $\Sigma$ as a function of $G$ as given in Eq.~(\ref{eq:FKselfcons2}), where different colors denote different frequencies. The physical $\Sigma$ and $G$ is then determined by the intersection of these curves with the line $\Sigma^{\pm}[G]$ obtained from Eq.~(\ref{eq:FKSigmaG}), cf.~black (full and dashed) curves in Fig.~\ref{fig:FKRainbow}. For $\nu>\nu^*$ the intersection of $\Sigma^+$ (black full line) with the colored lines yields the values for the physical $\Sigma$ and $G$, while for $\nu<\nu*$ the physical values are given by the crossing with $\Sigma^-$ (dashed line). One can clearly see, that exactly at the point where the change between the two branches of  $\Sigma$ occurs, $\delta\Sigma/\delta G$ diverges, yielding the above discussed vertex divergences in $\Gamma_c$.  
Let us emphasize, that this scenario represents a straightforward generalization of the corresponding one observed in zero space-time dimension (i.e., for the so-called one-point model) in Ref.~\onlinecite{Rossi2015}. 
In fact, our BM calculations allow for an extension of the one-point model results to the frequency domain: In the BM the one-point model evolution is realized separately for the single Matsubara frequencies as each of them crosses, one-by-one, the 
scale $\nu^*$. We should stress, however, that such straightforward extension is no longer applicable to the more complex cases described in the following sections.

\begin{figure*}[t!]
    \centering
  \begin{tabular}
      {llll}  \includegraphics[width=0.26\textwidth]{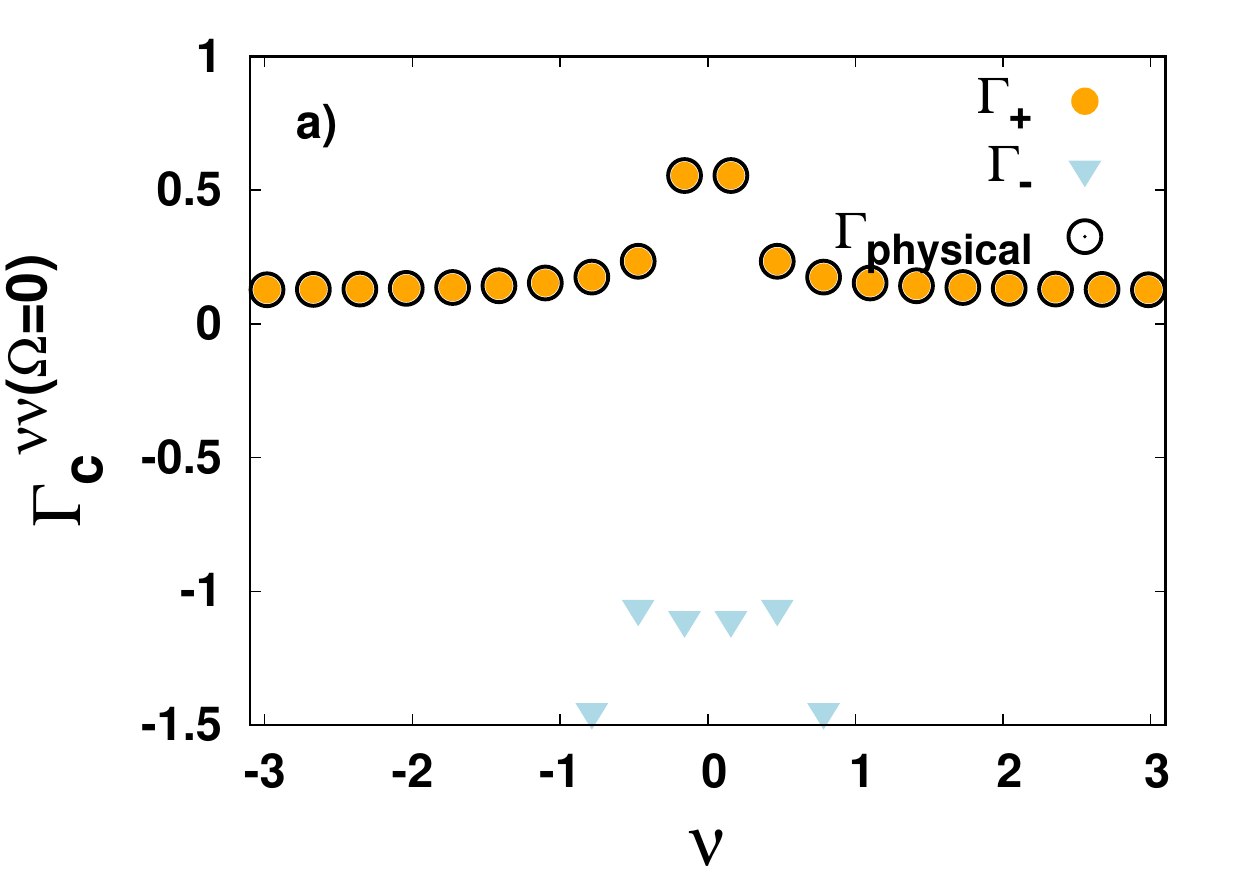} &
              \includegraphics[width=0.25\textwidth]{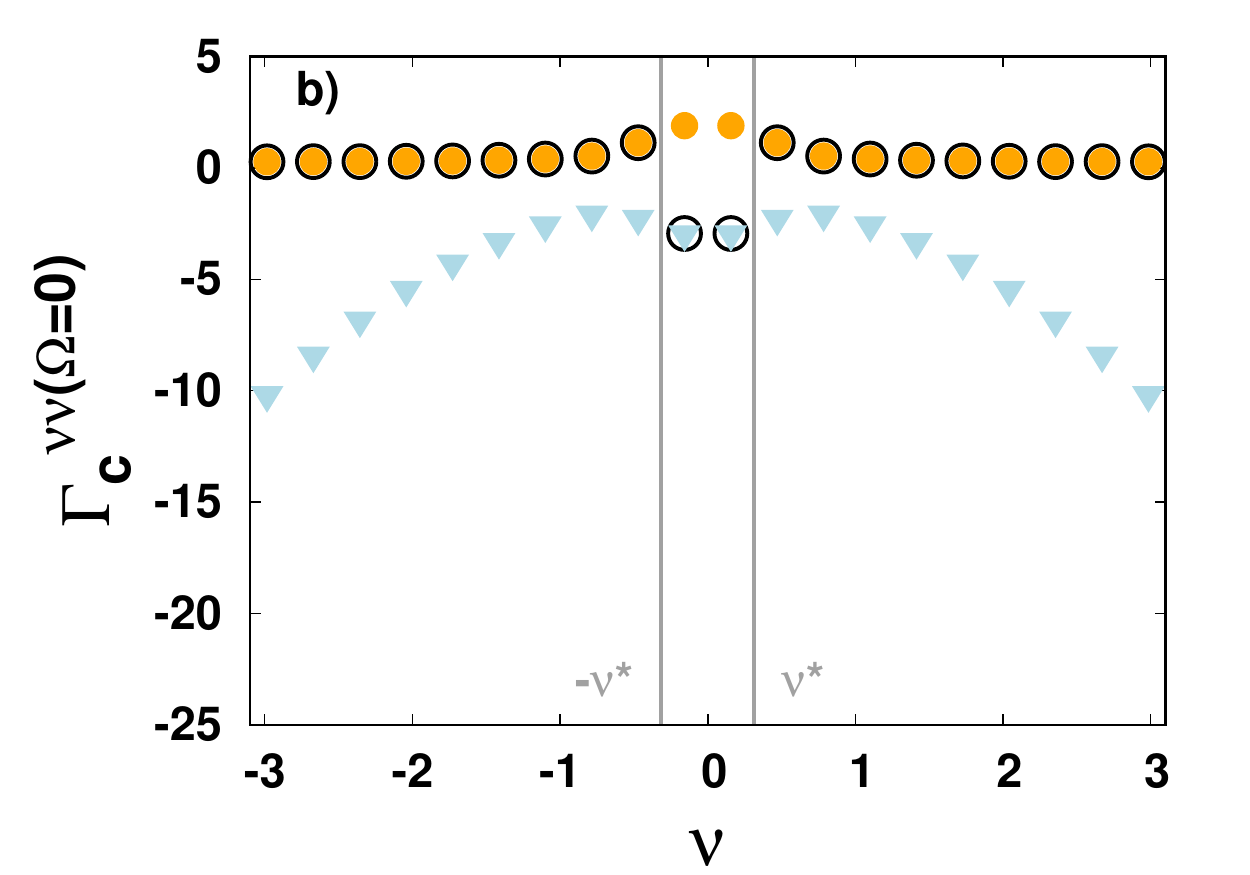} &
              \includegraphics[width=0.25\textwidth]{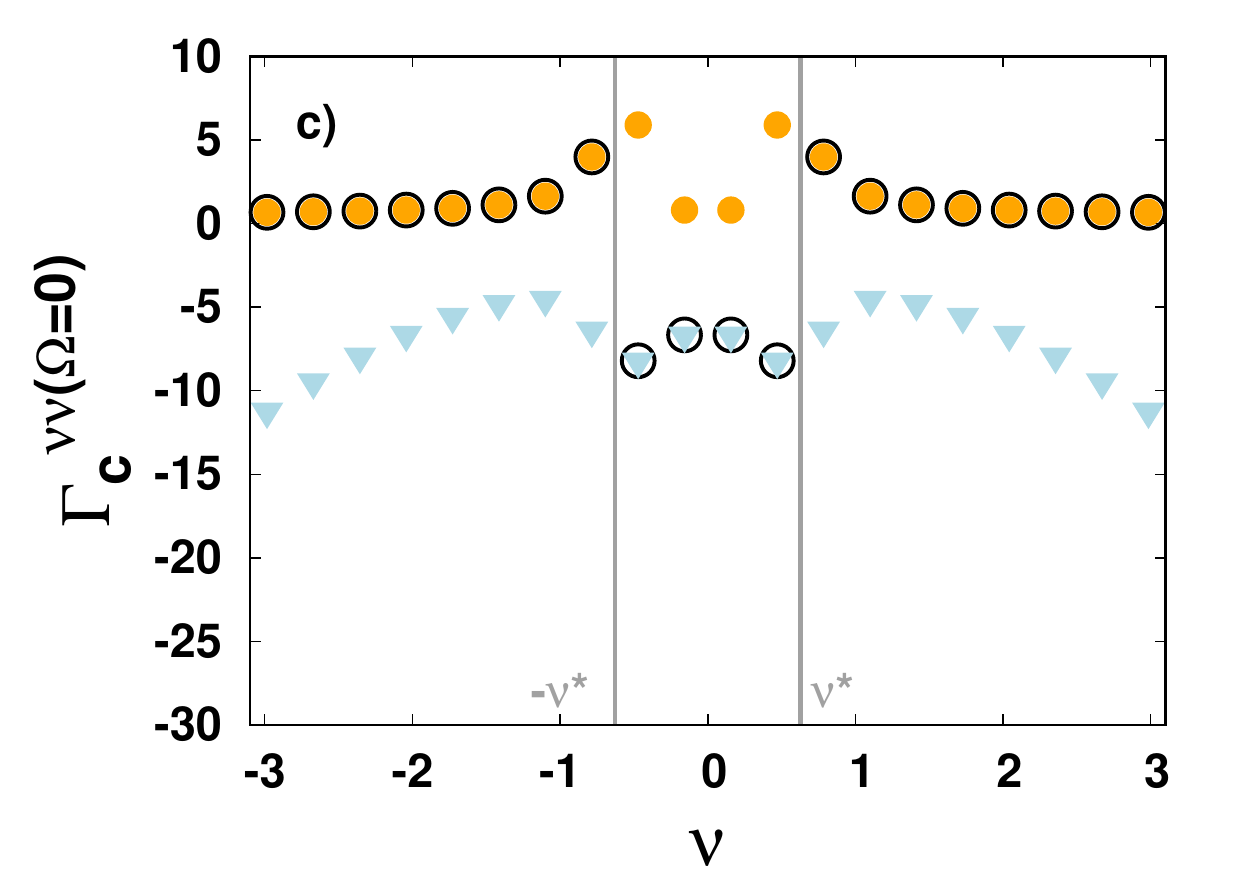} &
              \includegraphics[width=0.25\textwidth]{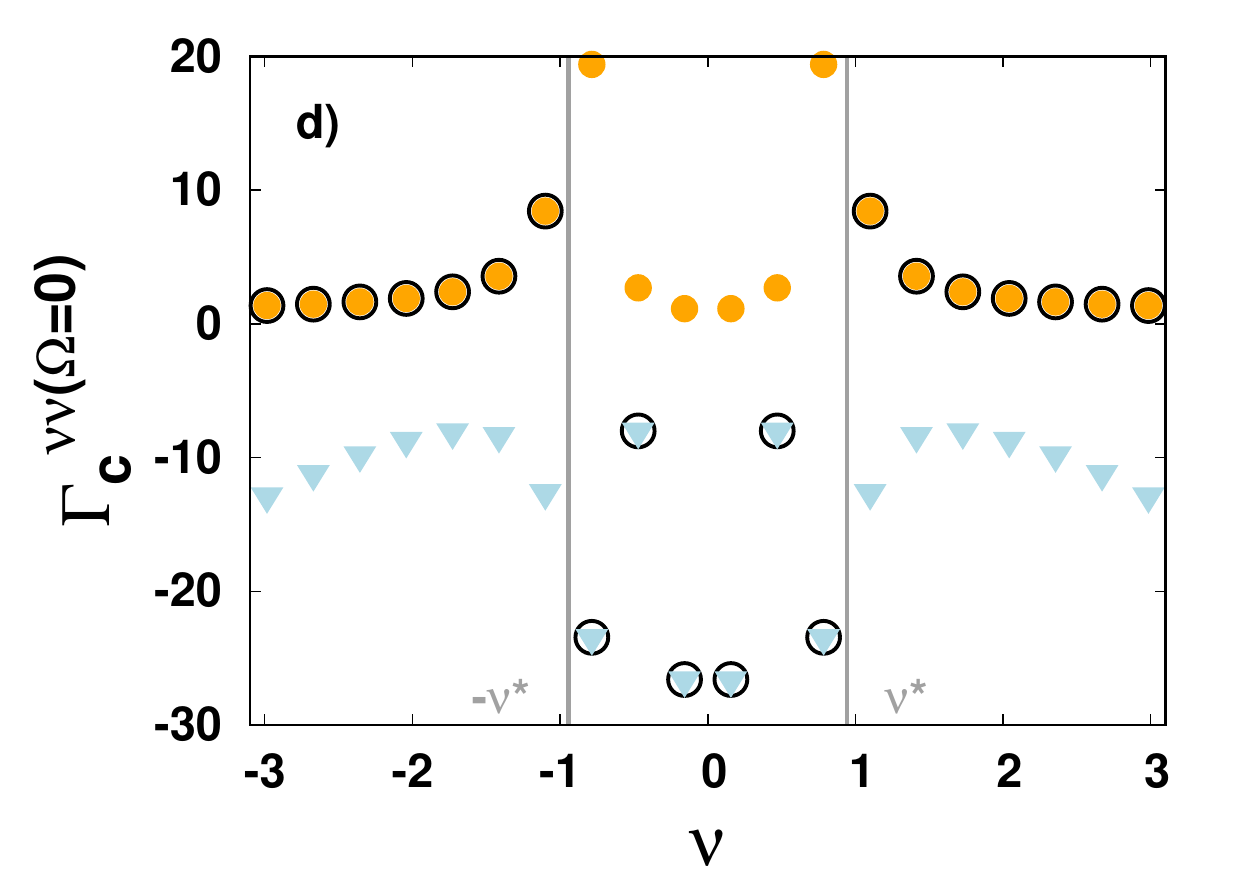}
  \end{tabular}
    \caption{Irreducible vertex in the charge channel, $\Gamma^{\nu\nu(\Omega=0)}_{\text{c}}$, of the BM model for different values of $W=0.7,1.0,1.5,2.0$ at $T=0.5$, plotted as a function of the fermionic Matsubara frequencies $\nu$.}
    \label{fig:FKVertexMaps}
\end{figure*}

As for the two-particle properties, in the special case of the BM model,  a completely analogous behavior of the
two branches of the irreducible vertex $\Gamma_{c,\pm}$ is found. For
$W<\widetilde{W}$ the physical $\Gamma_c$ is given by $\Gamma_{c,+}$ for all
Matsubara frequencies. For $W>\widetilde{W}$ this is true only for frequencies
$\nu>\nu^*(W)$ while for all Matsubara frequencies below this scale [i.e.,
$\nu<\nu^*(W)$] $\Gamma_{c,-}$ will represent the physical vertex function. This
situation is well reflected by our numerical data in Fig. \ref{fig:FKVertexMaps}
where the vertex functions $\Gamma_{c,+}^{\nu\nu(\Omega=0)}$,
$\Gamma_{c,-}^{\nu\nu(\Omega=0)}$ and their physical counterpart are shown for
four increasing values of $W$ at the same temperature $T=0.5$: One can clearly
see that, when $\pm\nu^*(W)$ (depicted by vertical lines) grows with $W$, $\Gamma_{c,-}$
rather than $\Gamma_{c,+}$ becomes the physical irreducible vertex in an
increasingly large frequency interval (centered at $\nu=0$). Note that, in
contrast to the self-energy, where no singular behavior is observed at the
switching point, the change of the branch occurs in $\Gamma_{c}$ via a
divergence of the vertex at exactly the frequency $\nu=\pm\nu^*$.

\subsubsection{Multivaluedness of correlation functions on the real frequency axis}
\label{sec:multivaluedreal}

\begin{figure}[t!]
    \centering
 \includegraphics[width=0.25\textwidth]{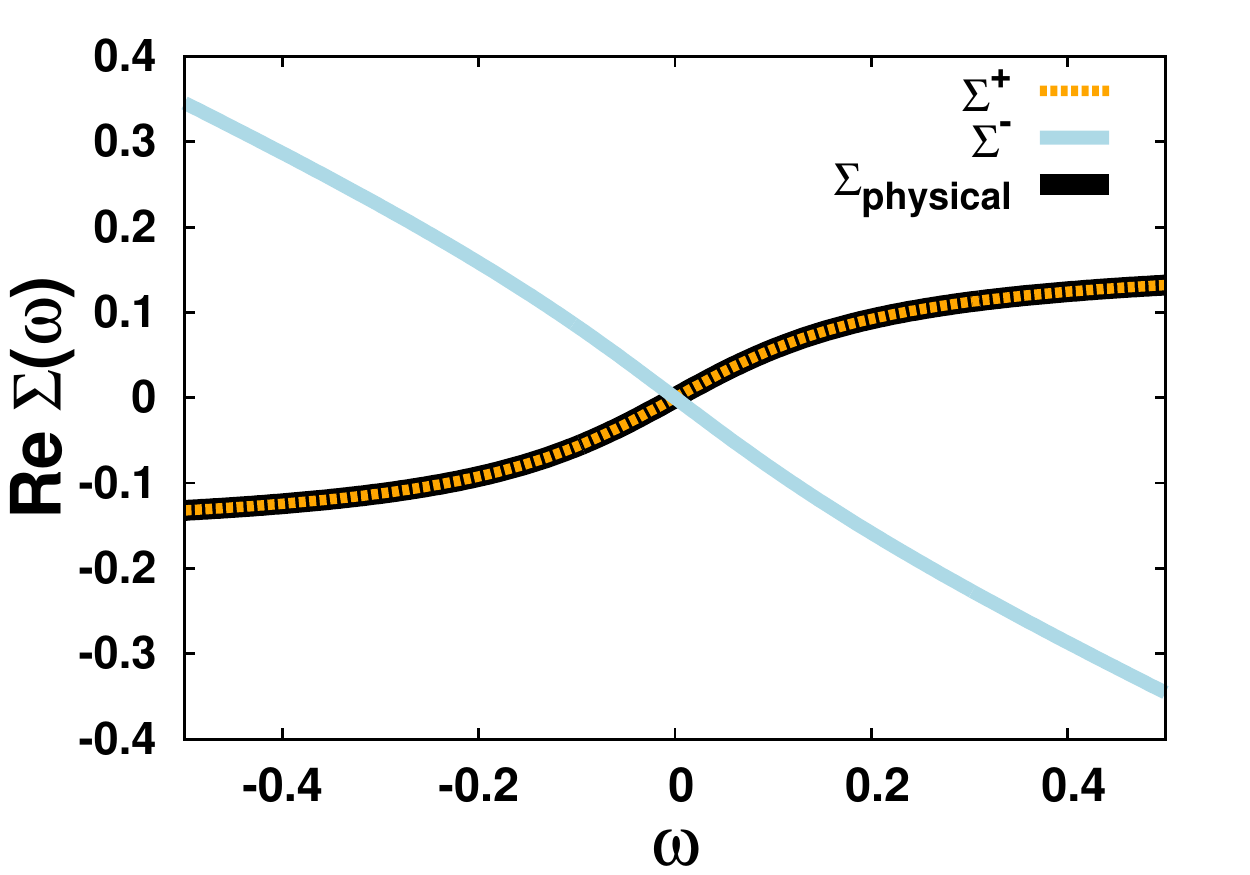}\includegraphics[width=0.25\textwidth]{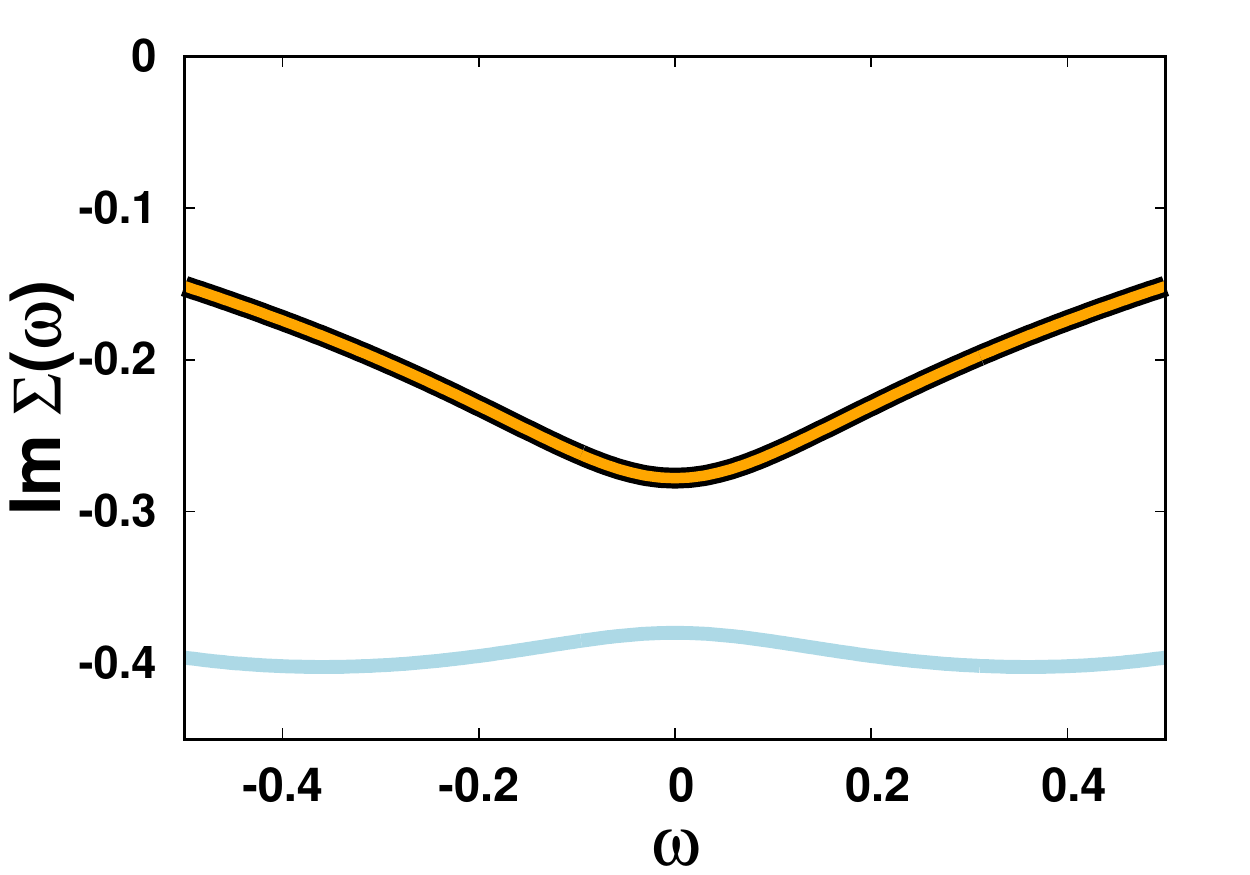}
    \includegraphics[width=0.25\textwidth]{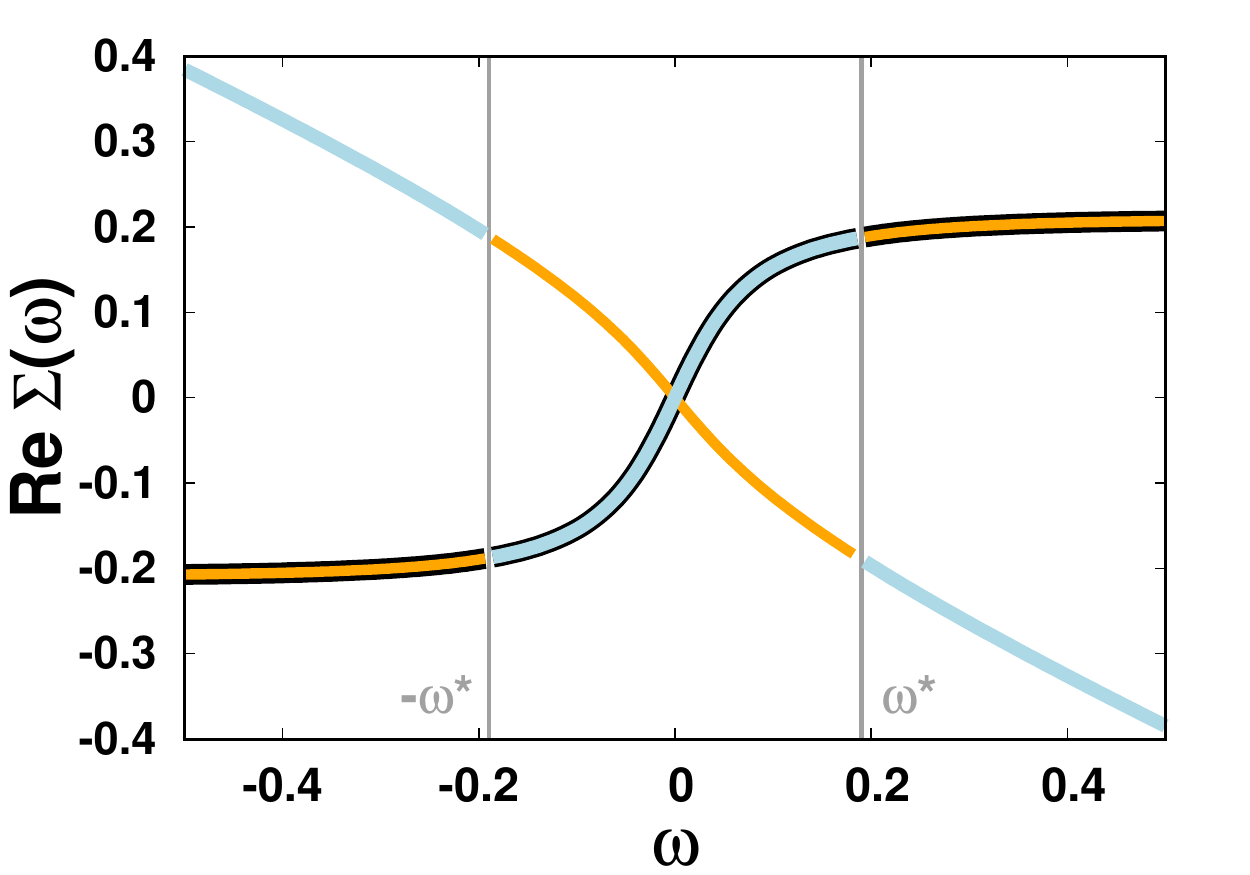}\includegraphics[width=0.25\textwidth]{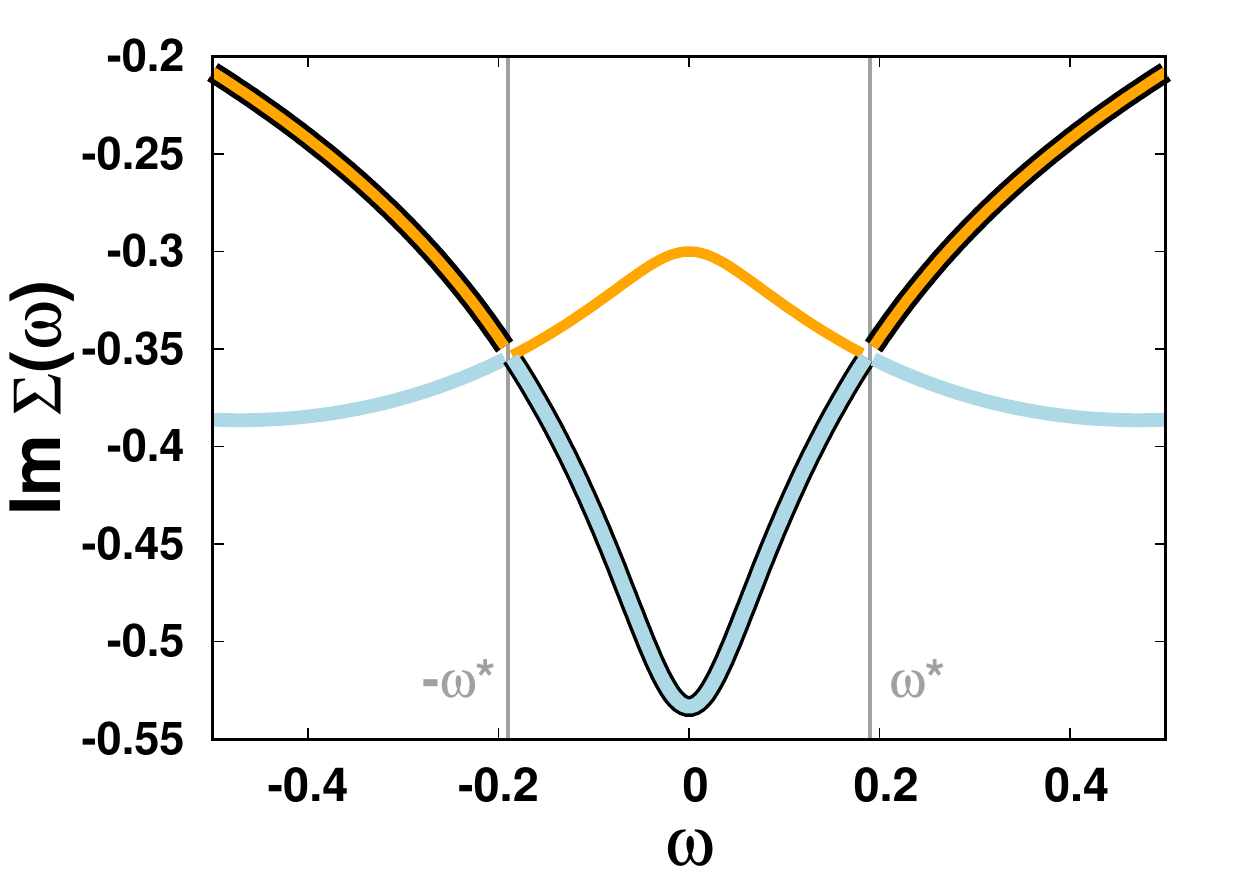}
    \caption{Real (left panels) and imaginary (right panels) parts of the DMFT self-energy of the BM model on the real frequency axis for $W\!=\!0.65\!<\!\widetilde{W}$ (top panels) and for $W=0.8> \widetilde{W}$ (bottom panels).}
    \label{fig:FKsigmaRe}
\end{figure}

Hitherto we have analyzed vertex divergences and their connection to the multivaluedness of the self-energy functional $\Sigma[G]$ for correlation function on the Matsubara axis. Evidently, it is of high interest whether these two related phenomena can be observed also for Green's functions (and the corresponding self-energy functional) on the {\sl real} frequency axis. That this is indeed the case is suggested by our numerical data in Fig.~\ref{fig:FKsigmaRe}: The upper panels show the real (left) and imaginary part (right) of the self-energy on the real frequency axis for $W<\widetilde{W}$. Evidently, the physical $\Sigma$ coincides with the perturbative $\Sigma^+$, obtained from Eq.~(\ref{eq:FKSigmaG}) when continued to the real axis, at all frequencies. The situation changes for $W>\widetilde{W}$ (lower panels of Fig. \ref{fig:FKsigmaRe}): Here, $\Sigma^+$ represents the physical self-energy only for $\omega>\omega^*$, while for smaller frequencies it is given by $\Sigma^-$. Differently from the two branches of the self-energy on the imaginary axis, however, $\Sigma^+$ and $\Sigma^-$ are not continuous at real frequencies as Re$[\Sigma^{\pm}]$ exhibit finite jumps at $\pm\omega^*$. Moreover, contrary to $\Gamma_c^{\nu\nu'(\omega=0)}$ on the Matsubara axis, we observe {\sl no} corresponding vertex divergences for real frequencies at the given value of $W$ (not shown).

In the following, we will analyze these differences in order to establish a connection between the corresponding phenomena observed for imaginary and real frequencies.
To this end, let us reconsider $\Sigma^{\pm}[G]$ in Eq.~(\ref{eq:FKSigmaG}): From a mathematical perspective, the two branches $\Sigma^+$ and $\Sigma^-$ originate from the two-valuedness of the square root function in the complex plane. In fact, $\Sigma^+$ corresponds to the first Riemannian surface, while $\Sigma^-$ describes the function on the second one. The two Riemannian surfaces are connected by a branch cut which is usually chosen to lie on the negative real axis. By crossing this branch cut, one switches from $\Sigma^+$ to $\Sigma^-$ (or vice versa) which corresponds -from a physical perspective- to moving from the perturbative to the non-perturbative branch. In fact, such a crossing occurs precisely when the imaginary part of the argument of the square root function in Eq.~(\ref{eq:FKSigmaG}), i.e., $\text{Im}[1+W^2G^2]$, vanishes, whereas at the same time the corresponding real part must be smaller than (or equal to) $0$, i.e., $\text{Re}[1+W^2G^2]\le 0$.

For imaginary frequencies, the above conditions are both automatically fulfilled by $1+W^2G^2\equiv 0$ [see Eq.~(\ref{eq:FKdivergence})] as Re$G(i\nu) \equiv 0$. Thus, the argument of the square root function moves from the first to the second branch via the origin of the complex plane, which implies a true divergence of the corresponding irreducible (charge) vertex $\Gamma_c$, as mentioned before and discussed in detail in Appendix \ref{sec:scaleimag}.

\begin{figure}
	\centering
		\includegraphics{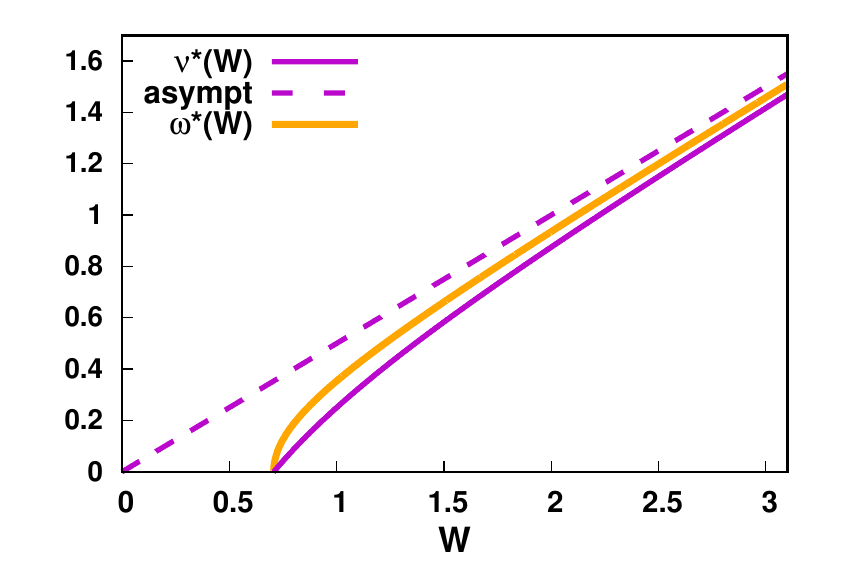}
	\caption{Comparison between the energy scales on the imaginary axis, $\nu^*(W)$, and the real frequency axis, $\omega^*(W)$, respectively, obtained from the DMFT solution of the BM model on a Bethe-lattice.}
	\label{fig:nu_omega}
\end{figure}

The situation, instead, is quite different for the self-energy (functional) on the real frequency axis. Specifically, the expression $1+W^2G^2$ will in general not vanish, since -even at half-filling- $G=G'+iG''$ typically consists of a finite real ($G'$) and a finite imaginary ($G''$) part (unlike the Matsubara Green's function where Re$G(i\nu) \equiv 0$ in the half-filling regime considered). Hence, we stick to the more general conditions for changing the branches of $\Sigma$, which explicitly read
\begin{subequations}
\label{subequ:branchcondreal}
\begin{align}
\label{equ:branchcondrealim}
 &\text{Im}\left[1+W^2(G'^2-G''^2)+i2G'G''\right]=2G'G''=0, \\
\label{equ:branchcondrealre}
 &\text{Re}\left[1+W^2(G'^2-G''^2)+i2G'G''\right]=\nonumber\\&\hspace{3cm}1+W^2(G'^2-G''^2)\le 0. 
\end{align}
\end{subequations}
According to Appendix \ref{sec:scalereal}, Eqs.~(\ref{subequ:branchcondreal}) imply $\text{Re}[G]=G'\equiv 0$ as necessary condition for passing from the perturbative to the non-perturbative branch of $\Sigma[G]$. This allows, together with the definition of $G$ in Eq.~(\ref{eq:FKGG0}) and the DMFT self-consistency condition for the Bethe lattice Eq.~(\ref{eq:FKselfcons2}), for the (analytic) determination of an energy scale $\omega^*(W)$, corresponding to the scale $\nu^*(W)$ on the imaginary axis. The explicit calculations are outlined in Appendix \ref{sec:scalereal} and yield:
\begin{equation}
 \label{equ:scalereal}
 \omega^*(W)=\frac{1}{2}\sqrt{\frac{2W^2-1}{2}}.
\end{equation}
Analogously to the behavior on the Matsubara axis, the physical $\Sigma$ is given by $\Sigma^+$ for $\lvert \omega\rvert > \omega^*(W)$ and by $\Sigma^-$ for $\lvert \omega\rvert < \omega^*(W)$. The energy scales for the onset of the non-perturbative behavior of the physical $\Sigma[G]$ are not exactly identical on the real [$\omega^*(W)$, Eq.~(\ref{equ:scalereal})] and on the imaginary [$\nu^*(W)$, Eq.~(\ref{equ:scale})] axis
, but they display a quite similar behavior (see Fig. \ref{fig:nu_omega}).
In particular, they coincide for the limiting cases $W\rightarrow\infty$ and $W=\widetilde{W}$, where $\omega^*(W)$ vanishes at exactly the same value $\widetilde{W}=1/\sqrt{2}$ as $\nu^*(W)$. 

This allows for a direct comparison between the real and the imaginary axis at $W=\widetilde{W}$. In fact, both $\Sigma(i\nu)$ and $\Sigma(\omega)$ are represented by $\Sigma^+$ for all frequencies except for $\nu=\omega=0$. In both cases the change of the physical branch is accompanied by a divergence of the respective irreducible vertex $\Gamma_c$ at zero frequency. Moreover, the stationary behavior observed in $G(i\nu)$ at $\nu=0$ (cf. inset in the right panel of Fig. \ref{fig:FKsigma} and the related discussion) is reflected, for $\nu^{*}=0$ in a change of slope of the {\sl real} part of the Green's function on the real axis, i.e., in $\frac{d}{d\omega}$Re$[G(\omega)]=0$, at $\omega=0$. The latter phenomenon has been already observed in a previous study in Ref.~\onlinecite{Janis2014} (see the inset in Fig. 8 therein).

These analogies demonstrate how the multivaluedness of $\Sigma$, the corresponding divergences and the existence of a stationary point in the Green's function observed on the Matsubara axis are also well reflected on the real axis at $W=\widetilde{W}$.

For $W>\widetilde{W}$ the situation is different. In particular, as it is discussed in detail in Appendix \ref{sec:scalereal}, the change of branches for $\Sigma$ does not occur any more at the origin of the complex plane. Consequently, this is not connected with any divergence of the corresponding irreducible vertex whose only singularity on the real axis, thus, occurs for $W\equiv\widetilde{W}$. A more detailed understanding of the full interrelation between the results on the real and the imaginary axes for $W>\widetilde{W}$ requires, however, further investigations which are beyond the scope of the present work.

\subsubsection{Impact on algorithmic developments}
\label{sec:algorithms}

We want to briefly comment here on the impact of the multivaluedness of the self-energy functional and the related divergences of $\Gamma_c$ on many-body algorithms based on the self-consistent determination of the self-energy from a given Green's function. To exemplify possible problems which might arise in such schemes we demonstrate them for the simple case considered in the three previous sections. To this end, along the lines of Ref.~[\onlinecite{Kozik2015}], we rewrite Eq.~(\ref{equ:sig2ndord}) in two different ways so that they represent iterative schemes for the determination of $\Sigma$ from a given $G$ (at a fixed value of $W$):
\begin{subequations}
\label{subequ:itscheme}
\begin{align}
 \label{equ:itscheme1}
\Sigma^{(N)}=\frac{W^2/4}{G^{-1}+\Sigma^{(N-1)}}=f_{1}\left(\Sigma^{(N)}\right),\\
 \label{equ:itscheme2}
 \Sigma^{(N)}=-G^{-1}+\frac{W^2/4}{\Sigma^{(N-1)}}=f_{2}\left(\Sigma^{(N)}\right).
\end{align}
\end{subequations}
One can easily check that both schemes are compatible with Eq.~(\ref{equ:sig2ndord}) by taking the limit $N\rightarrow\infty$, i.e., $\Sigma^{(N)},\Sigma^{(N-1)}\rightarrow \Sigma$, on both sides of the equations: This way one indeed retains the quadratic Eq.~(\ref{equ:sig2ndord}) for the determination of $\Sigma$ from $G$. Hence, both iterative schemes in Eq.~(\ref{subequ:itscheme}) have two fixed points, namely the two corresponding solutions given in Eq.~(\ref{eq:FKSigmaG}). By analyzing the magnitude of the derivative\cite{footnote3} of the functions $f_{1}\left(\Sigma\right)$ and $f_{2}\left(\Sigma\right)$ on the r.h.s. of Eqs. (\ref{subequ:itscheme}) w.r.t. $\Sigma$, one finds that in both cases only one of the two respective fixed points is attractive (stable) while the other is repulsive (unstable). Hence, a numerical iterative procedure which uses Eqs. (\ref{subequ:itscheme}) will converge only to the corresponding attractive (stable) fixed point of the respective scheme. For the iteration prescription in Eq.~(\ref{equ:itscheme1}) this attractive fixed point is given by the solution $\Sigma^+[G]$ while for Eq.~(\ref{equ:itscheme2}) the iteration converges to $\Sigma^-[G]$. 

The above considerations imply that, for frequencies $\nu<\nu^*(W)$ where $\Sigma^-$ is the physical solution, the scheme in Eq.~(\ref{equ:itscheme2}) has to be adopted, while for $\nu>\nu^*(W)$ Eq.~ (\ref{equ:itscheme1}) should be used. This is consistent with the switch in the solutions at the first Matsubara frequency reported in Ref.~[\onlinecite{Kozik2015}] for the AL case. A unique iterative scheme [i.e., Eq.~(\ref{equ:itscheme1})] can be thus used {\sl only} in the perturbative region $W<\widetilde{W}$ where the self-energy is given by the perturbative branch (here: $\Sigma^{+}$) at all frequencies. This observation is highly relevant for iterative methods based on a self-consistent determination of the self-energy from the full Green's function such as bold QMC\cite{boldQMC}, whose diagrammatic resummations are confronted with such formal ambiguities. In this respect, our findings also explain why -in certain cases- approaches based on $G_0$ rather than $G$ are preferable for the calculating the self-energy, as it is indeed the case for the iterated perturbation theory\cite{DMFTREV,DMFTMIT}.

\subsubsection{Distributed disorder}
\label{sec:distdisorder}

Before we extend our analysis of non-perturbative features in the BM to the case of the FK model in the next
subsection, we would like to comment briefly here on the relation of our
results, and in particular the occurrence of the irreducible vertex divergences,
to the nature of the quenched disorder considered. In fact, one may wonder if in the presence of {\sl distributed} disorder the same
divergences remain.

This is {\sl not} the case, as it can been proven for a
randomly chosen disorder variable (i.e., for the customary choice for the Anderson model): 
Assuming for instance a {\sl flat} and {\sl gapless} disorder distribution 
in the interval $[-\gamma/2,\gamma/2]$ we get (see Appendix \ref{app:Anderson})
\begin{equation}
G(z) = \frac{1}{\gamma}\log \left(\frac{G^{-1}_0(z)+\gamma/2}{G^{-1}_0(z)-\gamma/2}\right),
\label{eq:appGAnd}
\end{equation}
and a  {\it single-valued} determination for $\Sigma$
as a functional of $G$
\begin{equation}
\Sigma[G] = -G^{-1}+\frac{\gamma}{2\tanh (\gamma G/2)}.
\label{eq:appSigmaAnd}
\end{equation}
On the contrary, in the case of distributed disorder, $\Sigma$ becomes a non-analytic (local) functional of  $G_0$
\begin{equation}
\Sigma[G_0] = G^{-1}_0+\gamma{\log^{-1} \left(\frac{G^{-1}_0-\gamma/2}{G^{-1}_0+\gamma/2}\right)}.
\label{eq:appSigmaAndg0}
\end{equation} 
From Eq.~(\ref{eq:appSigmaAnd}) we get by differentiation w.r.t. $G$
\begin{equation}
\Gamma_{c,\pm}^{\nu\nu'(\Omega=0)}=\beta\delta_{\nu\nu'} \left ( \frac{1}{G^2}+\frac{\gamma^2}{4\sin^2 (\gamma \text{Im}[G]/2)}\right),
\label{eq:appGamma}
\end{equation}
where we have used that at half-filling $G$ is purely imaginary, i.e., $G=i\text{Im}[G]$ and $\sinh(ix)=i\sin(x)$.
The condition for a divergence of $\Gamma_{c,\pm}^{\nu\nu'(\Omega=0)}$
reads $\gamma \text{Im}[G]/2=n\pi$ where $n=0,\pm 1, \pm 2...$.  However, this condition is {\sl never} fulfilled,
since from Eq.~(\ref{eq:appGAnd}) one can straightforwardly obtain the constraint $\gamma |\text{Im}[G]|/2< \pi$.

In order to interpret the implications of the absence of a vertex divergence and a multivaluedness of $\Sigma[G]$, we note that no MIT and, hence, no gap in the spectrum exists in the distributed disorder model introduced above. 
Consistent with the interpretation of the vertex divergence as precursor of the MIT\cite{divergence,SchaeferPhD,georgthesis,Janis2014}, in the absence of a MIT none of these precursors are observed. However, when introducing a (large enough) gap in the disorder distribution, the MIT is restored, and, so, also the corresponding precursors (see Appendix \ref{app:Anderson}).

Let us also point out that, in the context of a distributed disorder model with a flat and gapless disorder,
vertex divergences could be only obtained by adding some source of dichotomic
disorder as in the case of the disordered FK model\cite{Janis2014} 
or, alternatively, in the presence of electron-lattice interaction\cite{DDSCiuFratDobro2016}. 
Finally, note that the situation will be different in the presence of long-range
interactions: The latter appear at the (extended) dynamical mean-field
level (EDMFT) as source of distributed local random energies, whose distribution becomes
{\sl bimodal} upon entering in the strongly correlated
regime near the Wigner crystallization\cite{DobroLongRange}. We therefore expect that vertex anomalies should
also be found in this case, consistent with their interpretation of precursors of the Mott transition. We also
note that the recent analysis of the Hedin's equations in the zero dimensional
limit\cite{Stan2015} provides further support to this speculation.

\subsection{DMFT divergences in the Falicov-Kimball model}
\label{sec:fk}

Let us now turn our attention to the FK model. Its DMFT single-particle Green's function for the mobile $\uparrow$-electrons reads
\begin{equation}
G(\nu)= \left ( \frac{\langle n_\downarrow \rangle}{G^{-1}_0(\nu)-\frac{U}{2}}+\frac{1-\langle n_\downarrow \rangle}{G^{-1}_0(\nu)+\frac{U}{2}}\right ),
\label{eq:FKG}
\end{equation}
where, analogously to the BM, $G_0^{-1}(\nu)=i\nu-\Delta(\nu)$ with $\Delta(i\nu)$ being the hybridization function of DMFT. Note that the chemical potential $\mu$ in $G_0(\nu)$  is, for the half-filled case considered here ($\langle n_{\uparrow}\rangle=1/2$ and $\mu=U/2$), exactly compensated by the corresponding Hartree term of the self-energy (the latter will, hence, appear without such Hartree contribution in the following). Moreover, one also has $\langle n_{\downarrow}\rangle=1/2$ immobile electrons per lattice site. This renders the single-particle Green's function of the FK model in Eq.~(\ref{eq:FKG}) completely equivalent to the corresponding one for the BM in Eq.~(\ref{eq:FKGG0}), if one considers $U=W$. In particular, also the self-energy $\Sigma^{\pm}[G]$ as given in Eq.~(\ref{eq:FKGSigma}) coincides for both models.

However, as we anticipated at the beginning of Sec.~\ref{sec:bmfk}, in spite of these similarities for one-particle quantities, there is a subtle difference between the BM and the FK, which appears at the level of the two-particle correlation functions. While in the BM the probability distribution of the disorder is static and, hence, {\sl independent} of the dynamics of the itinerant electrons in the system, in the FK model the disorder distribution is given by the density of the immobile $\downarrow$ electrons which are in thermal equilibrium with the mobile ones. This exemplifies the different (i.e., annealed) nature of the disorder in the FK model w.r.t. the BM (where one deals with a quenched disorder). Hence, when calculating the irreducible vertex $\Gamma_c$ by means of the functional derivative of $\Sigma^{\pm}[G]$ w.r.t. to $G$ [see Eq.~(\ref{equ:funcder})] one must consider that varying $G$ affects also $\langle n_{\downarrow}\rangle$ (but not the static disorder in the BM)\cite{FreericksRMP}. 
This implies that $\frac{\delta \langle n_{\downarrow}\rangle}{\delta G}\ne 0$, giving rise to an extra term in the irreducible vertex $\Gamma_c$ of the FK, which can be now expressed via a functional derivative [see Eq.~(49) in Ref.~\onlinecite{FreericksRMP}] as:
\begin{equation}
 \label{equ:fkderiv}
 \Gamma_{c,\pm}=\beta \frac{\delta \Sigma^{\pm}}{\delta G}+\beta \frac{\delta \Sigma^{\pm}}{\delta \langle n_{\downarrow}\rangle}\frac{\delta \langle n_{\downarrow}\rangle}{\delta G}.
\end{equation}
Note that, as we consider here only correlation functions for the mobile particles, we {\sl define} $\Gamma_c\equiv\Gamma_{\uparrow\uparrow}$, differently from the general definition given in Eq.~(\ref{equ:channeldefph}). An explicit evaluation of the functional derivatives in Eq.~(\ref{equ:fkderiv}) yields (see Appendix \ref{app:FK}):
\begin{align}
\Gamma_{c,\pm}^{\nu\nu'(\Omega=0)} &= \beta\delta_{\nu\nu'}\frac{\sqrt{1+U^2G^2(\nu)}\mp 1}{2G^2(\nu)\sqrt{1+U^2G^2(\nu)}}\nonumber\\+\beta&\frac{U^2}{4} C_{\pm} \frac{1}{\sqrt{1+U^2G^2(\nu)}}\frac{1}{\sqrt{1+U^2G^2(\nu')}},
\label{eq:DiagonalVertexFK}
\end{align}
where
\begin{equation}
 \label{equ:defineCpm}
 C_{\pm}=\frac{1}{1-K_{\pm}},\quad K_{\pm}=\sum_{\nu}\underset{-G^2\Gamma_{c,\pm}^{\text{BM}}/\beta}{\underbrace{\frac{\sqrt{1+U^2G^2}\mp1}{2\sqrt{1+U^2G^2}}}}.
\end{equation}
Here, an important remark is due: When considering $K$, the corresponding frequency sum is no longer well defined as the summand does not decay, but approaches $1$ for $\nu\rightarrow \infty$. 
In fact, as we will discuss in details below, once entered the non-perturbative regime ($U>1/\sqrt{2}$), the physical correct definition of the coefficient $K$ requires the inclusion of a frequency-dependent switch of signs.

In the meanwhile, we start by noticing that the first summand on the r.h.s. of Eq.~(\ref{equ:fkderiv}) is completely equivalent to the irreducible vertex of the BM in Eq.~(\ref{eq:FKGamma}). Hence, it exhibits exactly the same divergences of $\Gamma_c$ and a related energy scale $\nu^*(U)$, as discussed in the previous section. Moreover, because of the similarity in the denominators, also the second summand on the r.h.s of Eq.~(\ref{equ:fkderiv}) diverges exactly at the same parameters as the first one, without any overall cancellation of the divergence. Thus, all results of Sec. \ref{sec:BMdiv} are applicable also to the corresponding divergences in the FK model with only two exceptions:

(i) The degeneracy of the singular eigenvector is lifted. We show this by considering the following explicit expression for $\chi_{c,\pm}$:
\begin{align}
 \label{eq:FKchi}
 \chi_{c,\pm}^{\nu\nu'(\Omega=0)}&=\delta_{\nu\nu'} E_{\text{BM}}(\nu) \nonumber\\
 +\frac{\beta}{U^2} (1 \mp &\sqrt{1+U^2G^2(\nu)}) (1 \mp \sqrt{1+U^2G^2(\nu')}),
\end{align}
where the term in the first line is equivalent to the corresponding expression for the BM in Eq.~(\ref{eq:BMchi}). However, due to the additional term in the second line of Eq.~(\ref{eq:FKchi}) the eigenvectors  (and eigenvalues) are not automatically the same for both models. In particular, as discussed in the previous section [see Eq.~(\ref{eq:BMchi}) and below] for the BM, $E_{\text{BM}}(\bar{\nu})=E_{\text{BM}}(-\bar{\nu})$ represents an eigenvalue to an eigenvector of the {\sl degenerate} subspace spanned by $V_{0}(\bar{\nu})=A \delta_{\nu\bar{\nu}} + B\delta_{\nu(-\bar{\nu})}$.  This degeneracy allows us to choose $A$ and $B$ in such a way that $V_{0}(\bar{\nu})$ represents also an eigenvector to the second line of Eq.~(\ref{eq:FKchi}). To this end, we notice that the latter term is symmetric under the transformation $\nu'\rightarrow -\nu'$. Hence, a total anti-symmetric vector $V_{0}(-\nu)=-V_{0}(\nu)$ will always correspond to an eigenvector of the second line in Eq.~(\ref{eq:FKchi}) to the eigenvalue $0$. 
Thus, for the full $\chi_{c,\pm}$ of the FK model an eigenvector is given by the anti-symmetric combination
\begin{equation}
 \label{equ:eigv}
 V_{0,\bar{\nu}}(\nu)=\frac{1}{\sqrt{2}}\left[\delta_{\nu\bar{\nu}}-\delta_{\nu(-\bar{\nu})}\right],
\end{equation}
for each Matsubara frequency $\bar{\nu}$. As for the BM, the related eigenvalue $E_{\text{BM}}(\bar{\nu})$ vanishes in the case that $\bar{\nu}=\nu^*(U)$. We anticipate here that this anti-symmetric form is exactly the same as that obtained in the atomic limit of the Hubbard model (discussed in the next section).

(ii) A second difference regarding the vertex divergence (associated with the condition $1+U^2G^2=0$) between the BM and the FK model concerns the switching between the two solutions $\Sigma^{\pm}$ and $\Gamma_{c,\pm}$ at the frequency $\nu=\nu^*(U)$. For the {\sl one-particle} self-energy we observe exactly the same situation for both models: While for $\nu>\nu^*(U)$ the physical self-energy is given by $\Sigma^+(\nu)$ it corresponds to $\Sigma^-(\nu)$ for $\nu<\nu^*(U)$. The same also hold for the irreducible vertex $\Gamma_{c,\pm}$ of the BM and, hence, for the corresponding first line part of the corresponding vertex of the FK model [see first line in Eq.~(\ref{eq:DiagonalVertexFK})]. The frequency dependent part of the second line of this correlation function, however, is independent of the chosen branch which just affects the constant prefactor $C_{\pm}$ [see Eq. (\ref{equ:defineCpm})].

Here, we can further elaborate on the definition and the meaning of the expressions for $C_{\pm}$ given by Eq.~(\ref{equ:defineCpm}) above. While these expressions are formally derived from Eqs.~(\ref{equ:fkderiv})-(\ref{eq:DiagonalVertexFK}), we have already noted that $C_{-}$ is actually ill-defined, because of the non-convergent Matsubara summation in $K_{-}$. This is, in fact, a manifestation of an important difference between the simpler situation of the BM and the one captured by the FK. In fact, as the coefficient $K$ (appearing in $C$) is defined through a summation of {\sl all} Matsubara frequencies, a switching between the two determinations of, e.g.., $\Gamma_{c,\pm}^{\text{BM}}$ occurs now also {\sl inside} a frequency sum: The correct definition of the frequency independent coefficient $C$ requires that  in the internal summation on the r.h.s. of Eq.~ (\ref{equ:defineCpm}) a sign-switch should be made at $\nu=\nu^*(U)$. Thus, while, for $U<\widetilde{U}$, $C_+$ represents indeed the physical solution for the frequency-independent prefactor, for $U>\widetilde{U}$ neither $C_+$ nor $C_-$ match the corresponding physical $C$. This implies that, while in the FK the definition of the two determinations of the self-energy $\Sigma_{\pm}$ remain valid as in the BM, the same does not hold for the vertex function, due to the frequency non-local relations which define $C$. Let us point out that this frequency non-locality can be directly traced back to the functional derivative $\delta\langle n_\downarrow\rangle$ in Eq.~(\ref{equ:fkderiv}): The density itself is defined as a sum over {\sl all} frequencies of the Green's functions and, hence, represents a highly frequency non-local functional of $G$. 

Let us emphasize here also the consistency of our results with previous calculations of the irreducible $ph$-vertex for the FK model\cite{footnote2} as presented in Refs.~\onlinecite{Shvaika2001,Freericks2000,Ribic2016}.

We now turn to the analysis of the terms of $\Gamma_{c,\pm}$ (and $\chi_{c,\pm}$) in the second line(s) of Eq(s).~(\ref{eq:DiagonalVertexFK}) [and (\ref{eq:FKchi})], not present in the corresponding expressions for the BM, in more detail. The FK model exhibits -in addition to frequency-localized divergences of $\Gamma_{\text{c},\pm}$ determined by the condition $1+U^{2}G^{2}=0$- another kind of singularities very different from the first ones. This happens when the frequency-independent prefactor $C_{\pm}$ [see Eq.~(\ref{equ:defineCpm})] itself diverges. These divergences in the phase-diagram occur also along several lines\cite{Ribic2016}, at first glance very similar to those of the first divergence. Their positions, however, are (for a given $T$) always at larger interactions compared to the singularities of the first type. In fact, by a closer inspection of Eq.~(\ref{equ:defineCpm}), one can see that $K_{\pm}$ diverges at the first singularity as $\sqrt{1+U^2G^2}$ becomes $0$ and, hence, $C_{\pm}$ becomes $0$. The {\sl divergence} of $C_{\pm}$ instead occurs only later, when $K_{\pm}$ reaches $1$. At this point, however, no additional simultaneous singularity in the first line of Eq.~(\ref{eq:DiagonalVertexFK}) is observed.

More important, to be noted, the nature of such new kind of divergences is qualitatively different from the previous ones: Due to the frequency independence of $C_{\pm}$,  they are {\sl not} localized in frequency, i.e., at these singularities $\Gamma_{\pm}$ diverges at {\sl all} Matsubara frequencies. The qualitative difference between these two kinds of divergences in the FK model is reflected in a corresponding different structure of the singular eigenvectors of the  generalized susceptibility\cite{Ribic2016}. In fact, according to the discussion in Sec.~\ref{sec:2b1}, a divergence of $\Gamma_c$ at {\sl all} frequencies corresponds to an eigenvector of $\chi_c$ which has finite weight at {\sl all} frequencies. In the following, we will prove this statement for the specific case of the second type of divergences in the FK model by explicitly determining the form of the singular eigenvector. To enhance the readability of our notation we define:
\begin{equation}
 \label{equ:defF}
 {\cal F}_{\pm}(\nu)=\frac{1\mp\sqrt{1+U^2G^2(\nu)}}{U},
\end{equation}
which allows us to express $\chi_c$ in the following way:
\begin{equation}
 \label{equ:rewritesusc}
 \chi_{c}^{\nu\nu'(\Omega=0)}=E_{\text{BM}}(\nu)\delta_{\nu\nu'}+\beta {\cal F}(\nu){\cal F}(\nu'),
\end{equation}
where we omit the $\pm$ dependence of $\chi_c$, $E_{\text{BM}}$ and ${\cal F}$ by assuming that for each frequency the corresponding physical branch is chosen (see also Refs.~[\onlinecite{Shvaika2001,Ribic2016}], where the corresponding quantities are expressed in a unique way in terms of the self-energy). Let $V(\nu)$ be an eigenvector of $\chi$ to the eigenvalue $\lambda$, then the following eigenvalue equation holds
\begin{equation}
 \label{equ:ev2equ}
 E_{\text{BM}}(\nu)V(\nu)+\beta P {\cal F}(\nu)=\lambda V(\nu),
\end{equation}
where $P=\sum_{\nu'} {\cal F}(\nu')V(\nu')$ is a (for the moment unknown) constant. From Eq.~(\ref{equ:ev2equ}) one can now express the eigenvector $V(\nu)$ in the following form
\begin{equation}
 \label{equ:ev0}
 V(\nu)=\beta P\frac{{\cal F}(\nu)}{\lambda-E_{\text{BM}}(\nu)}.
\end{equation}
The factor $P$ can be now determined by multiplying this equation with ${\cal F}(\nu)$ and summing over $\nu$: As $P$ then appears linearly on both sides of the equation and can be, hence, eliminated (for $P\ne 0$, otherwise $V(\nu)$ would correspond to a total antisymmetric eigenvector which is, however, associated with the first divergence). This leads to the following relation which implicitly defines the eigenvalue(s) $\lambda$:
\begin{equation}
 \label{equ:detev}
 1=\beta\sum_{\nu}\frac{{\cal F}^2(\nu)}{\lambda-E_{\text{BM}}(\nu)},
\end{equation} 
Setting $\lambda=0$ leads exactly to the condition $K_{\pm}=1$ in Eq.~(\ref{equ:defineCpm}) which corresponds to the occurrence of the second type of vertex divergence. The above calculations thus prove that the eigenvector $V(\nu)$ in Eq.~(\ref{equ:ev0}), which has finite weight at {\sl all} Matsubara frequencies, is indeed related to this second type of divergences.

Let us also briefly comment on possible divergences in the particle-particle vertex $\Gamma_{pp}$. As is shown in the literature\cite{Shvaika2001,Ribic2016}, the latter is completely equivalent to the first part of the particle-hole one, i.e., to the expression in the first term of Eq.~(\ref{eq:DiagonalVertexFK}). However, one has to perform the frequency shift $\Omega\rightarrow\Omega-\nu-\nu'$ for the first ($\delta_{\nu\nu'}$-like) contribution to $\chi_{\text{c}}$ (see Sec. \ref{sec:2b1}) and take $\Omega=0$ {\sl after} this transformation. For this reason no divergences are found for the (zero transfer frequency) particle-particle vertex of the FK model at finite temperatures.

Summarizing the results of this Section, we have found for both the BM and the FK model vertex divergences, located at a {\sl single} given frequency $\pm \nu$ for interaction values $U(W)>\widetilde{U}(\widetilde{W})$. At finite $T$, these divergences occur along infinitely many curves in the phase diagram(s), which, for $T=0$, display a unique accumulation point in the metallic region. Remarkably, (i) the fully localized nature of these divergences, as well as (ii) of the corresponding eigenvectors of the generalized susceptibility, together with the (iii) temperature independence of the corresponding DMFT (CPA) single-particle Green's function are all manifestations of the emergence of a {\sl single} underlying energy scale $\nu^*(U)$ [$\nu^*(W)$]. The latter marks the progressive onset of the non-perturbative physics in the low energy sector of the spectral properties of the system already in the metallic phase at $\widetilde{U}<U_{\text{MIT}}$ [$\widetilde{W}<W_{\text{MIT}}$]. The presence of this energy scale is also reflected at the one-particle level in the non-single-valuedness of the self-energy functionals $\Sigma^{\pm}[G]$ for $U>\widetilde{U}$ [$W>\widetilde{W}$].

Moreover, but only in the FK model, a {\sl second} kind of divergences appear\cite{Ribic2016}, with qualitatively different properties: these divergences, as well their eigenvector, are not localized in frequency and, hence, not associated with a unique energy scale.

As we will see, despite the simplification of the interaction provided by the disordered models, they already provide a rather non-trivial description of the irreducible vertex divergences, which will also apply -to a significant extent- to the more complex cases treated in the next sections.

\section{Atomic Hubbard model}
\label{sec:atomic}

To make further steps in our progressive understanding of the irreducible vertex divergences, we will now consider another situation where analytical derivations are still feasible: the atomic limit of the Hubbard model (AL). In this case only two energies are present, i.e., the Hubbard interaction parameter $U$ and the temperature $T$. The latter sets the energy scale for the system which can be, thus, described in terms of the single parameter $U/T$. Differently from the BM and the FK model, in the AL the physics is enriched by the presence of an $\uparrow\downarrow$ susceptibility subject to the SU(2) symmetry of the system, which also characterizes the more realistic situation of the Hubbard model on a lattice. 
In the following we will again stick to the case of half-filling $\langle n_\uparrow \rangle=\langle n_\downarrow \rangle = \frac{1}{2}$. While this corresponds to the most correlated situation, let us remark that it represents a somewhat ``special'' situation, in the sense that the diagrammatic expansion up to second order of the self-energy (but -of course- not of the vertex functions!) is already exact\cite{Logan2014}, if performed in terms of the {\sl bare} interaction.

For the AL one- and two-particle Green's functions can be calculated analytically\cite{RVT, Pairault_2000, Hafermann_2009, georgthesis}. In particular, the generalized susceptibilities introduced in Sec. \ref{sec:2b1} can be obtained exactly in this limit. On the other hand, as the AL corresponds to a fully interacting system the determination of the irreducible vertices by a functional derivative of $\Sigma$ w.r.t. $G$ {\sl cannot} be performed analytically, different from what we did in Sec. \ref{sec:bmfk} for the BM and the FK model. In particular, although the physical one-particle Green's function has exactly the same form as the one for the BM/FK model in Eq.~(\ref{eq:FKGG0}) with $G_0(i\nu)=\frac{1}{i\nu}$, the latter relation does {\sl not} represent the (complete) functional $G[G_0]$. In fact, for the AL, one should rather calculate the Green's function and the corresponding self-energy in the most general way, i.e., in the presence of symmetry breaking fields, as discussed in Sec.~\ref{sec:2b2}. This will lead to highly non-local frequency dependencies, more complex than those observed for the functional derivative of $\langle n_\downarrow\rangle$ w.r.t. to $G$ in the FK model. 
Neglecting this aspect would lead to the incorrect prediction that the behavior of the vertex divergences in the BM and the atomic limit is exactly the same.

For the AL there are, in fact, in general no explicit expressions for the irreducible vertices $\Gamma_r$ currently available. However, as discussed in Sec.~\ref{sec:2b1} crucial information about divergences of $\Gamma_r$ can be obtained from the corresponding eigenvalues and eigenvectors of $\chi_r$. In the following, we will stick to the calculation of the generalized susceptibility according to Eq.~(\ref{eqn:ph}) and analyze its eigenvalues and eigenvectors. As most of the expressions are rather lengthy we present here just the main final results and refer the reader to the specific literature\cite{RVT, Pairault_2000, Hafermann_2009, georgthesis} for the technical details.

\begin{figure}[t!]
	\centering
		\includegraphics{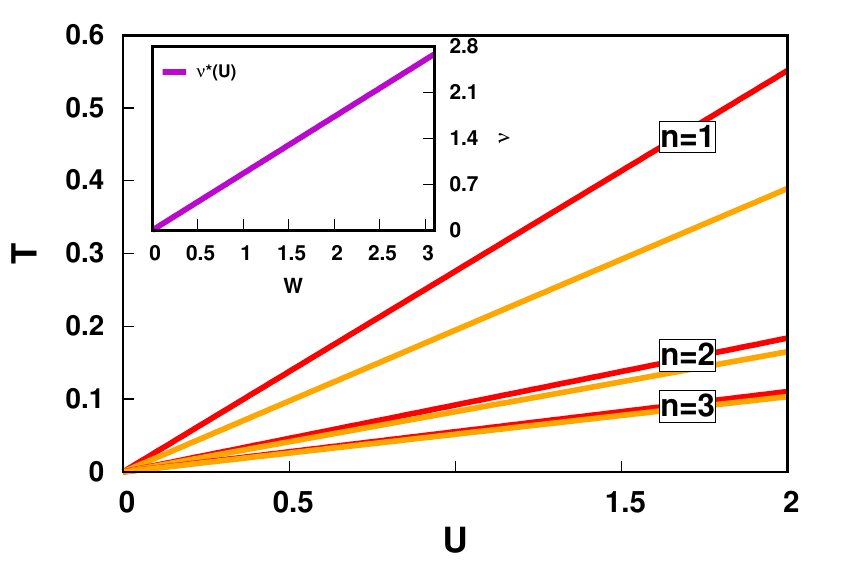}
	\caption{Main panel: Phase diagram for the atomic limit: The divergence lines for $\Gamma_c$ associated with the energy scale $\nu^*(U)$ are shown in red and classified according to the number of the Matsubara frequency at which the divergence takes place. The divergence lines of second kind, appearing in $\Gamma_{\text{si}}$, are depicted in orange. Inset: energy scale $\nu^*$ for the AL.}
	\label{fig:phase_diagram_AL}
\end{figure}

We start our analysis with the generalized charge susceptibility $\chi_c^{\nu\nu'(\Omega=0)}$ in the atomic limit. Its contributions can be classified into three distinct groups: (i) The first one depends on $\nu^2$ and $\nu'^2$ and is, hence, symmetric under the transformation $\nu^{(\prime)}\rightarrow -\nu^{(\prime)}$. (ii) The second one is proportional to $\delta_{\nu\nu'}$ and (iii) the third one is proportional $\delta_{\nu(-\nu')}$. This suggests the following ansatz for an eigenvector of $\chi_c$:
\begin{equation}
\label{equ:eval}
 V_{c,\bar{\nu}}(\nu)=\frac{1}{\sqrt{2}}\left[\delta_{\nu\bar{\nu}}-\delta_{\nu(-\bar{\nu})}\right],
\end{equation}
where $\bar{\nu}=(2\bar{n}-1)\pi T$ is an arbitrary fixed fermionic Matsubara frequency. When applying $\chi_c^{\nu\nu'(\Omega=0)}$ to this eigenvector all symmetric terms of type (i) vanish so that only the $\delta$-like contributions of type (ii) and (iii) contribute to the result. This reduces the eigenvalue problem to a two-by-two subspace of the full $\chi_c^{\nu\nu'(\Omega=0)}$ spanned by the frequencies $\bar{\nu}$ and $-\bar{\nu}$. Considering, moreover, the symmetry relation $\chi_c^{(-\nu)(-\nu')(\Omega=0)}=\chi_c^{\nu\nu'(\Omega=0)}$ simplifies our problem to the eigenvalue-problem of the following two-by-two block matrix\cite{Gunnarsson2016}
\begin{equation}
 \label{equ:twobytwo}
 \begin{pmatrix}\chi_c^{\bar{\nu}\bar{\nu}(\Omega=0)} & \chi_c^{(-\bar{\nu})\bar{\nu}(\Omega=0)} \\ \chi_c^{(-\bar{\nu})\bar{\nu}(\Omega=0)} & \chi_c^{\bar{\nu}\bar{\nu}(\Omega=0)} \end{pmatrix}.
\end{equation}
The corresponding eigenvalue to the antisymmetric eigenvector in Eq.~(\ref{equ:eval}) is then simply given by the difference between the diagonal and the out-of-diagonal elements of this matrix, i.e., $\chi_c^{\bar{\nu}\bar{\nu}(\Omega=0)}-\chi_c^{(-\bar{\nu})\bar{\nu}(\Omega=0)}$. Adopting the explicit expression for $\chi_c^{\nu\nu'\omega}$, which can be found, e.g., in Refs. \onlinecite{RVT,georgthesis}, the eigenvalue equation then reads 
\begin{equation}
\label{equ:eveq}
 \sum\limits_{\nu'}\chi_{\text{c}}^{\nu\nu'\Omega=0}V_{c,\bar{\nu}}(\nu')=\underbrace{\beta\frac{\bar{\nu}^{2}-\frac{3U^{2}}{4}}{(\bar{\nu}^{2}+\frac{U^{2}}{4})^{2}}}_{\lambda_{\text{c},\bar{\nu}}}V_{c,\bar{\nu}}(\nu),
\end{equation}
so that the analytic condition for a vanishing eigenvalue $\lambda_{\text{c},\bar{\nu}}\overset{!}{=}0$ can be expressed as
\begin{equation}
 \bar{\nu}=\frac{\pi}{\beta}(2\bar{n}-1)=\frac{\sqrt{3}{U}}{2}\quad\Leftrightarrow\quad\frac{{T}}{{U}}=\frac{\sqrt{3}}{2\pi}\frac{1}{2\bar{n}-1}.
 \label{eqn:ratio}
\end{equation}
The corresponding divergence lines are plotted in red color in the phase diagram shown in the main panel of Fig.~\ref{fig:phase_diagram_AL}. 
Evidently, Eq.~(\ref{eqn:ratio}) defines an energy scale associated with the vanishing of the eigenvalues of $\chi_c$ and a corresponding divergence of $\Gamma_c$:
\begin{equation}
 \label{equ:scaleal}
 \nu^*(U)=\frac{\sqrt{3}}{2}U,
\end{equation}
which is depicted in the inset of Fig. \ref{fig:phase_diagram_AL}. The irreducible charge vertex will, hence, diverge whenever one of its fermionic arguments matches this scale and the divergence will occur at exactly this frequency. Hence, in this respect, the situation is analogous to the one we have observed for all vertex divergences in the BM and for the  (frequency localized) divergences of first kind in the FK model. In particular, the emergence of an energy scale associated with the vertex divergences in the AL is induced by the very same ingredients, which we found for the BM and the FK model: (i) The locality of the corresponding eigenvector and (ii) the temperature independence of the single-particle Green's function. This suggests that also the corresponding non-single-valuedness of the self-energy functional should take place. While, as mentioned before, a direct analytical proof of this assumption is not easily feasible for the AL, its validity has been demonstrated in Ref.~[\onlinecite{Kozik2015}] by explicit numerical calculations following a scheme analogous to the one proposed in Eqs.~(\ref{subequ:itscheme}). 

In the AL, moreover, another type of divergences of $\Gamma_c$ and associated eigenvectors and (vanishing) eigenvalues exist, which share some similarities with the second divergences of the FK model: As in the latter case the corresponding eigenvectors have finite weight at {\sl all} Matsubara frequencies and, hence, the corresponding irreducible vertex $\Gamma_c$ diverges, globally, at all fermionic Matsubara frequencies if one of the related eigenvalues becomes $0$. As in the FK model, divergences of this type always occur {\sl after} the ones of the first type, i.e., at larger values of $U/T$. 

Remarkably, at the very same values $U/T$ where the second divergences of $\Gamma_c$ appear, one finds also a divergence of $\Gamma_{\text{pair}}$ in the AL. As discussed previously this is not present in the FK model, and it indeed indicates a richer structure of the AL w.r.t. the former two models. In the following we will, hence, focus on this new aspect of the second kind of divergences. Let us mention that, for technical convenience, we consider the so-called parquet-singlet channel $\Gamma_{\text{si}}=2\Gamma_{\text{pair}}+\Gamma_{\uparrow\uparrow}$ instead of $\Gamma_{\text{pair}}$: As $\Gamma_{\uparrow\uparrow}$ does not exhibit any divergences at $\Omega=0$ there is indeed a one-to-one correspondence between singularities of $\Gamma_{\text{pair}}$ and $\Gamma_{\text{si}}$. While the details about the formal structure of these vertex functions and, in particular, the related susceptibilities can be found in Refs. \onlinecite{RVT,georgthesis}, we report here just the eigenvector related to a vanishing eigenvalue of $\chi_{\text{si}}$:
\begin{equation}
 \label{equ:AL2ev}
 V_{\text{si}}(\nu)=2B\cos\left(\frac{\beta{B}}{2}\right)\sqrt{\frac{2B}{\beta[\beta{B}-\sin(\beta{B})]}}\frac{1}{\nu^{2}-B^{2}},
\end{equation}
where $B$ is given by
\begin{equation}
 \label{equ:defB}
 B=U\frac{\sqrt{-1+3e^{\beta U/2}}}{2\sqrt{1+e^{\beta U/2}}}.
\end{equation}
Let us stress that $V_{\text{si}}(\nu)$ in Eq.~(\ref{equ:AL2ev}) represents an eigenvector of $\chi_{\text{si}}$ only if the corresponding eigenvalue vanishes. The condition for the vanishing of an eigenvector is 
\begin{equation}
\label{equ:div2cond}
 \beta U=-\frac{\tan(\beta B/2)}{2\beta B},
\end{equation}
which corresponds to a transcendental equation for the quantity $\beta U=U/T$ and defines the slopes of the divergence lines in the corresponding $U-T$ phase diagram, depicted in Fig.~\ref{fig:phase_diagram_AL} in orange. As these divergences are associated with an eigenvector with spectral weight at all Matsubara frequencies [Eq.~(\ref{equ:AL2ev})], they are {\sl not} associated with a single energy scale in the system. Remarkably, upon reducing the temperature ($\beta\rightarrow\infty$) both types of divergences come closer to each other as it can be also inferred by rewriting Eq. (\ref{equ:div2cond}) in the form $\frac{\beta B}{2}=\arctan\left(\frac{2\beta^{2}U}{B}\right)$ and considering that $\arctan(\infty)=\frac{\pi}{2}(2n+1),n\in\mathbb{Z}$ where $n$ denotes the different branches of the multivalued $\arctan$-function.
Eq. (\ref{eqn:ratio}) is then indeed recovered when taking into account that for $\beta\rightarrow\infty$, $B\rightarrow\frac{\sqrt{3}}{2}U$.

\section{Hubbard model}
\label{sec:hubbard}

When turning to the DMFT study of the vertex divergences occurring in the Hubbard model, 
the possibility of deriving  analytical expressions is strongly limited. Hence, the data we
present in this section are essentially numerical: these will provide an extension to the {\sl whole}
phase-diagram of the (half-filled) Hubbard model of the first studies of Refs.~[\onlinecite{divergence,YangarXiv,Gunnarsson2016}].
The comparison
to the explicit analytic derivations of the previous sections, moreover, will allow for a rigorous interpretation of our results in
a large portion of the phase-diagram, where the nature of the divergences occurring in the DMFT solution of the Hubbard model 
turns out to be fundamentally similar to those we could precisely classify in the case of the disorder and atomic-limit models. Eventually, 
valuable insight will be gained also in the most interesting parameter region, where differences w.r.t. the results of Secs.~III-IV emerge through a detailed inspection of such discrepancies. 

\subsection{DMFT results}
\label{sec:hubb_dmft}

In this subsection, we present our numerical DMFT analysis of the 
vertex divergences in the Hubbard model. As anticipated, our starting point is
the preceding result of Ref.~\onlinecite{divergence}.
In particular, there, it was demonstrated that at certain loci of the phase-diagram, which represent {\sl two} lines in the $(U,T)$-plane, the irreducible vertex in
the charge $\Gamma^{\nu\nu'\Omega=0}_\text{c}$ as well as the
particle-particle up-down 
$\Gamma^{\nu\nu'\Omega=0}_{\text{pp},\uparrow\downarrow}$ channels diverge and -upon passing these parameter values- show an evident 
sign-change for $\Gamma$ at low fermionic frequencies. For large 
couplings, the two divergence lines can be traced up to the AL, where their slopes
coincide with the (semi-)analytic results obtained in the previous section. Lowering the temperature the
lines start to bend backwards (re-entrance), ``getting around'' the critical endpoint of the
MIT (see also the discussion of Fig. \ref{fig:pd} below).

\noindent
\begin{figure}[ht!]
    \centering
    \includegraphics[width=0.45\textwidth]{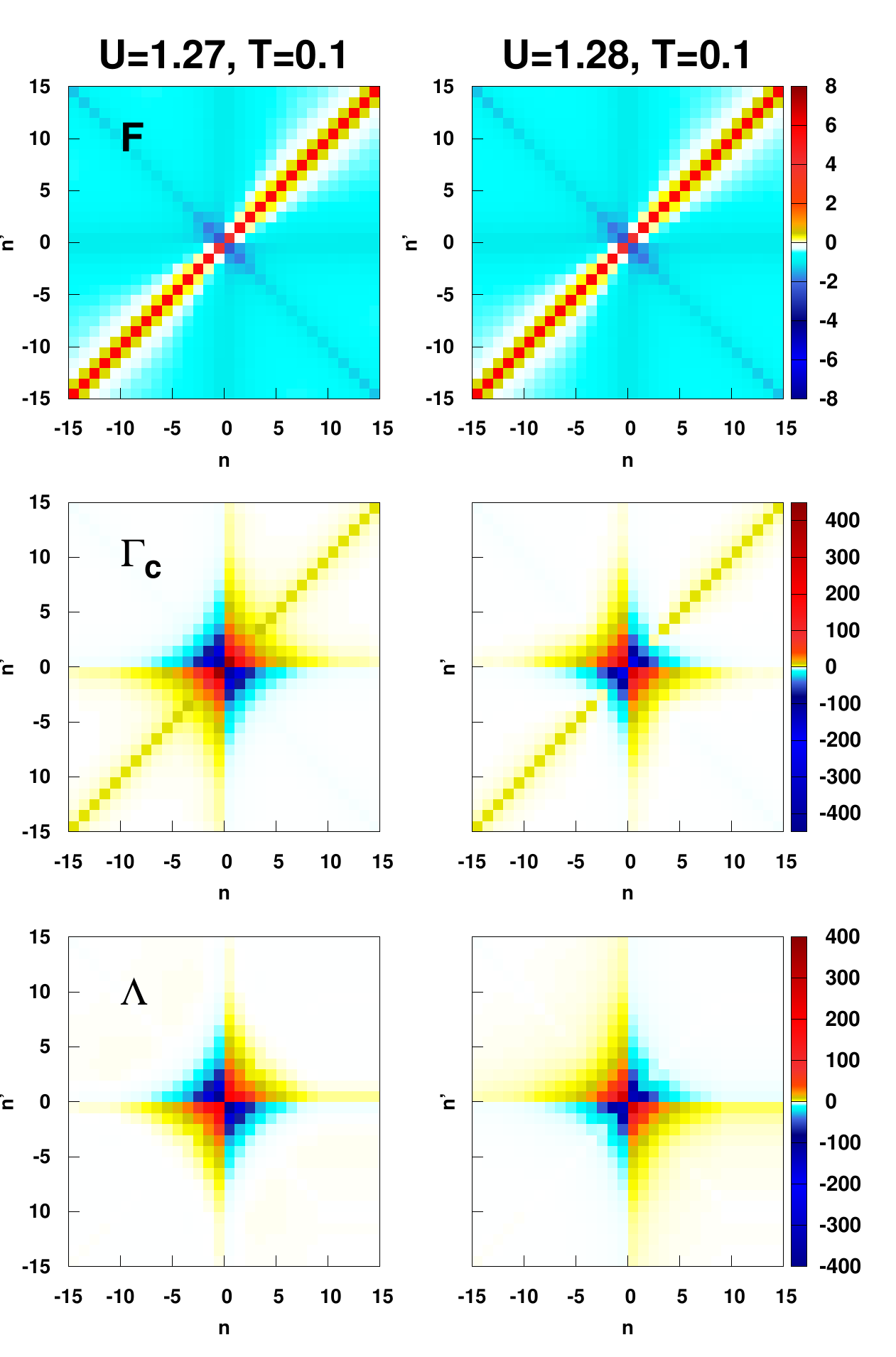}
    \caption{Upper panel: Matsubara frequency density-plot of the full (local) scattering amplitude  $F^{\nu\nu'(\Omega=0)}$ of the DMFT solution of the Hubbard model, on both sides of the first divergence (red) line of Fig.~\ref{fig:pd} at $T=0.1$. Middle/Lower panels: the same for 2PI functions in the charge $\Gamma_{\text{c}}^{\nu\nu'(\Omega=0)}$ and in all channels $\Lambda^{\nu\nu'(\Omega=0)}$, respectively. Note that, for a better readability, in all cases the bare interaction term $U$ has been subtracted.}
    \label{fig:fglambda}
\end{figure}

\begin{figure*}[ht!]
    \centering
    \includegraphics[width=1.0\textwidth]{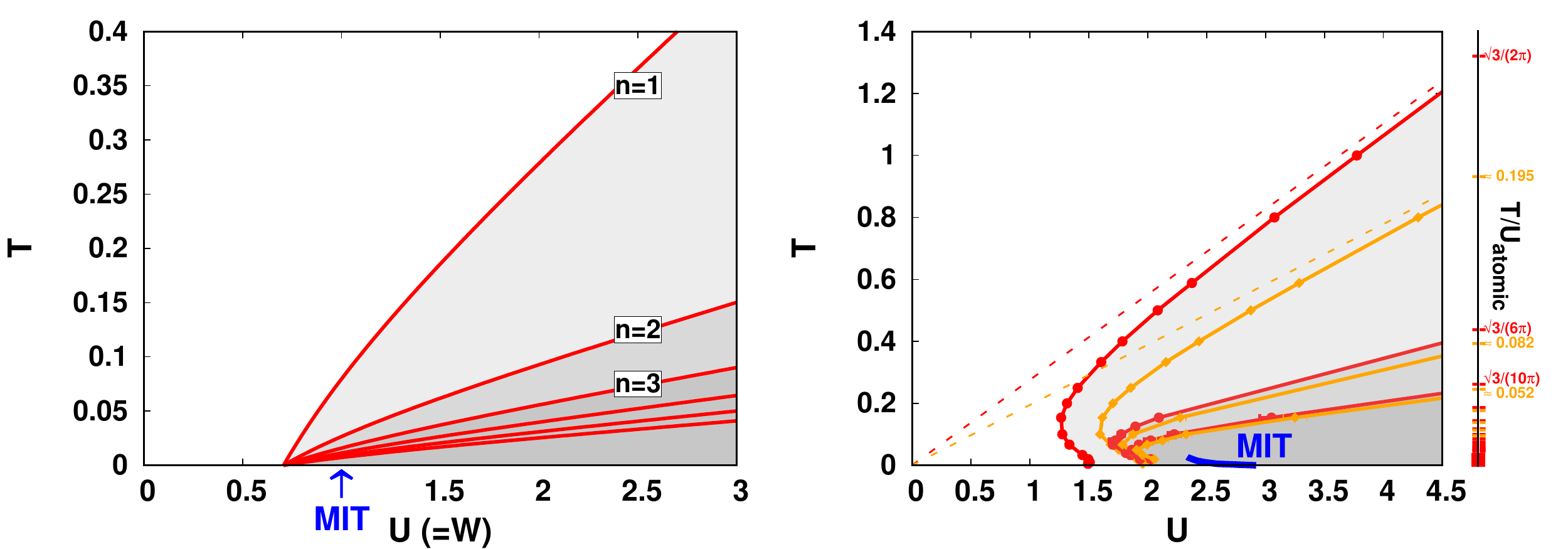}
    \caption{Left panel: DMFT phase-diagram of the irreducible vertex divergences (of the first kind) of the BM/FK model (see Sec. III), replotted for a comparison against the Hubbard model results discussed in this section. Right panel: DMFT phase diagram showing the landscape of irreducible vertex divergences surrounding the Mott-Hubbard MIT in the half-filled unfrustrated Hubbard model. The blue line indicates the MIT\cite{Bluemer_PhD}. The red and orange lines show the points where $\Gamma^{\nu\nu'(\Omega=0)}_\text{c}$ diverges at low frequencies, whereas an additional divergence of $\Gamma^{\nu\nu'(\Omega=0)}_{\text{pp},\uparrow\downarrow}$ takes place simultaneously to the one of $\Gamma^{\nu\nu'(\Omega=0)}_\text{c}$ (only) at the orange lines. On the right-hand side, the (exact) values of the ratio $T/U$ for the atomic limit ($t=0$) are listed.} 
    \label{fig:pd}
\end{figure*}

We also recall that the reported divergences are intrinsically different \cite{divergence} from those
occurring at physical phase transitions, e.g. at the Mott-Hubbard
transition at $T\!=\!0$. The latter appear indeed at the level of
the full two-particle scattering amplitude ($F$), which is related to the generalized susceptibility
via
\begin{equation}
 F^{\nu\nu'(\Omega=0)}=-\frac{1}{G^{2}(\nu)}\left(\chi_{\text{c}}^{\nu\nu'\Omega=0}-\chi_{0}^{\nu\nu'\Omega=0}\right)\frac{1}{G^{2}(\nu')},
\end{equation}
while
the ones discussed here regard the irreducible vertex functions $\Gamma_{\text{c}}$ and $\Gamma_{\text{pp}}$.

In fact, our divergences are deeply rooted in the diagrammatics, as it turns out that they appear, simultaneously, also in the fully two-particle irreducible (2PI) vertex-function $\Lambda$, extracted through the inversion of the parquet equations\cite{Bickersbook,RVT,Gunnarsson2016}.
 
The considerations above are nicely illustrated by the DMFT data shown in Fig.~\ref{fig:fglambda}, which complete the preceding analysis of Ref.~[\onlinecite{divergence}] of the first divergence line, see also \onlinecite{SchaeferPhD}. They show a Matsubara fermionic frequency density-plot of the local DMFT vertex functions of the Hubbard model across the first divergence line in the phase-diagram, classified in terms of their increasing two-particle irreducibility: from the full scattering amplitude $F$ (upper panels) to the irreducible vertex in the charge channel $\Gamma_c$ (middle) to the fully 2PI vertex $\Lambda$ (bottom panel). 
In particular, consistent with Ref.~\onlinecite{divergence}, we work at fixed $T=0.1$, selecting two values of $U$ slightly lower ($U=1.27$) and slightly larger ($U=1.28$) than the interaction $\widetilde{U}$ where the first divergence of $\Gamma_c$ occurs at this $T$. 
Our numerical data clearly show that {\sl no} effect of divergences can be traced in the full scattering function $F$, evidently convergent for all Matsubara frequencies. On the contrary (note also the different color scales!), the effects of the proximity to a vertex divergence located between the two $U$ values are determining -with a very similar pattern- the low-frequency structure of the vertex functions irreducible in the charge-channel ($\Gamma_c$) and in all channels ($\Lambda$), respectively.

The DMFT results for the divergences in the {\sl whole} phase-diagram are reported in
Fig.~\ref{fig:pd} (right panel).  Our new data for the positions of the vertex divergences makes the DMFT ``map''
of irreducible vertex divergences surrounding the Mott-Hubbard MIT significantly richer. Specifically, the blue
line marks the Mott-Hubbard MIT of DMFT and the red lines indicate the points in the phase
diagram where $\Gamma^{\nu\nu'(\Omega=0)}_\text{c}$ diverges, whereas at
the orange lines {\it both}  $\Gamma^{\nu\nu'(\Omega=0)}_\text{c}$ and
$\Gamma^{\nu\nu'(\Omega=0)}_{\text{pp},\uparrow\downarrow}$ diverge
simultaneously. Furthermore, the values listed in red/orange on the right-hand side of Fig.~\ref{fig:pd} are the ratios $T/U$ for which the irreducible vertices in the AL diverge (see Sec. \ref{sec:atomic}):  
It can be seen that the slopes of the extrapolated divergence lines of $\Gamma_{\text{c}}$ and $\Gamma_{\text{pp}}$ of DMFT coincide with
the corresponding ratios in Eq.~(\ref{eqn:ratio}) for $\bar{n}=0$ and Eq.~(\ref{equ:div2cond}) for the smallest solution of this (transcendental) equation.

Thus, the extension of the work of Ref.~\onlinecite{divergence} results in a 
highly non-trivial divergence phase-diagram for the Hubbard model (Fig.~\ref{fig:pd}), as the two lines already reported in Ref.~\onlinecite{divergence} are evidently 
not the {\sl only} ones where a vertex divergence takes place. 
By approaching the MIT from the metallic phase, one observes several
eigenvalues of the generalized charge and particle-particle
susceptibilities progressively passing through zero (``singular'' eigenvalues) at certain values of
$(\tilde{T},\tilde{U})$. These determine the corresponding divergences
of the irreducible vertices as well as a sign
change of their low-frequency structure on the two sides of the
divergence line\cite{divergence}. Their positions in the phase diagram are marked -in the usual notation- by red and orange lines. By moving further into the non-perturbative parameter regime, because of the high density of divergence lines, the extraction 
of the irreducible vertex functions becomes more challenging, and we could determine with sufficient numerical 
accuracy only the first seven divergences of $\Gamma_{\text{c}}$ and $\Gamma_{\text{pp}}$. However, by exploiting the one-to-one large-$U$ correspondence of the divergence lines in the Hubbard phase diagram
with the infinitely many divergences of the exact atomic limit solution (see Sec.~IV and rightmost scale in Fig~.\ref{fig:pd}), we can infer a presence of
an {\sl infinite} number of (red and orange) divergence lines over the whole phase diagram.
 Finally, we should also remark that the irreducible vertex in the (predominant)
spin channel $\Gamma_{\text{s}}^{\nu\nu'\Omega}$ does not exhibit any
low-frequency divergences in the whole parameter region considered.

As for the low-temperature regime ($\beta \gtrsim 100$), the numerical treatment becomes significantly harder. Hence, in addition to HF-QMC, we have performed CT-QMC calculations in the hybridization expansion\cite{Gull,Parragh2012,markusthesis}.
The latter does not suffer from the finite-size problems, neither the ones of ED (bath discretization) nor the ones of HF-QMC (Trotter time-discretization), and can more easily access lower temperatures. 
Let us however note that if a qualitative change in the shape of the divergence lines would take place at exponentially small temperature scales, even the CT-QMC analysis might miss it. The presence of an exponentially small temperature scale characterizes for instance the physics of some multi-orbital models \cite{spin-freezing}, for which the Fermi-liquid coherence temperature \cite{coherence} could not be reached by CT-QMC calculations.
Bearing these limitations in mind, the results for the first two divergence lines are compatible with a $T\rightarrow{0}$ extrapolation
of $\widetilde{U}(T=0)\approx 1.45$ and $\approx 1.95$, respectively (for details about the low-temperature data see Appendix~\ref{app:lowT} and Fig.~\ref{fig:lowT} and \ref{fig:bethe} therein): In both cases $\widetilde{U} <U_{c2} \sim 3$ of the MIT, and, therefore,
 the line terminates well inside the metallic regime, where a well defined, coherent quasi-particle peak is visible in the 
DMFT spectral functions.

Looking at our divergence results for the whole phase diagram, one will -- first of all --
 find a confirmation for the heuristic
interpretation proposed in Ref.~\onlinecite{divergence} of the
irreducible vertex singularities as {\sl non-perturbative precursors} of the Mott-Hubbard transition, linked\cite{Gunnarsson2016} to a gradual suppression of the physical charge susceptibility and the opening of a spectral gap when approaching the MIT.
Besides this rather generic consideration, however, it remains a problem to gain deeper understanding of the 
origin of this impressive manifestation of the breakdown of perturbation theory around the MIT, of its interrelation
 with other theoretical aspects of the non-perturbative physics, and -- if they exist -- of its effects in observable quantities.  
To this aim, in the next subsection, we will proceed by performing a detailed comparison of the Hubbard model data of Fig.~\ref{fig:pd} (right panel) with those
of simplified models reported in Secs.~III-IV. 

\subsection{Interpretation of the results}

\begin{figure}[t!]
    \centering
    \includegraphics[width=0.40\textwidth]{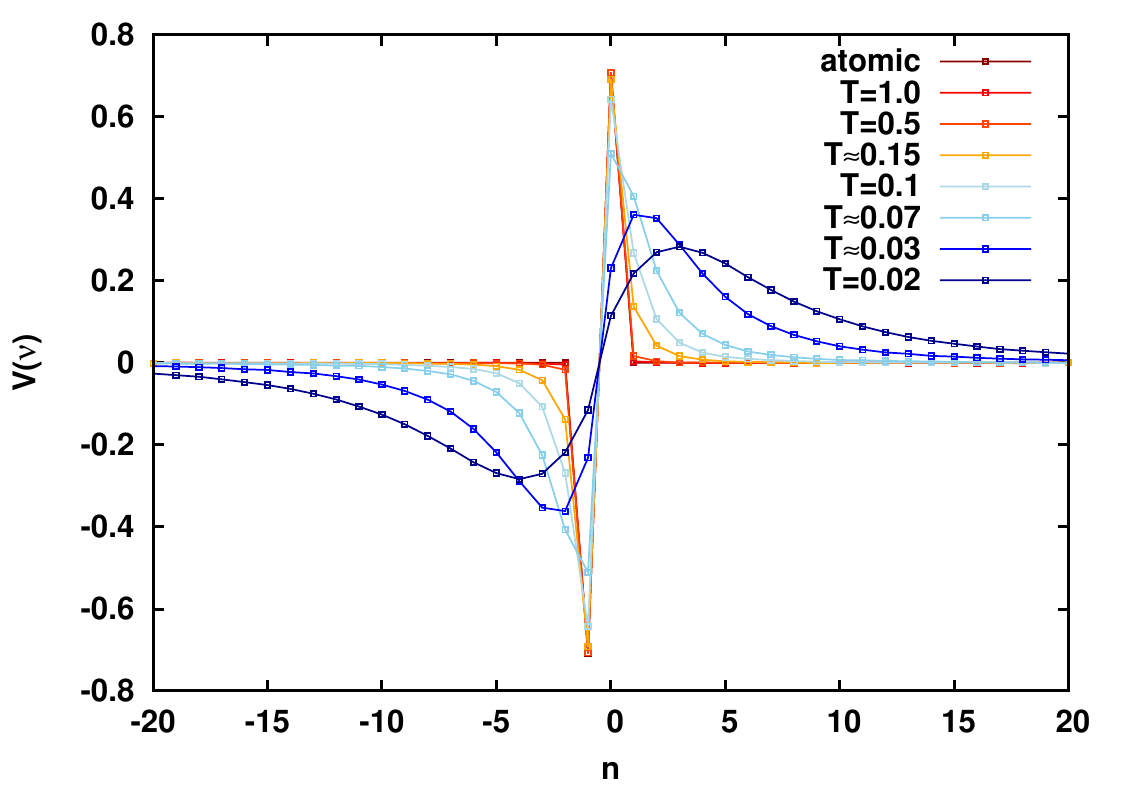}
    \caption{Evolution of the frequency structure of the eigenvector of $\left(\chi_{\text{c}}/\chi_{0}\right)^{\nu\nu'(\Omega=0)}$ (see also footnote \onlinecite{chi0}) associated with the first zero eigenvalue for different temperatures along the first divergence line of Fig.~\ref{fig:pd}.}
    \label{fig:ev_d}
\end{figure}

Despite the high degree of complexity displayed by the many vertex divergence lines surrounding the MIT of the Hubbard model
(Fig.~\ref{fig:pd}, right panel),  these can be classified -- in a large portion of the phase diagram -- in a similar, simple framework
like that of the disordered/atomic-limit model.
In particular, by comparing our results for the Hubbard model (Fig.~\ref{fig:pd}, right panel) with those of the disordered models
(Fig.~\ref{fig:pd}, left panel) and of the atomic limit (Fig.~\ref{fig:pd}, rightmost scale), we note immediately a qualitatively similar behavior of both (red and orange) kind of divergence lines in all three cases in the region of large $U$ and $T$. This corresponds to the parameter region, where the relation between
$\widetilde{U}$ and $\widetilde{T}$ is approximately linear in the Hubbard model. 
On the other hand, by following the divergence lines towards the low-$T$ regime, one observes {\sl two} relevant discrepancies, between the
the Hubbard and the BM/FK/atomic results: (i) an evident re-entrance is displayed by all the lines, (ii)
 the divergence lines do no longer accumulate at a unique point in the $T\rightarrow 0$ region (Hubbard model).\\
This twofold outcome of the visual comparison between the phase-diagrams of Sec.~V and Secs.~III-IV (disordered models, AL) is also supported, quantitatively, by the 
analysis of the evolution of the singular eigenvector $V(\nu)$ along the divergence lines. For the sake of conciseness, we discuss here 
the frequency structure of the eigenvector corresponding to the first (red) divergence line in Fig.~\ref{fig:pd} (right panel).  
 Specifically, Fig.~\ref{fig:ev_d} shows
the evolution of the singular eigenvector (associated with a zero eigenvalue\cite{SchaeferMSc,chi0}) of 
$\chi_{\text{c}}^{\nu\nu'\Omega=0}$ by reducing gradually the value of $U$ and $T$ from the strong-coupling/high-$T$ limit, 
following the first divergence line of $\Gamma_{\text{c}}^{\nu\nu'\Omega=0}$.  
One immediately notices that, within the region
where the divergence curve in the phase diagram can be approximated well by a
straight line, the frequency shape of eigenvector of $\chi_{\text{c}}^{\nu\nu'\Omega=0}$ is
very close to  the purely frequency-localized form of Eqs.~(\ref{equ:eigv}) and (\ref{equ:eval}). This means that its frequency shape is exactly the 
same of the singular eigenvector identifying the first (red) divergence line in the BM/FK and atomic limit phase-diagrams.

On the other hand,
for lower temperatures, the frequency structure of $V(\nu)$ in the Hubbard model deviates from the
delta-peaked one of Eqs.~(\ref{equ:eigv}) and (\ref{equ:eval}). In particular, the discrepancy gets progressively larger by decreasing ($U,T$) and
starts to become appreciable in correspondence
to the re-entrance of the curve in the phase-diagram.
This marks evidently a qualitative difference w.r.t.~the BM/FK/atomic case, where the frequency-structure of  $V(\nu)$
 does not change along the entire (red) divergence lines.

On the basis of the above comparisons, a precise insight into the divergences occurring in the large $U$ region of the Hubbard model
can readily  be gained. There, the shape of the lines and the frequency structure of the corresponding singular 
eigenvector are the same as in the BM/FK/atomic model. This means that the same scenario applies as in those cases: each (red) divergence line can be associated with a fixed Matsubara frequency, $\nu = (2n-1) \pi T$,  and all of them will be (almost perfectly) collapsing to a {\sl unique} line via a rescaling factor equal to the corresponding Matsubara index $2n-1$.  We  deduce, as a consequence, that the multiple divergence lines in the large ($U$,$T$) limit of the Hubbard model are also a manifestation of an underlying {\sl single} energy scale $\nu^*(U)$. The latter will be well approximated in this regime by the corresponding atomic limit expression, given in Eq.~(\ref{equ:scaleal}), which also implies -to a large extent- that all consideration made about the multivaluedness of $\Sigma[G]$ (and of the related Luttinger-Ward functional) are also applicable here. Somewhat similar considerations can be drawn also for the second kind of
divergences (orange lines), which for large $U$ and $T$ resembles in many respects, the corresponding  divergences of second kind in the FK model/atomic limit: In both cases, they always occur after the divergences of first kind, they are associated with a non-localized singular eigenvector, and, thus, appear as global divergences of the whole irreducible vertex functions at each pair of fermionic Matsubara frequencies $\nu$ and $\nu'$. However, as already noted in Sec.~III~B, in the Hubbard model, as well as in the atomic limit, such divergences of second kind occur simultaneously in {\sl two channels}, while they were confined to the particle-hole sector in the case of the FK model.

The theoretical interpretation becomes obviously more challenging in the low-$T$ regime, where all the divergence lines display a pronounced bending in the Hubbard-phase diagram: Here, our numerical results  for the Hubbard model deviate qualitatively from those of the BM/FK/atomic case. Thus, we cannot certainly invoke the same scenario as in the large ($U$,$T$) limit.  

This low-$T$ regime, on the other hand, is one of the most interesting in the DMFT phase-diagram of the Hubbard model: It corresponds to the correlated metallic phase on the left-side of the Mott-Hubbard MIT, whose low-energy physics is -- differently from the case of BM/FK/atomic-limit -- dominated by the presence of a narrow coherent quasiparticle (Kondo) peak. In particular, an important question naturally arises about the persistence of  low-energy excitations of (supposed) Fermi-liquid nature in a parameter regime where the perturbation theory breakdown has already taken place, as signaled by the presence of several divergence lines. In other words, the question is, how the highly {\sl non-perturbative} divergences which we found for $U<U_{\text{MIT}}$ can coexist -for the same parameters- with an essentially Fermi-liquid-like physics at low energy? Does a non-perturbative Fermi liquid exist? Or was the assumption of having Fermi-liquid excitations for all $U< U_{c2}$ (at $T=0$) not fully correct? 

Because of the lack of analytic expressions for the vertex divergences and for the associated singular eigenvector in this regime, we cannot rigorously answer these questions yet. However, the discrepancies we observe w.r.t. the FK/BM case are already suggesting possible alternative scenarios.  If we focus, for the sake of simplicity, on the first kind  of (red) divergence lines, we note that the condition ensuring the existence a {\sl single} energy scale $\nu^*$ below which the self-energy becomes {\sl non-perturbative}, i.e. a pure locality of the singular eigenvector (see Sec.~III), is progressively violated by reducing $T$ and $U$. Hence, the existence of an single underlying $\nu^*(U)$, such as that shown in the inset of Fig.~\ref{fig:phase_diagram_FK} and in Fig.~\ref{fig:phase_diagram_AL}, should be disregarded in the correlated metallic regime of the Hubbard model. In fact, we note that the presence of such a scale would have been particularly hard to reconcile with the Fermi-liquid nature of the low-frequency self-energy  in this parameter regime. 

Speculating further on these considerations, one of the simplest scenarios to be considered may be the following: By increasing $U$ the breakdown of perturbation theory cannot happen as in the BM/FK/atomic limit cases starting from the low energy sector (i.e., from $\omega^*=0$ at $U=\widetilde{U}$), because the perturbative nature of such a sector must be somewhat ``protected'' by the quasi-particle peak excitations. In the DMFT for the Hubbard model, such protection would occur through a Kondo screening of the auxiliary AIM, evidently missing in the other model cases (BM,FK, atomic-limit).
However, nothing would prevent a perturbative breakdown to occur initially at finite real frequencies at a given $\widetilde{U}$, where the first divergence is found, and then to gradually extend its influence towards lower and higher frequencies. In this scenario, at $T=0$ only at $U=U_{c2}$, i.e., at the MIT, the non-perturbative features of finite frequency would eventually reach the Fermi level. Then, only when the QP excitations are gone, the underlying scenario would become again similar to the one described in Secs.~III-IV. At the same time, the apparent contradiction of observing a Fermi-liquid behavior in a non-perturbative regime, would be solved by the fact that the low-frequency/low-energy region would remain perturbative until the MIT occurs. 
Obviously, it is at the moment unclear how the {\sl two} energy scales can be formally related to the vertex divergences in Matsubara space.

While a detailed study of the realization of the abovementioned scenario, or of even more complicated ones, is beyond the scopes of the present work, it is important to notice that some hints for the existence of an intermediate scale for the occurrence of non-perturbative features in the strongly  correlated metal described by DMFT might  already be traced in some of the peculiar spectral and transport-property features of the DMFT solution for the half-filled Hubbard model. Here, we are referring to the occurrence of kinks\cite{Byczuk2007,Raas2009,Toschi2009,Held2013} in the spectral function and in the specific heat. Such kinks, in fact, appear at finite frequency and temperature, precisely in the correlated metallic regime of the Hubbard model, where the irreducible vertex divergences are also found. 
In particular, within a scenario with two energy scales, one could suppose that the lowest one might be associated with the kink energy, as both of them would vanish exactly at the MIT.

\section{Conclusions and outlook}
\label{sec:conclu}

In this paper, we have presented a systematic quantum many-body investigation of the divergences of the irreducible vertex functions, which represent {\sl one} of the several manifestations of the perturbation-expansion breakdown in correlated electron theory.
By considering the intrinsically non-perturbative DMFT solutions for correlated electronic Hamiltonians of increasing 
complexity (from the disordered BM and FK, over the atomic limit case to the Hubbard model), we were able to extend significantly the results of previous studies\cite{divergence}. In particular, exploiting the analytical derivations for the disordered/atomic limit cases, we could: (i)  demonstrate the existence of {\sl infinitely} many divergences in all phase-diagrams of the models considered; (ii) classify these divergences in terms of the (frequency localized/not localized)  properties of the corresponding singular eigenvectors and of the existence of a {\sl single} underlying energy scale, delimiting the frequency range affected by non-perturbative features; (iii) precisely demonstrate the link of the vertex divergences to other manifestations of the perturbation-theory breakdown, such as the multivaluedness of the self-energy (and, hence, of the Luttinger-Ward) functionals.

All these findings apply to  a certain extent also to the {\sl infinitely} many divergences lines of the phase-diagram of the Hubbard model, at least in the strong-coupling, high-$T$ regime, and provide possible interpretative keys for the most interesting region of the Hubbard DMFT phase diagram, the low-$T$ strongly correlated metallic regime. There, the qualitative modification of the frequency dependence of the singular eigenvectors seems to preclude the explanation in terms of a unique energy scale, marking the most important discrepancy w.r.t. the physics of the disordered/atomic-limit models. We ascribe this difference to the presence of a coherent quasi-particle peak in the DMFT spectral functions of this parameter regime: The Kondo screening at work in the auxiliary AIM of DMFT should ``protect'' the low-frequency physics from the non-perturbative divergences, and allows for the stabilization of the Fermi-liquid physics in DMFT solutions until the Mott-Hubbard MIT is reached. However, as nothing prohibits the non-perturbative breakdown to first manifest at {\sl finite} frequencies (where Kondo screening work less effectively) in the Hubbard model case, we speculate that a minimal model with at least {\sl two} characteristic energy scales might capture the essence of the irreducible vertex divergences in the Hubbard model. The lower one of such energy scales might be then related to the finite-energy kinks, whose emergences have been demonstrated in the same correlated metallic regime of our interest.  

The progress in the understanding of the non-perturbative physics and its several manifestation in quantum many-body theory, as reported in this paper, is of possible impact in different respects and potentially inspiring for future studies. In particular, from a more algorithmic perspective, the discussion of the validity of the Feynman diagrammatics in non-perturbative regime can play a central role in the future developments of diagrammatic extensions of DMFT. Particularly important information is encoded in the fact that the multiple irreducible vertex divergence lines can be
  the simple manifestation of an {\sl single} energy scale, and that these divergences are {\sl never} appearing in the full scattering amplitude $F$, but that -- at the same time -- they are rooted down to the most fundamental building-block of the parquet equations, the fully two-particle irreducible vertex function. Also of importance, especially for the bold diagrammatic QMC algorithms, is the definition of the frequency/energy range, controlled by non-perturbative effects, where the standard resummation of diagrammatic (skeleton) series can lead to unphysical results. 

As for the physics, our combined analysis of different disordered and the Hubbard models, support the proposed\cite{divergence} interpretation of the occurrence of irreducible vertex divergences as {\sl non-perturbative precursors} of the Mott MIT in correlated systems. In the case of different distributed disordered models, in fact, the absence of the MIT determines the disappearance of all vertex divergences. As for the Hubbard model, moreover, the Mott-Hubbard MIT cannot be reached from the perturbative regime without crossing several divergence lines (the exact number likely depending on the model details). In general, the divergences of the vertex functions appear to be triggered by a separation of energy scales, which, in the model considered, freezes the charge fluctuations and leads ultimately to the opening of a spectral gap.

Our understanding of the highly non-perturbative features of the MIT physics, however, is far from being complete for the case of the Hubbard model, where the quasi-particle physics most likely competes with the reduction of charge fluctuations at low frequencies.

Hence, further effort should be made in this direction, possibly further combining information extracted by analytic and numerical calculations. Eventually, extensions to include the effect of non-local correlations and interactions, e.g.,  by means of the Dynamical Cluster Approximation and the EDMFT should be also considered. In the former case, the recently proposed relation of the vertex divergences with the formation of an RVB cluster state and pseudogap spectral features \cite{Gunnarsson2016} might be tested against the existence of {\sl infinitely} many divergence lines and the existence of different non-perturbative energy scales for different momenta at the Fermi level. In the latter case, we expect that systems with long-range interactions might indeed display qualitatively similar divergence features.

A full understanding of all non-perturbative aspects\cite{divergence,YangarXiv,Janis2014,Kozik2015,EderComment,Stan2015,Rossi2015,Ribic2016,Gunnarsson2016} of the (non-relativistic) quantum many-body theory will certainly require considerable additional efforts. These efforts might be, however, highly rewarding in the perspective of gaining the total control of the theory of correlated electron systems in their most elusive, but phenomenologically interesting, parameter regions.  \\

\textit{Acknowledgments.} We are indebted to P. Gunacker for benchmarking our  CT-QMC vertex calculations at the lowest temperature. We thank S. Andergassen, J. LeBlanc, M. Capone, C. Taranto, A. Tagliavini, N. Wentzell for insightful discussions and P. Chalupa and M.-T. Philipp for carefully reading our manuscript. We thank A. Berger, P. Romaniello, F. Werner and L. Reining for organizing the workshop ``Multiple Solutions 
in Many-Body Theories'' (Paris, June 2016) which led to illuminating discussions in the context of the subject of this paper. We acknowledge support from the Austrian Science Fund
(FWF) through the Doctoral School ``Building Solids for Function''  (TS, FWF project W1243), through the project ``Quantum Criticality in Strongly Correlated Magnets''  (AT, TS, GR; FWF project I-610-N16) and the SFB ViCoM  (FWF project F41). G.S. acknowledges financial support from the research unit FOR 1346 of the Deutsche Forschungsgemeinschaft. Calculations were performed on the Vienna Scientific
Cluster (VSC).

\appendix


\section{Technical details of the Hubbard model DMFT calculations}
\label{app:DMFT}
As for the results for the Hubbard model in DMFT presented in Sec. \ref{sec:hubbard}, most of the quantities of interest have been calculated by means of a Hirsch-Fye Quantum-Monte-Carlo solver. The accuracy of the results has been checked by a comparison with corresponding exact diagonalization calculations. 
Specifically, for the vertex functions shown in the main text, the Hirsch-Fye quantum Monte Carlo solver has been used with a grid of $40\times40$ positive fermionic frequencies and a Trotter-discretization of $\Delta\tau=\frac{\beta}{N_{\beta}}=\frac{1}{16}$ down to an inverse temperature of $\beta=10.0$ and 
$80\times 80$ positive fermionic frequencies and a Trotter-discretization of $\Delta\tau=\frac{\beta}{N_{\beta}}=\frac{1}{50}$ for all lower temperature calculations, respectively.
\\\\
The CT-QMC data was obtained from the hybridization expansion (CT-HYB) solver of the w2dynamics package (version 0.5, git 901b8669, see Refs.~\onlinecite{markusthesis,Parragh2012}) with 512 CPU cores \`{a} 100,000 sweeps \`{a} 1000 steps. The relevant estimator was the two-particle Green's function, which was measured separately for the particle-hole channel, $G^{\nu\nu^\prime\Omega}_\mathrm{ph}$, and particle-particle channel, $G^{\nu\nu^\prime\Omega}_\mathrm{pp}$.  $2\beta$ fermionic frequencies were used for both $\nu$ and $\nu^\prime$, while the bosonic transfer frequency $\Omega$ was pinned to zero.  To avoid transformation bias, the measurement was directly performed in frequency space by use of a non-equidistant fast Fourier transform\cite{markusthesis}.
As the error bars of $G^{\nu\nu^\prime\Omega}$ are essentially constant in frequency in CT-HYB, while its value drops as the inverse of the frequency, care must be taken when calculating $\Gamma$: for a given run, as we enlarge the frequency box, the inversion will eventually be dominated by the high-frequency noise, rendering the result unreliable.  To circumvent this, we start from a small frequency box in the inversion and subsequently add more frequencies until the eigenvalue of interest is well-converged\cite{markusthesis}.

\section{Divergence lines at low temperatures}
\label{app:lowT}
\begin{figure}[t!]
  \includegraphics[width=0.4\textwidth]{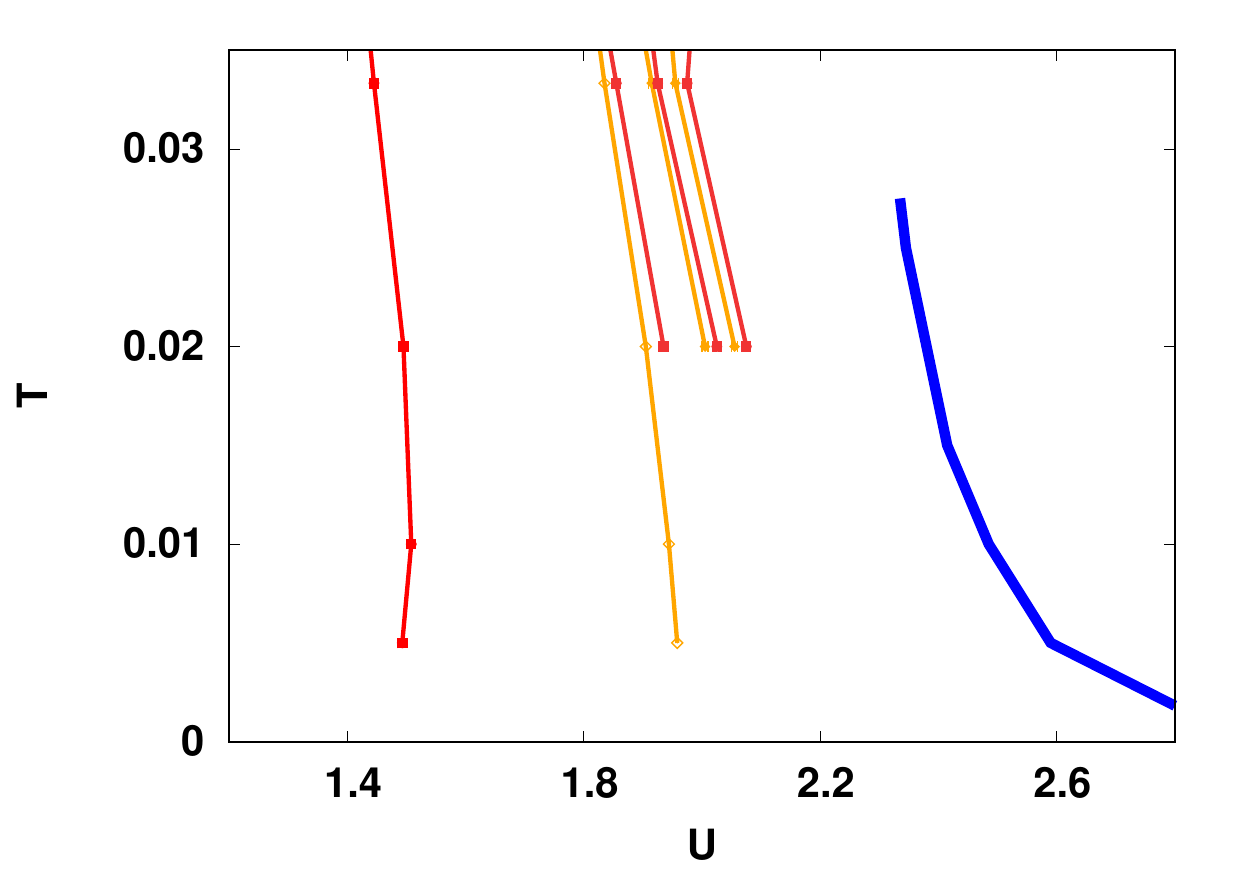}
  \caption{Low temperature dependence of the first seven divergence lines in the Hubbard model for the 2D square lattice. The blue line represents the Mott-Hubbard transition.}
  \label{fig:lowT}
\end{figure}

\begin{figure}[b!]
  \includegraphics[width=0.4\textwidth]{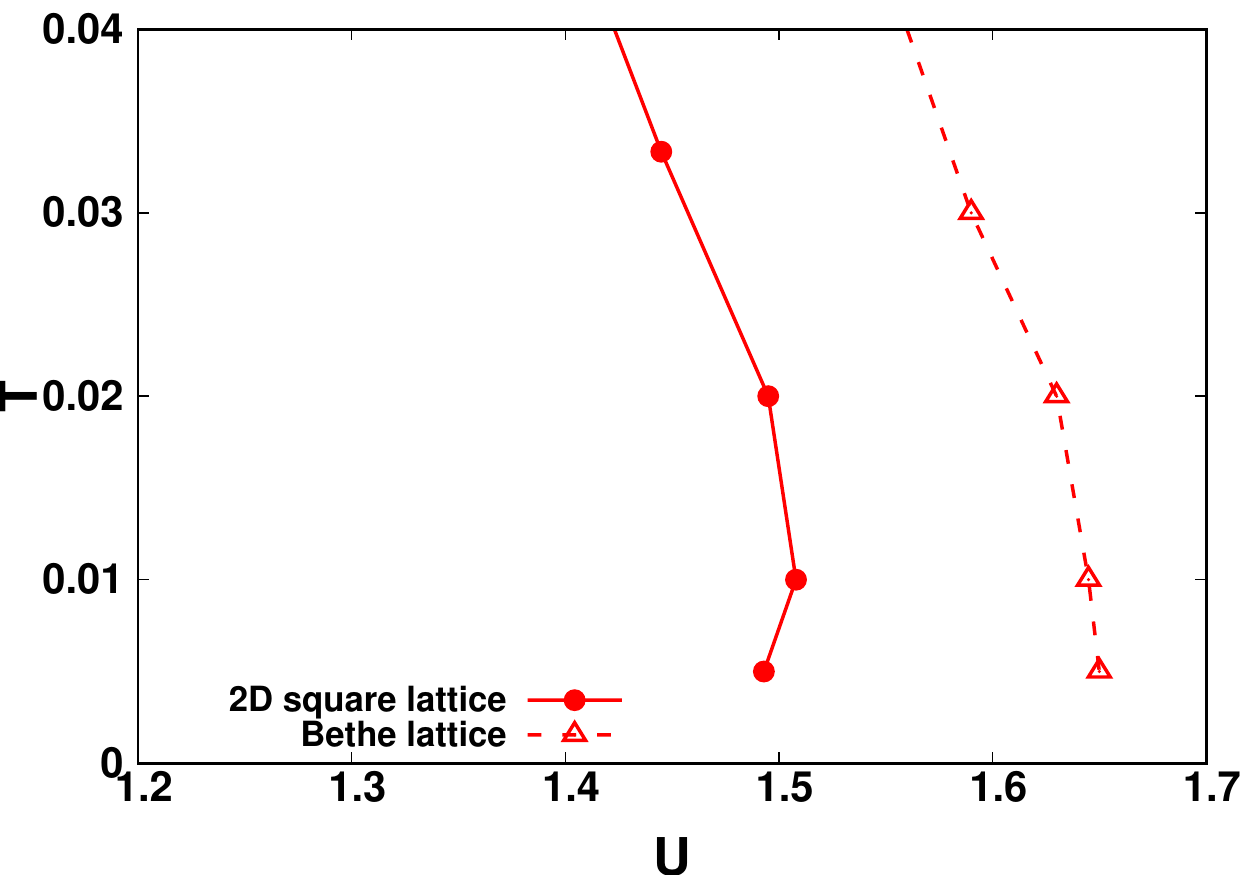}
  \caption{Low temperature dependence of the first divergence line in the Hubbard model for the 2D square lattice (solid line) and the Bethe lattice (dashed line).}
  \label{fig:bethe}
\end{figure}

As in reference to the data discussed in Sec.~\ref{sec:hubb_dmft} of the main text, calculations in the low temperature regime ($T \lesssim 0.01$) become significantly hard, and, there, Hirsch-Fye and ED algorithms become less performant. For these temperature points, hence, CT-QMC calculations have been performed. 
Fig.~\ref{fig:lowT} shows the first seven divergence lines in the low temperature regime. The results for the first two divergence lines indicate that the $T\rightarrow{0}$ extrapolation
yields $\widetilde{U}(T=0)\approx 1.45$ and $\approx 1.95$, respectively.
\\\\
At closer inspection, one also notes a
slight bending of the lines towards weak-coupling for the lowest
temperatures considered ($T < 0.01$). The latter feature, however, is likely to be ascribed to the 
van Hove singularity of the non-interacting density of states in the
two-dimensional Hubbard model, which enters the DMFT self-consistency
cycle. Its low-$T$ influence on the
divergence lines can be inferred by comparing with analogous calculations
performed on a Bethe lattice with the same second moment of the DOS (D = 1), where this ``backbending'' feature
disappears and the lines follow a straight almost vertical path towards $T\rightarrow 0$ (see Fig.~\ref{fig:bethe}).



\section{Falicov-Kimball model}
\label{app:FK}The aim of this Appendix is to derive Eq. (\ref{eq:DiagonalVertexFK}) starting from Eq. (\ref{equ:fkderiv})

We have to perform the functional derivative appearing in Eq. (\ref{equ:fkderiv}) using expressions for $\Sigma$ and $G$
which are valid out of half-filling. In term of the pseudospin energy $\epsilon=\frac{U}{2}(2 n_\downarrow-1)$
where $\downarrow$ are the immobile particles we get

\begin{equation}
G=\frac{G^{-1}_0+\epsilon}{G^{-2}_0-\frac{U^2}{4}}
\label{equ:AppFKGSigma}
\end{equation}
and
\begin{equation}
\Sigma[G_0]=\frac{\frac{U^2}{4}G_0+\epsilon}{1+\epsilon G_0}
\label{equ:AppFKFKSigmaG0}
\end{equation}
which reduces to Eqs. (\ref{eq:FKGSigma},\ref{eq:FKSigmaG0}) respectively at half-filling ($\epsilon\rightarrow 0$).
Notice that out of half-filling $\Sigma$ is still unique functional of $G_0$. 
Eliminating $G_0$ from Eqs. (\ref{equ:AppFKGSigma},\ref{equ:AppFKFKSigmaG0}) we get
\begin{equation}
\Sigma^{\pm}[G]= \frac{\pm\sqrt{1+U^2G^2+4\epsilon G}-1}{2G}.
\label{eq:AppFKFKSigmaG}
\end{equation}
Again even out of half-filling we have two determination for the self-energy as a functional of $G$.

The explicit functional
dependence of the parameter $\epsilon$ on $G_0$ is known in the grand-canonical ensemble \cite{FreericksRMP}.
Using our notation with $\mu=0$ meaning half-filling, from Eqs. (24-28) of Ref. \onlinecite{FreericksRMP} we get
\begin{equation}
\epsilon=\frac{U}{2}\frac{Z_0(\mu+U/2)-e^{-\beta E}Z_0(\mu-U/2)}{Z_0(\mu+U/2)+e^{-\beta E}Z_0(\mu-U/2)}
\label{equ:AppFKeps}
\end{equation}
where $E$ is the difference between the energy level of the immobile electrons and the chemical potential and
\begin{equation}
Z_0(\mu)=2e^{\beta\mu/2}\Pi_\nu\frac{G^{-1}_0(\nu)}{i\nu}
\label{equ:AppFKZ0}
\end{equation}
and $\nu=(2n+1)\pi/\beta$ is a fermionic Matsubara frequency ($n=0,\pm 1,\pm 2 ...$).

From Eqs. (\ref{eq:AppFKFKSigmaG})-(\ref{equ:AppFKZ0}) we notice that 
$\Sigma^{\pm}(\nu)$ is a non-local  functional of $G(\nu^\prime)$ (in frequency) and the non-locality is due to the 
$\epsilon$ term which can be understood as a functional of $G(\nu^\prime)$ trough the relation 
between $G$ and $G_0$.

Let us rewrite the $\Omega=0$ part of the vertex function from Eq. (\ref{equ:fkderiv})
\begin{equation}
\Gamma_{c,\pm}^{\nu\nu'(\Omega=0)} = \delta_{\nu\nu'} \Gamma^{\pm}_{BM}(\nu)+
\beta \frac{\partial \Sigma^{\pm}(\nu)}{\partial \epsilon } \frac{\partial  \epsilon}{\partial G(\nu^\prime)}.
 \label{equ:AppFKfkderiv}
\end{equation}
where
\begin{equation}
 \label{equ:AppFKGammaBM}
\Gamma^{\pm}_{BM}(\nu)=\beta\frac{R(\nu)\mp 1}{2G^2(\nu)R(\nu)}
\end{equation}
and $R(\nu)=\sqrt{1+U^2G^2(\nu)}$. In Eq. (\ref{equ:AppFKfkderiv}) the partial derivative
of $\Sigma^{\pm}$ w.r.t. $\epsilon$ should be taken 

Let us evaluate the two factors appearing in the last summand of Eq. (\ref{equ:AppFKfkderiv}).
From Eq. (\ref{equ:AppFKeps}) we get
\begin{equation}
\frac{\partial \Sigma^{\pm}(\nu)}{\partial \epsilon }=\pm \frac{1}{R(\nu)}.
\end{equation}

Having in mind that $\epsilon$ is a known functional of $G_0$ through Eqs. (\ref{equ:AppFKeps},\ref{equ:AppFKZ0}),
we express
\begin{equation}
 \frac{\partial  \epsilon }{\delta G(\nu)}=\sum_{\nu^\prime} \frac{\partial  \epsilon }{\delta G^{-1}_0(\nu^\prime)} \frac{\partial  G^{-1}_0(\nu^\prime) }{\partial G(\nu)}.
\end{equation}
From the definition of $G_0$ we resort to $G$ and $\Sigma$ obtaining

\begin{align}
\frac{\partial  G^{-1}_0(\nu^\prime) }{\delta G(\nu)}&=
\delta_{\nu,\nu^\prime}(-\frac{1}{G^2(\nu^\prime)}+\Gamma^{\pm}_{BM}(\nu^\prime)/\beta)+\nonumber\\
&+\frac{\partial \Sigma(\nu^\prime)}{\partial \epsilon}\frac{\partial \epsilon}{\partial G(\nu)}
\end{align}
therefore
\begin{equation}
 \frac{\partial  \epsilon }{\delta G(\nu)}=C^{\pm}(-\frac{1}{G^2(\nu)}+\Gamma^{\pm}_{BM}(\nu)/\beta)\frac{\partial \epsilon}{\partial G^{-1}_0(\nu)}.
\label{equ:AppFKdepsdG}
\end{equation}
where
\begin{equation}
C^{\pm}=\frac{1}{1-
	\sum_{\nu^\prime}\frac{\partial \epsilon}{\partial G^{-1}_0(\nu^\prime)} \frac{\partial \Sigma(\nu^\prime)} {\partial \epsilon}}.
\end{equation}

Now we evaluate the last factor in Eq. (\ref{equ:AppFKdepsdG}) using Eqs. (\ref{equ:AppFKeps},\ref{equ:AppFKZ0}).
By setting $\epsilon=U(A-B)/(2(A+B))$ with $A=Z_0(\mu+U/2)$ and $B=e^{-\beta E}Z_0(\mu-U/2)$
we get
\begin{equation}
\delta \epsilon = U\frac{AB}{(A+B)^2}\delta \log (\frac{A}{B}).
\label{equ:AppFKdeltaeps}
\end{equation}
In terms of $G^{-1}_0$ Eq. (\ref{equ:AppFKdeltaeps}) reads
\begin{equation}
\frac{\partial \epsilon}{\partial G^{-1}_0(\nu)}=\frac{U}{4}(1-\frac{4\epsilon^2}{U^2})\frac{\partial }{\partial G^{-1}_0(\nu)} \log \left(e^{\beta E}\frac{Z_0(\mu+U/2)}{Z_0(\mu+U/2)} \right).
\label{equ:AppFKdepsdG01}
\end{equation}
Using Eq. (\ref{equ:AppFKZ0}) we get
\begin{equation}
\frac{\partial \epsilon}{\partial G^{-1}_0(\nu)}=-\frac{U^2}{4}(1-\frac{4\epsilon^2}{U^2})\frac{1}{G^{-2}_0(\nu)-U^2/4}
\label{equ:AppFKdepsdG03}
\end{equation}
At half filling ($\epsilon=0$) taking into account the explicit form of $\Sigma$ Eq. (\ref{equ:AppFKFKSigmaG0}) we obtain 
the following simple expression
\begin{equation}
\frac{\partial \epsilon}{\partial G^{-1}_0(\nu)}=-\Sigma(\nu) G(\nu)
\label{equ:AppFKdepsdG02}
\end{equation}

Now we use Eqs. (\ref{equ:AppFKdepsdG02},\ref{equ:AppFKdepsdG},\ref{equ:AppFKfkderiv}) 
together with Eqs. (\ref{eq:AppFKFKSigmaG},\ref{equ:AppFKGammaBM}) to get
\begin{equation}
\Gamma_{c,\pm}^{\nu\nu'(\Omega=0)} = \delta_{\nu\nu'} \Gamma^{\pm}_{BM}(\nu)+
\beta \frac{U^2}{4} \frac{C^{\pm}}{R(\nu)R(\nu')}
 \label{equ:AppFKGammaC}
\end{equation}
which is Eq. (\ref{eq:DiagonalVertexFK}) of the main text.
An important fact that should be stressed is that the relations obtained in this appendix 
{\it are not lattice specific}. The specific form of the lattice enters in the self-consistency equation that determines the 
Green's function entering in Eq. (\ref{eq:DiagonalVertexFK}).

\section{Analytic determination of energy scales in the BM (and FK) model(s)}
\label{sec:scaleanalytic}

In this section we present details about the analytic derivation of the energy scales $\nu^*(W)$ and $\omega^*(W)$ in the BM (and the FK) model(s) on a Bethe lattice. This is achieved by combining the general definition of the one-particle Green's function for these models in Eq.~(\ref{eq:FKGG0}) [and Eq.~(\ref{eq:FKG})] with the corresponding DMFT self-consistency condition for the Bethe lattice in Eq.~(\ref{eq:FKselfcons2}) which yields
\begin{equation}
 \label{equ:betheG}
 G(z)=\frac{1}{2}\left(\frac{1}{z-t^2G(z)-\frac{W}{2}}+\frac{1}{z-t^2G(z)+\frac{W}{2}}\right).
\end{equation}
For the results presented in this paper we set $t=1/2$ as discussed in Sec. \ref{sec:2a}. In principle, Eq.~(\ref{equ:betheG}) represents a third-order equation for $G(z)$ and could be, hence, solved analytically. However, as the resulting expressions are cumbersome, we will instead use this relation directly in its explicit form given above, in order to extract the relevant energy scales on the real and on the Matsubara axis.

Originally, the energy scale $\nu^*(W)$ on the Matsubara axis has been derived from the requirement Eq.~(\ref{eq:FKdivergence}) for the divergence of the irreducible vertex. However, as we discuss in Sec.~\ref{sec:multivaluedreal}, the general condition for the existence of such a scale in the BM is that the argument of the square root in Eq.~(\ref{eq:FKSigmaG}),~i.e. $1+W^2G^2$, crosses the negative imaginary axis (representing the branch cut of the two-valued square root function), i.e., $\text{Im}[1+W^2G^2]=0$ and $\text{Re}[1+W^2G^2]\le 0$, see Eqs. (\ref{subequ:branchcondreal}). Below, we will present all our analytical derivations for $\nu^*(W)$ and $\omega^*(W)$ from this more general starting point, highlighting all possible mathematical and physical conditions for the existence of such scales as well as their implications in the most transparent way. In the following we will adopt the notation introduced in Sec. \ref{sec:multivaluedreal}, i.e., $G(z)=G'(z)+iG''(z)$ where $G(z)$ denotes the real and $G''(z)$ the imaginary part of $G(z)$.    

\subsection{\texorpdfstring{Energy scale $\nu^*(W)$ on the Matsubara axis}{Energy scale on the Matsubara axis}}
\label{sec:scaleimag}

On the Matsubara axis at half-filling the Green's function is purely imaginary, i.e., $G(i\nu)\equiv iG''(i\nu)$, and anti-symmetric, i.e., $G(-i\nu)\!=\!-G(i\nu)$. Hence, we can restrict our discussion to the case $\nu>0$ where $G''(i\nu)\!<\!0$. From $G'(i\nu)\!\equiv\!0$ it immediately follows that $\text{Im}[1+W^2G^2]\!=\!0$, while the second condition for the physical self-energy changing from $\Sigma^+$ to $\Sigma^-$, $\text{Re}[1+W^2G^2]\le 0$, has to be further investigated.

In the main text we have determined the energy scale by the requirement $1+W^2G^2=0$ (which, in fact, implies a vertex divergence). This implicitly corresponds to the assumption that the expression $1+W^2G^2$ crosses the negative imaginary axis [i.e., the branch cut of the square root function in Eq.~(\ref{eq:FKSigmaG})] {\sl only} at the origin of the complex plane. This is equivalent to the property 
\begin{equation}
 \label{equ:condgimag}
 1+W^2G^2=1-W^2G''^2\ge 0,
\end{equation}
of the (DMFT solution of the) Matsubara Green's function of the BM. In the following we will prove at Eq.~(\ref{equ:condgimag}) is indeed correct for all $\nu$.

Eq.~(\ref{equ:condgimag}) is equivalent to the condition $\lvert G(i\nu)\rvert \le 1/W$. To demonstrate the validity of the latter relation, let us first note that $\left|G(i\nu)\right|$ decays as $1/\nu$ for $\nu\rightarrow\infty$. Hence, either (i) $\lvert G(i\nu)\rvert$ is a strictly monotonous function and assumes its maximum value at $\nu=0$, or (ii) $\left|G(i\nu)\right|$ takes its maximum at some finite value $\bar{\nu}$, i.e., $dG(i\nu)/d\nu\lvert_{\nu=\bar{\nu}}=0$. 

For (i) we just have to set $\nu=0$ in Eq.~(\ref{equ:betheG}). This gives us a third order equation in $G$ with the three solutions $G_1\equiv 0$ and $G_{2,3}=\mp i\sqrt{t^2-W^2/4}/t^2$ where we will consider only the $-$ sign (since $\nu>0$). For $W<W_{\text{MIT}}=2t$ ($=1$ in our units) the physical solution is given by $G_2$, while for $W>W_{\text{MIT}}=$ $G_2$ would become real and, hence, $G_1=0$ represents the physical solution in this case. This is consistent with the vanishing of the Matsubara Green's function (at $T=0$) in the Mott insulating phase at zero (Matsubara) frequency. Now it is straightforward to show\cite{footnote4} that $\lvert G_2 \rvert  \le 1/W$ for $W<W_{\text{MIT}}$ which implies $\lvert G(i\nu)\rvert<1/W$ for all $\nu$ as $G$ has been assumed to be monotonous.  

(ii) In order to find out whether $\lvert G(i\nu)\rvert$ exhibits a local maximum which might possibly lead to a violation of Eq.~(\ref{equ:condgimag}) we differentiate Eq.~(\ref{equ:betheG}) with respect to $\nu$ and set $dG/d\nu=0$ in the resulting expression. This yields straightforwardly the condition
\begin{equation}
 \label{equ:condGimagmax}
 i\nu-t^2G=i\frac{W}{2}.
\end{equation}
Reinserting Eq.~(\ref{equ:condGimagmax}) in the r.h.s. of Eq.~(\ref{equ:betheG}) yields $G=-i/W$, which represents then the maximal value of $\lvert G \rvert$ in the entire frequency regime (note that there is found only one stationary point). Hence, also in this case $\lvert G(i\nu)\rvert<1/W$ for all $\nu$, which completes our proof.

The above discussion implies that the condition for changing the physical branch of the self-energy from $\Sigma^+$ to $\Sigma^-$ is indeed given by $1+W^2G^2=0$. Moreover, the latter relation is obviously fulfilled at the point where $\lvert G \rvert$ takes its maximal value $i/W$. The corresponding frequency at which this happens can be easily obtained by inserting this value for $G$ into Eq.~(\ref{equ:condGimagmax}) which yields
\begin{equation}
 \label{equ:matcond}
 \nu=\frac{2W^2-4t^2}{4W}\overset{t=\frac{1}{2}}{=}\frac{2W^2-1}{4W}=:\nu^*(W),
\end{equation}
which is equivalent to Eq.~(\ref{equ:scale}) in the main text. Let us recall, that, as this equation has been derived from $G=-i/W$, it is valid only for $\nu>0$. Hence, $\nu^*(W)<0$ indicates the vanishing of the scale which occurs at $W=\widetilde{W}=\sqrt{2}t(=1/\sqrt{2}$). This, however, also means that, for $W<\widetilde{W}$, $\lvert G(i\nu)\rvert$ is monotonous while above this threshold it exhibits the maximum $1/W$ at $\nu=\nu^{*}(W)$.

We can now sum up the above results in the following way: For $W>\widetilde{W}$ there exists an energy scale $\nu^*(W)$ at which the physical self-energy switches from $\Sigma^+$ to $\Sigma^-$ in Eq.~(\ref{eq:FKSigmaG}) as it is discussed in the main text. Moreover, at exactly this frequency $\lvert G(i\nu)\rvert$ exhibits a local maximum. Finally, also the physical irreducible vertex $\Gamma_c^{\nu\nu(\omega=0)}$ changes at $\nu^*(W)$ from $\Gamma^+$ to $\Gamma^-$. This happens via a divergence, since the changing of the branches occurs exclusively for $1+W^2G^2=0$. Let us also mention that a further consequence of this condition is that the change of branches of the self-energy occurs continuously on the Matsubara axis.
 
For $W<\widetilde{W}$ instead the scale $\nu^*(W)$ vanishes and, hence, the physical $\Sigma$ is defined by $\Sigma^+$ for all frequencies, $G(\nu)$ becomes (for $\nu>0$ and $\nu<0$) a monotonous function and no vertex divergences are observed.

\subsection{\texorpdfstring{Energy scale $\omega^*(W)$ on the real axis}{Energy scale on the real axis}}
\label{sec:scalereal}

On the real frequency axis neither the real nor the imaginary part of the Green's function $G(\omega)$ are identically $0$. Hence, the first condition Eq.~(\ref{equ:branchcondrealim}) for changing the physical self-energy from $\Sigma^+$ to $\Sigma^-$ is not automatically fulfilled. In fact, for the fulfillment of this relation either $G'=0$ or $G''=0$ should hold. The latter, however, leads to $1+W^2G^2=1+W^2G'^2>0$, which violates the second condition Eq.~(\ref{equ:branchcondrealre}) for changing the physical $\Sigma$ from $\Sigma^+$ to $\Sigma^-$. Hence, we have to require $G'\equiv 0$ at the frequency $\omega^*(W)$ where the change of branches occurs. To determine the value of $\omega^*(W)$ we, hence, replace $z\rightarrow \omega(=\omega^*)$ and $G=iG''$ in Eq.~(\ref{equ:betheG}). Then this complex relation represents actually a set of {\sl two} independent equations, i.e., one for its real and one for its imaginary part, which can be solved for the two variables $G''$ and $\omega=\omega^{*}$ as a function of $W$ (and $t$). The real part of the equation yields
\begin{equation}
 \label{equ:condrealaxisre}
 (t^2G'')^2=-\omega^2+\frac{W^2}{4}.
\end{equation} 
As $\omega^2>0$ this sets already an upper limit for the scale $\omega^*(W)$, i.e., $\lvert\omega^{*}\rvert\le W/2$. In the imaginary part of Eq.~(\ref{equ:betheG}) we can now express $G''$ through $\omega$ and $W$ via Eq.~(\ref{equ:condrealaxisre}). This yields the energy scale
\begin{equation}
 \label{equ:scalerealaxis}
 \omega^*(W)=\frac{1}{2}\sqrt{W^2-2t^2}\overset{t=\frac{1}{2}}{=}\frac{1}{2}\sqrt{\frac{2W^2-1}{2}}.
\end{equation}
Like the scale $\nu^*(W)$ on the imaginary axis, $\omega^*(W)$ vanishes at $W=\widetilde{W}=\sqrt{2}t(=1/\sqrt{2})$. 

By reinserting the expression for $\omega^*$ from Eq.~(\ref{equ:scalerealaxis}) into Eq.~(\ref{equ:condrealaxisre}) $G''(\omega^*)$ is determined as $G''(\omega^*)=-\frac{1}{\sqrt{2}t}(=-\sqrt{2})$. Remarkably, the value of $G''(\omega^*)$ is independent of $W$ and, considering that $G'(\omega^*)=0$, this $W$-independence holds for the full $G(\omega^*)$ as long as $W>\widetilde{W}$.

Inserting the expression for $G(\omega^*)$ into the condition for the changing of the branches of $\Sigma$ we obtain $1+W^2G^2(\omega^*)=1-\frac{W^2}{2t^2}=1-2U^2$. Obviously, for $U>\sqrt{2}t$ ($U>1/\sqrt{2}$), $1+W^2G^2(\omega^*)<0$, i.e. the branch cut of the square root function in Eq.~(\ref{eq:FKSigmaG}) is crossed at some point on the negative real axis of the complex plane (excluding $0$) rather than at the origin, as it was always the case for the corresponding expression on the Matsubara axis. This observation has two consequences: (i) It leads to a discontinuity of the Re$[\Sigma^+(\omega)]$ and Re$[\Sigma^-(\omega)]$, i.e. the (real parts of the) two solutions do not cross at one point (as it was the case for $\Sigma^+$ and $\Sigma^-$ on the Matsubara axis, see Figs. \ref{fig:FKsigma} and \ref{fig:FKsigmaRe}). (ii) The change of branches of the physical vertex function from $\Gamma^+$ to $\Gamma^-$ does {\sl not} occur via a divergence as the denominator in Eq.~(\ref{eq:FKGamma}) does not become $0$.

The only exception to the above described scenario occurs exactly at the point $W=\widetilde{W}$. Here, $1+W^2G^2(\omega^*=0)\equiv 0$ and, therefore, the situation is completely analogous to the behavior on the imaginary axis: the two solutions for $\Sigma^+$ and $\Sigma^-$ cross at the single point $\omega^*=0$ and the vertex diverges. Hence, on the real axis, a vertex divergence can be observed in the BM only for $W=\widetilde{W}$ at $\omega^*=0$ as it has been already stated in Ref. \onlinecite{Janis2014}.

\section{Disordered models}
\label{app:Anderson}
\begin{figure}[ht!]
  \includegraphics[width=5cm]{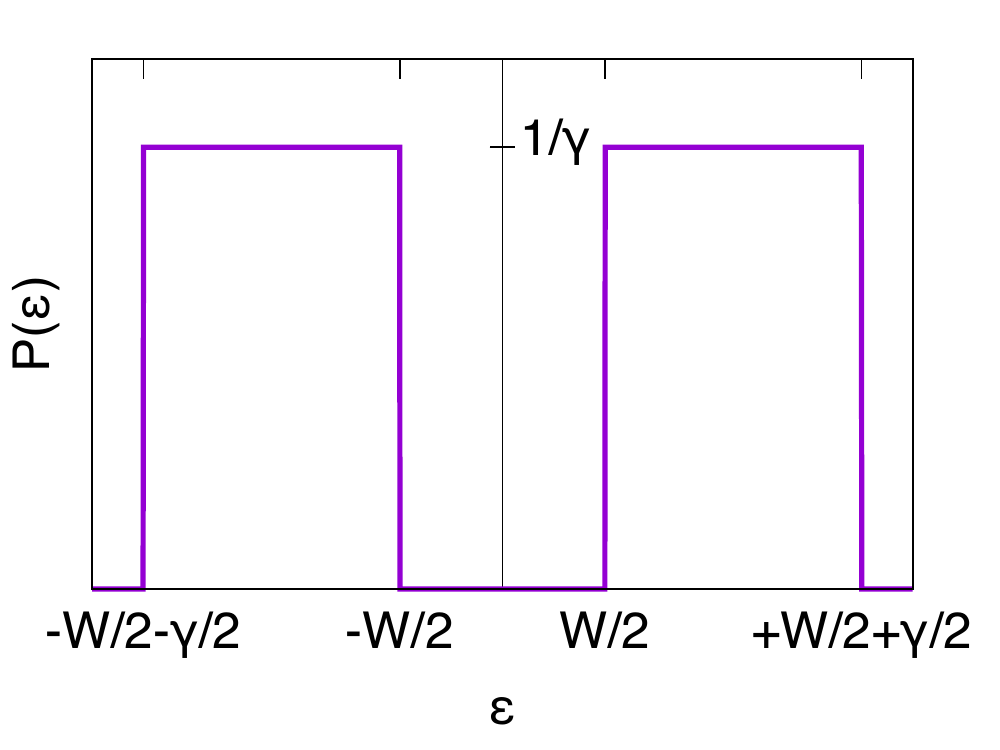}
  \caption{Distributed disorder with gap $W$.}
  \label{fig:2boxes}
\end{figure}

Here we consider models of distributed disorder for which the disorder variables $\epsilon_i$ in Eq.~(\ref{eqn:FKhubb})
are distributed according to the bimodal distribution $P(\epsilon) \ne 0$ (and constant) only when $W^2/4<\epsilon^2<(W+\gamma)^2/4$ as depicted in Fig. \ref{fig:2boxes}.

This distribution has the following relevant limits

i) The binary mixture discussed in section \ref{sec:fk}\\as $\gamma\rightarrow 0$
 
ii) The uniform distribution as $W\rightarrow 0$.

The local Green's function can be calculated as:
\begin{equation}
G = \int d\epsilon P(\epsilon) \frac{1}{G^{-1}_0-\epsilon}
\end{equation}
where $G_0$ is the same as defined below Eq.~(\ref{eq:FKGG0}).
Performing the integral we get
\begin{equation}
G = \frac{1}{\gamma} \log \left ( \frac{G^{-1}_0-W/2}{G^{-1}_0+W/2} \frac{G^{-1}_0+W/2+\gamma/2}{G^{-1}_0-W/2-\gamma/2} \right )
\label{eq:appGloc}
\end{equation}
By inverting the previous equation we get $G^{-1}_0$ as a function of $G$
\begin{equation}
G_0 = -\frac{1}{W(1+W/\gamma)\tanh (\gamma G/2)} D_{\pm}(G)
\label{eq:appG0}
\end{equation}
where
\begin{equation}
D_\pm[G]=  \left ( 1 \mp \sqrt{1+4\frac{W}{\gamma}(1+\frac{W}{\gamma})\tanh ^2 (\gamma G/2)}\right ).
\label{eq:appDefD}
\end{equation}
From $\Sigma=G^{-1}_0-G^{-1}$ we get
\begin{equation}
\Sigma^{\pm}[G] =-G^{-1}-W(1+W/\gamma)\tanh (\gamma G/2)/D_\pm(G).
\label{eq:appSigma}
\end{equation}
It is interesting to explicitly consider in Eq.~(\ref{eq:appSigma}) the two aforementioned limits:

(i) The BM limit $\gamma\rightarrow 0$. We must expand Eq.~(\ref{eq:appSigma}) to the first order in $\gamma$
obtaining Eq.~(\ref{eq:FKSigmaG})
\begin{equation}
\Sigma^{\pm}[G] = -\frac{1}{2G} (\pm \sqrt{1+W^2G^2}-1).
\label{eq:appSigmaBM}
\end{equation}
In this limit it follows from Eq.~(\ref{eq:appGloc}) that $\Sigma$ as a function of $G_0$ has a single determination
\begin{equation}
\Sigma[G_0] = \frac{W^2}{2} G_0
\label{eq:appSigmaG0BM}
\end{equation}
as in Eq.~(\ref{eq:FKSigmaG0}).

(ii) The uniform distribution (Anderson model) as $W\rightarrow 0$ gives instead  single determination for $\Sigma$
as a function of $G$ (the $+$-branch of $D_\pm$ of Eq. (\ref{eq:appDefD}) should be taken in this case):
\begin{equation}
\Sigma[G] = -G^{-1}+\frac{\gamma}{2\tanh (\gamma G/2)}.
\label{eq:appSigmaAnd1}
\end{equation}
Instead from Eq.~(\ref{eq:appGloc}) $\Sigma$ as a function of $G_0$ is a non-analytic function
\begin{equation}
\Sigma[G_0] = G^{-1}_0+\gamma{\log^{-1} \left(\frac{G^{-1}_0-\gamma/2}{G^{-1}_0+\gamma/2}\right)}.
\label{eq:appSigmaAnd2}
\end{equation} 

We thus see that in order to have multiple solutions for the self-energy (as a functional of $G$) 
we must have a gapped disorder distribution.
At the same time, a gapped disorder distribution is also necessary to give a Mott transition within CPA.
To study the Mott transition and the vertex divergence within this disordered model we specialize the 
previous equation to the Bethe lattice. In this case $G^{-1}_0=i\nu - G(i\nu)/4$ 
(the half bandwidth is our energy unit).
Due to the particle-hole symmetry $G(\nu)=iG''(\nu)$, which vanishes near the MIT. Thus, we can expand Eqs. (\ref{eq:appG0}) and (\ref{eq:appDefD}) yielding
\begin{equation}
\frac{1}{4}G'' = \frac{W\gamma}{2}(1+W/\gamma)\frac{G''}{1\mp[1-W(1+W/\gamma )\frac{\gamma^2 G''^2}{2})]}.
\label{eq:MIT1}
\end{equation}
Taking the $+$-sign in Eq.~(\ref{eq:MIT1}) gives a finite value $G''=-2/\sqrt{\gamma}$, therefore, to obtain a MIT, we must choose the $-$-sign, 
which corresponds to $\Sigma^{+}$ in Eq.~(\ref{eq:appSigma}). In this case a MIT is possible ($G''=0$) if $\gamma=\gamma_{\text{MIT}}$ where
\begin{equation}
\gamma_{\text{MIT}}=\frac{1-W^2}{W}.
\label{eq:appgammaMIT}
\end{equation}

Now we calculate the equal frequency vertex function $\Gamma_{c,\pm}^{\nu\nu'(\Omega=0)}=\beta\delta_{\nu\nu'} \Gamma_D(\nu)$ 
where the diagonal part is defined as $\Gamma_D(\nu)=\frac{d \Sigma(\nu)}{d G(\nu)}$. From Eq.~(\ref{eq:appSigma})
we get 
\begin{eqnarray}
\Gamma^{\pm}_D&=&G^{-2}-W\gamma(1+W/\gamma)\frac{1}{2\cosh^2 (\gamma G/2) D_\pm(G)}+\nonumber\\
&+&W(1+W/\gamma)\tanh (\gamma G/2)\frac{D^\prime_\pm(G)}{D^2_\pm(G)}.
\label{eq:appGamma1}
\end{eqnarray}
\begin{figure}[t!]
  \includegraphics[width=5cm]{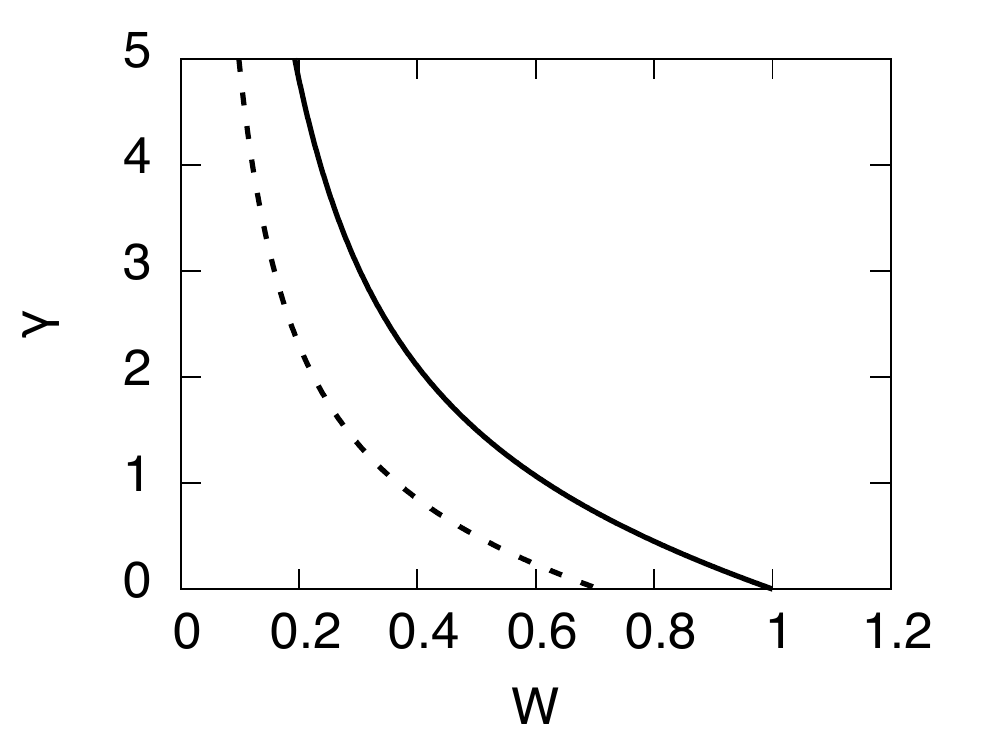}
  \caption{Phase diagram of the disorder model. Dashed line: vertex divergence, solid line: MIT.}
  \label{fig:gammaWPD}
\end{figure}
The divergence of $\Gamma^{\pm}_D$ is due to the term in the second line of Eq.~(\ref{eq:appGamma1}) and occurs
when the derivative of $D_\pm$, $D^\prime_\pm(G)$ diverges. Such a divergence occurs when the 
term in the square-root of Eq.~(\ref{eq:appDefD}) vanishes at zero frequency.
Then the divergence of the equal frequency vertex necessarily occurs when the two solutions $\Sigma^{\pm}$
coincide (at zero frequency). The way in which this occurs is explained in details in the main text, here we determine the critical $\gamma_c$ for the Bethe lattice case.
Taking the $G_0$ appropriate for the Bethe lattice, from Eq.~(\ref{eq:appG0}) when the two solutions merge we have 
\begin{equation}
\frac{1}{4}G'' = W(1+W/\gamma) \tan (\gamma G''/2).
\end{equation}
This equation can be satisfied with a non-zero imaginary part of the Green's function provided that
$\gamma>\gamma_c$ where
\begin{equation}
\gamma_c=\frac{1/2-W^2}{W}.
\label{eq:appgammac}
\end{equation}

The lines of the MIT [Eq.~(\ref{eq:appgammaMIT})] and of vertex divergence [Eq.~(\ref{eq:appgammac})] are plotted
in figure \ref{fig:gammaWPD}. The vertex divergence is always a precursor of the MIT and does not occur at $W=0$
where the MIT is absent.


\begin{references}
\bibitem{ppsusc} A physical susceptibility is usually defined in such a way that it describes fluctuations of a corresponding observable {\sl around} its average (expectation) value. That is why in the third line of Eq.~(\ref{eqn:ph}) a product of non-interacting Green's function is subtracted. For the particle-particle channel the expectation value of the physical observable, i.e., the pairing amplitude $c_\uparrow c_\downarrow +c^{\dagger}_\downarrow c^{\dagger}_\uparrow$, is $0$ in the normal phase. Hence, the subtraction in Eq.~(\ref{eqn:ph}) has to be canceled which is achieved by subtracting the bare susceptibility in Eq.~(\ref{equ:channeldefpp}).
\bibitem{noteMarkus} In the generic case (e.g., out-of-half-filling), where the matrix $\chi$ becomes non-Hermitian in the fermionic Matsubara frequencies, the existence of a spectral representation is not guaranteed. In this situation, however, similar arguments can be also made by exploiting the singular value decomposition of $\chi$ rather than its spectral representation in terms of its eigenvectors.
\bibitem{note1} As we are at half-filling $G$ is a purely imaginary function of the Matsubara frequencies and $G(-i\nu)=-G(i\nu)$. Hence, according to Eq.~(\ref{eq:FKGamma}) if $\Gamma_{c,\pm}$ diverges for a given positive Matsubara frequency $\nu$ it also does for the corresponding negative one $-\nu$.
\bibitem{footnote3}
For a function $f(x)$ with $f(x^{*})=x^{*}$, the fixed point $x^{*}$ is stable (attractive) if $\left|f'(x^{*})\right|<1$ and unstable (repulsive) if $\left|f'(x^{*})\right|>1$.
\bibitem{footnote2}{In these works, however, no ambiguities were reported, because the irreducible vertex was always directly expressed in terms of the physical self-energy. In this respect, Eq.~(\ref{eq:DiagonalVertexFK}) should be regarded more as a symbolic notation for the calculation of $\Gamma_c$, to be interpreted along the lines discussed above.}
\bibitem{chi0} In order to increase the numerical stability of the calculations, the singular eigenvector of $(\chi_{\text{c}}/\chi_{0})^{\nu\nu'\Omega=0}$ is plotted. This, however,
 is always very similar to the one of $\chi_{\text{c}}$ and even coincides with it in the atomic limit.
\bibitem{footnote4}
For this consider that $0 \le \left(\frac{W}{2t^{2}}-\frac{1}{W}\right)^{2}$ which yields $\frac{1}{t^{2}}-\frac{W^{2}}{4t^{4}} \le \frac{1}{W^{2}}$ and, hence, $\frac{1}{t^{2}}\sqrt{t^{2}-\frac{W^{2}}{4}}=\left|G_{2}\right|\le\frac{1}{W}$.
\end{references}
\end{document}